\documentclass[iop,8pt]{emulateapj}
\slugcomment{{\sc Accepted to ApJ:} October 11, 2017}
\usepackage{scrextend} 

\usepackage{natbib}
\usepackage{color}

\usepackage{amsmath,amssymb}  
\makeatletter
\newsavebox\myboxA
\newsavebox\myboxB
\newlength\mylenA

\newcommand{\mime}{m_{\rm i}/m_{\rm e}}
\newcommand{\teti}{T_{\rm e}/T_{\rm i}}
\newcommand{\betai}{\beta_{\rm i}}
\newcommand{\comp}{c/\omega_{\rm pe}}

\newcommand*\xoverline[2][0.75]{%
    \sbox{\myboxA}{$\m@th#2$}%
    \setbox\myboxB\null
    \ht\myboxB=\ht\myboxA%
    \dp\myboxB=\dp\myboxA%
    \wd\myboxB=#1\wd\myboxA
    \sbox\myboxB{$\m@th\overline{\copy\myboxB}$}
    \setlength\mylenA{\the\wd\myboxA}
    \addtolength\mylenA{-\the\wd\myboxB}%
    \ifdim\wd\myboxB<\wd\myboxA%
       \rlap{\hskip 0.5\mylenA\usebox\myboxB}{\usebox\myboxA}%
    \else
        \hskip -0.5\mylenA\rlap{\usebox\myboxA}{\hskip 0.5\mylenA\usebox\myboxB}%
    \fi}
\makeatother
\usepackage{csvsimple}
\usepackage{booktabs}
\usepackage{longtable}
\usepackage{color}
\DeclareFontFamily{OT1}{pzc}{}
\DeclareFontShape{OT1}{pzc}{m}{it}{<-> s * [1.10] pzcmi7t}{}
\DeclareMathAlphabet{\mathpzc}{OT1}{pzc}{m}{it}
\usepackage{graphicx}   
\usepackage{bm}
\usepackage{verbatim}   
\usepackage{color}         
\usepackage{subfigure}  
\usepackage{xcolor}
\colorlet{linkequation}{blue}
\usepackage[colorlinks]{hyperref}    
\hypersetup{
	colorlinks = true,
	linkcolor = {blue},
	citecolor = {blue}
	}
	\newcommand*{\SavedEqref}{}
		\let\SavedEqref\eqref
		\renewcommand*{\eqref}[1]{%
  		\begingroup
    		\hypersetup{
      			linkcolor=linkequation,
      			linkbordercolor=linkequation,
    			}%
    		\SavedEqref{#1}%
  	\endgroup
	}

\usepackage{booktabs}
\newcommand{\ra}[1]{\renewcommand{\arraystretch}{#1}}
	
\begin{document}
\title{Electron and proton heating in trans-relativistic magnetic reconnection }
\author{Michael E. Rowan,$^1$ Lorenzo Sironi,$^2$ and Ramesh Narayan$^1$}
\affil{$^1$Harvard-Smithsonian Center for Astrophysics, 
60 Garden Street, Cambridge, MA 02138, USA
\\
$^2$Department of Astronomy, Columbia University, 550 W 120th St, New York, NY 10027, USA}
 \email{E-mail: michael.rowan@cfa.harvard.edu}
\begin{abstract} 
Hot collisionless accretion flows, such as the one in Sgr A$^{*}$ at our Galactic center, provide a unique setting for the investigation of magnetic reconnection. Here, protons are non-relativistic while electrons can be ultra-relativistic.
By means of two-dimensional particle-in-cell simulations, we investigate electron and proton heating in the outflows of trans-relativistic reconnection  (i.e.,  $\sigma_w\sim 0.1-1$, where the magnetization $\sigma_w$ is the ratio of magnetic energy density to enthalpy density). 
For both electrons and protons, we find that heating at high $\beta_{\rm i}$ (here, $\beta_{\rm i}$ is the ratio of proton thermal pressure to magnetic pressure) is dominated by adiabatic compression (``adiabatic heating''), while at  low $\beta_{\rm i}$ it is accompanied by a genuine increase in entropy (``irreversible heating'').  
For our fiducial $\sigma_w=0.1$, the irreversible heating efficiency at $\beta_{\rm i}\lesssim 1$ is nearly independent of the electron-to-proton temperature ratio $T_{\rm e}/T_{\rm i}$ (which we vary from $0.1$ up to $1$), and it asymptotes to $\sim 2\%$ of the inflowing magnetic energy in the low-$\beta_{\rm i}$ limit. 
Protons are heated more efficiently than electrons at low and moderate $\beta_{\rm i}$  (by a factor of $\sim7$), whereas the electron and proton heating efficiencies become comparable at $\beta_{\rm i}\sim 2$ if $T_{\rm e}/T_{\rm i}=1$, when both species start already relativistically hot. We find comparable heating efficiencies between the two species also in the limit of relativistic reconnection ($\sigma_w\gtrsim 1$).
Our results have important implications for the two-temperature nature of collisionless accretion flows, and may provide the sub-grid physics needed in general relativistic MHD simulations.
\end{abstract}


\maketitle

\section{Introduction} \label{sec:introduction}
The ultra-low-luminosity source at the center of the Milky Way, Sagittarius A$^{*}$ (Sgr A$^{*}$), is thought to be powered by accretion onto a supermassive black hole. 
Sgr A$^{*}$ radiates well below the Eddington limit and there is strong evidence that the accreting gas can be described as an advection-dominated accretion flow (ADAF, also referred to as a radiatively inefficient accretion flow, RIAF) \citep{Narayan1994,Narayan1995,Narayan1995a,Abramowicz1995,Narayan2008,Yuan2014}.
In ADAFs, the disk is geometrically thick and optically thin.  Additionally, the plasma is predicted to be two-temperature for several reasons:  
first, in the ADAF configuration, the density of accreting gas is low enough that Coulomb collisions between electrons and protons are extremely rare on accretion timescales, so that the species become thermally decoupled.  Second, electrons radiate more efficiently than protons.  Lastly, relativistic electrons are heated less than non-relativistic protons when subjected to the same adiabatic compression.  
For all these reasons, the plasma is expected to be two-temperature, with protons significantly hotter than electrons \citep{Narayan1995,Yuan2003}.  

Despite the above arguments, the two-temperature gas may be driven to a single-temperature state by kinetic processes, such as reconnection and instabilities \citep{Quataert2002,Riquelme2012,Riquelme2015,Sironi2015,Sironi2015a,Werner2016}.  
To capture the effects of these plasma processes, one requires a fully-kinetic description, which can be achieved via numerical techniques such as particle-in-cell (PIC) simulations.  
In principle, such \textit{ab initio} simulations can be used to provide the necessary sub-grid physics that, to date, cannot be captured in magnetohydrodynamic (MHD) simulations \citep[e.g.,][]{Ressler2015,Ressler2017,Ball2016,Ball2017, Chael2017, Sadowski2017}.

In supermassive black hole accretion flows, the ratio of ion thermal pressure to magnetic pressure, 
\begin{align}
\beta_{\rm i} = \frac{8 \pi n_0 k_{\rm B} T_{\rm i} }{B_{0}^{2}},
\end{align}
(where $n_0$ is the ion number density, $k_{\rm B}$ is Boltzmann's constant, $T_{\rm i}$ is the ion temperature, and $B_{0}$ is the magnitude of the magnetic field) 
 is expected to vary in the disk midplane in the range $\beta_{\rm i} \sim 10$ -- $30$ \citep[See Fig. 1 of ][]{Sadowski2013}.  However, in plasma far above and below the midplane, the ``corona,'' the system is expected to be magnetically dominated, such that $\beta_{\rm i}\lesssim 1$.
Here, the dissipation of magnetic energy via reconnection can result in particle heating, acceleration, and bulk motion.

Even in the magnetized corona, the magnetization,
\begin{align} \label{eq:sigmai}
\sigma_{\rm i} = \frac{B_{0} ^{2}}{4 \pi n_0 m_{\rm i} c^{2}},
\end{align}
is generally small, i.e.,~$\sigma_{\rm i} \lesssim~1$.
Electron heating by reconnection in the non-relativistic limit ($\sigma_{\rm i} \ll 1$)  has been studied extensively, both theoretically and by means of PIC simulations, in the context of the solar wind, Earth's magnetotail, and laboratory plasmas \citep{Hoshino2001,Jaroschek2004,Loureiro,Schoeffler2011,Schoeffler2013,Shay2014,Dahlin2014,Daughton2014,Li2015b, Haggerty2015, Numata2015, Le2016, Li2017}.  Though less commonly studied, relativistic reconnection (i.e., ~$\sigma_{\rm i} \gg~1$) in electron-proton plasmas has also received some attention in recent years \citep{Sironi2015c,Guo2016a}.  

The collisionless plasma in hot accretion flows around black holes provides a peculiar environment for reconnection, since $\sigma_{\rm i} \lesssim 1$, a regime that falls between the well-studied non-relativistic and ultra-relativistic regimes.  For $\beta_{\rm i}\sim 1$ and $\sigma_{\rm i} \lesssim 1$, protons are generally non-relativistic, yet electrons can be ultra-relativistic.  This territory remains largely unexplored, in terms of both simulation and theory, and studies have only recently begun to probe reconnection in this parameter regime \citep{Melzani2014,Werner2016}.

The aim of this work is to explore particle heating via magnetic reconnection in the trans-relativistic regime $\sigma_{\rm i} \lesssim 1.$ 
We study heating in the outflows of anti-parallel reconnection (i.e., in the absence of a guide field perpendicular to the alternating fields) by means of fully-kinetic PIC simulations, choosing inflow parameters appropriate for the coronae of collisionless accretion flows.  We present the electron and proton heating as a function of mass ratio (up to the physical value), inflow magnetization, ion plasma $\beta_{\rm i}$ and temperature ratio $T_{\rm e}/T_{\rm i}$. 

We show that heating in the high-$\beta_{\rm i}$ regime is primarily dominated by adiabatic compression (we shall call this contribution ``adiabatic heating''), while for low $\beta_{\rm i}$ the heating is genuine, in the sense that it is associated with an increase in entropy (``irreversible heating''). At our fiducial $\sigma_{\rm i}\sim 0.1$, we find that for $\beta_{\rm i}\lesssim 1$ the irreversible heating efficiency is independent of $T_{\rm e}/T_{\rm i}$ (which we vary from $0.1$ up to $1$). For equal electron and proton temperatures, the fraction of inflowing magnetic energy converted to electron irreversible heating at realistic mass ratios decreases from $\sim 1.6\%$ down to $\sim 0.2\%$ as $\beta_{\rm i}$ ranges from $\beta_{\rm i}\sim 10^{-2}$ up to $\beta_{\rm i}\sim 0.5$, but then it increases up to $\sim 3\%$ as $\beta_{\rm i}$ approaches $\sim2$. Protons are heated much more efficiently than electrons at low and moderate $\beta_{\rm i}$  (by a factor of $\sim7$), whereas the electron and proton heating efficiencies become comparable at $\beta_{\rm i}\sim 2$ if $T_{\rm e}/T_{\rm i}=1$, when both species start already relativistically hot. We find comparable heating efficiencies between the two species also in the limit of relativistic reconnection, when the magnetization exceeds unity. The unifying feature of these two cases (i.e., high magnetization, and high $\betai$ at low magnetization) is that the scale separation between electrons and protons in the reconnection outflows approaches unity, so the two species behave nearly the same. Motivated by our findings, we propose an empirical formula (Eq.~\ref{eq:fit}) that captures the magnetization and plasma-$\beta_{\rm i}$ dependence of the electron heating efficiency (normalized to the overall electron + proton heating efficiency) over the whole range of magnetization and $\beta_{\rm i}$ that we explore.

We also measure the inflow speed (i.e., the reconnection rate) as a function of the flow conditions, finding that for our fiducial magnetization $\sigma_w=0.1$ it decreases from $v_{\rm in}/v_{\rm A} \approx 0.08$ down to $0.04$ as $\beta_{\rm i}$ ranges from $\beta_{\rm i}\sim 10^{-2}$ up to $\beta_{\rm i}\sim 2$ (here, $v_{\rm A}$ is the Alfv\'en speed). Similarly, the outflow speed saturates at the Alfv\'{e}n velocity for low $\beta_{\rm i}$, but it decreases with increasing $\beta_{\rm i}$ down to $v_{\rm out}/v_{\rm A}\approx 0.7$ at $\beta_{\rm i}\sim2.$
The inflow (outflow, respectively) speed is independent of $T_{\rm e}/T_{\rm i}$ at low $\beta_{\rm i}$, with only a minor tendency for lower (higher, respectively) speeds at  larger $T_{\rm e}/T_{\rm i}$ in the high-$\beta_{\rm i}$ regime.

The organization of the paper is as follows.  
In Section \ref{sec:setup}, we provide details about the simulation setup and parameters.
In Section \ref{sec:technique}, we discuss our technique for extracting from PIC simulations the heating efficiencies.
In Section \ref{sec:results}, we discuss the dependence of the reconnection rate, the outflow speed and the electron and proton heating efficiencies on the flow conditions.
We conclude in Section \ref{sec:conclusion}, with a summary and discussion.
 
\section{Simulation setup}
\label{sec:setup}
\begin{figure}
		\centering
		\includegraphics[width=0.5\textwidth,clip,trim=0cm 0cm 0cm 0.6cm]{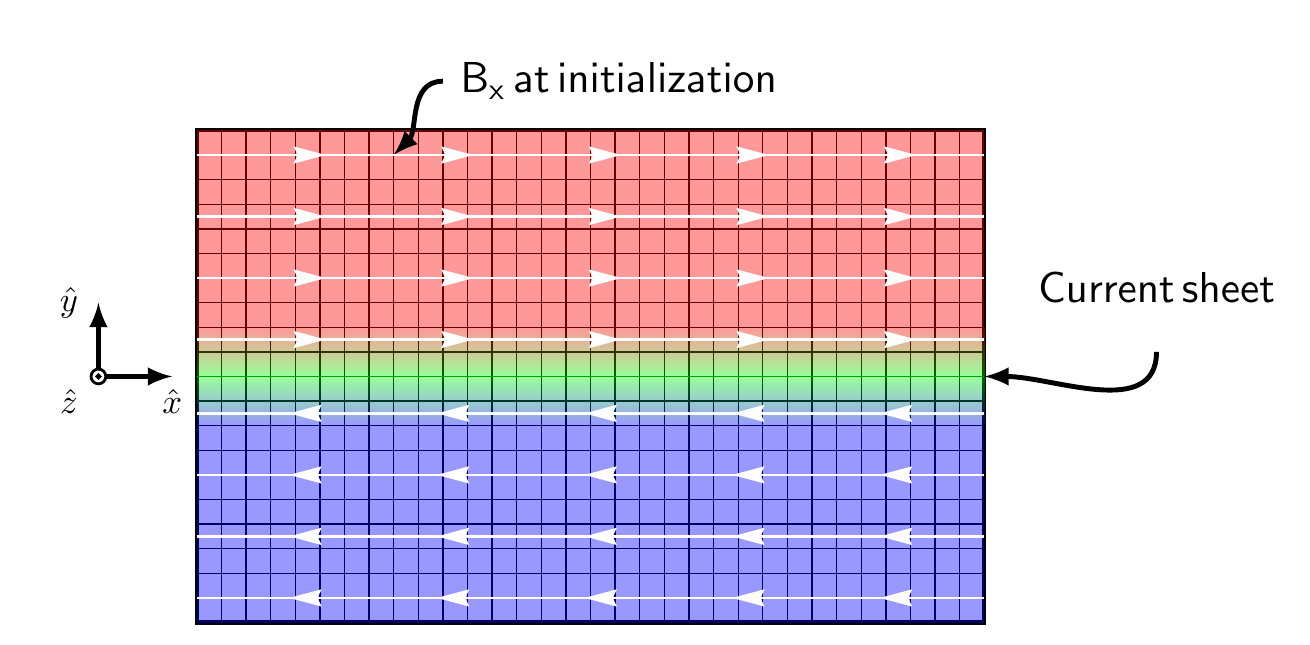} \\
			\caption{Schematic depiction of the reconnection layer initial configuration.  Red and blue regions show magnetic field lines of opposite polarity.  A hot, over-dense component of plasma (green region) balances the magnetic pressure outside the current sheet. \label{fig:2drecbox} \\} 
\end{figure}

We use the electromagnetic PIC code \texttt{TRISTAN-MP} to perform fully-kinetic simulations of reconnection \citep{Buneman1993,Spitkovsky2005}. We employ two-dimensional (2D) simulations, but all three components of velocity and electromagnetic fields are tracked.
Our setup is similar to that described in \citet{Sironi2014}. The initial field configuration is illustrated in Fig.~\ref{fig:2drecbox}.  From the red to the blue region, the polarity of the inflow magnetic field reverses, as shown by the white arrows.  An out-of-plane current,  in the green region, satisfies Ampere's law for the curl of the magnetic field. 
The reconnection layer is initialized in Harris equilibrium, with a magnetic field profile $\mathbf{B}=B_{0} \tanh(2 \pi y/\Delta)\, \mathbf{\hat{x}}$. We focus on anti-parallel reconnection, postponing the study of guide field effects to a future work. The field strength is parameterized via the magnetization, 
\begin{align} 
\sigma_{w} &= \frac{B_{0}^{2}}{4 \pi w},
\label{eq:sigmaw}
\end{align}
where $B_{0}$ is the magnitude of the  magnetic field in the inflow region, $w=(\rho_{\rm e} + \rho_{\rm i}) c^{2} + \hat{\gamma}_{\rm e}  u_{\rm e} + \hat{\gamma}_{\rm i} u_{\rm i}$ is the enthalpy density per unit volume, and 
$\rho_{\rm e}=m_{\rm e}n_0$, $\rho_{\rm i}=m_{\rm i}n_0$, 
$\hat{\gamma}_{\rm e}$, $\hat{\gamma}_{\rm i}$, and 
$u_{\rm e}$, $u_{\rm i}$
are the rest mass densities, adiabatic indices, and internal energy densities, respectively, of electrons and protons. Here, $n_{0}$ is the electron number density in the inflow region, $m_{\rm e}$ and $m_{\rm i}$ are the electron and proton masses.
The definition of magnetization in Eq. \ref{eq:sigmaw} reduces to Eq. \ref{eq:sigmai} in the limit of non-relativistic temperatures, but for relativistic particles the enthalpy in $\sigma_{w}$ properly accounts for the relativistic inertia.

In all runs, we set the current sheet thickness to be $\Delta=40\,c/\omega_{\rm pe}$, where $c/\omega_{\rm pe}$ is the electron skin depth,
\begin{align}
\omega_{\rm pe}=\sqrt{\frac{4 \pi n_{0} e^{2}}{m_{\rm e}}}\left( 1 + \frac{\theta_{\rm e}}{\hat{\gamma}_{\rm e} - 1}\right)^{-1/2}
\end{align}
is the electron plasma frequency. Here, $\theta_{\rm e}=k_{\rm B} T_{\rm e}/m_{\rm e}c^2$ is the dimensionless electron temperature, whereas $e$ is the electric charge.
The size of the computational domain in the $x$ direction is $L_{x}=4318\,c/\omega_{\rm pe},$ which is large enough to resolve both electron and proton heating physics (see Appendix \ref{sec:lxconvergence}, where we study the convergence of our results with respect to the domain size).  
While $L_{x}$ in units of $c/\omega_{\rm pe}$ remains fixed across our simulations, the domain size in units of the proton skin depth
\begin{align}\label{eq:skindepth}
\frac{c}{\omega_{\rm pi}}\! \approx \!\frac{c}{\omega_{\rm pe}} \!\sqrt{\frac{m_{\rm i}}{m_{\rm e}}}\! \left( \!1\!+ \!\frac{\theta_{\rm e}}{\hat{\gamma}_{\rm e} - 1}\right)^{-1/2}\!
\left( \!1 \!+ \!\frac{\theta_{\rm i}}{\hat{\gamma}_{\rm i} - 1}\right)^{1/2},
\end{align} 
increases as electrons become more relativistic (see Tab.~\ref{tab:params}). Here, $\theta_{\rm i}=k_{\rm B} T_{\rm i}/m_{\rm i}c^2$ is the dimensionless proton temperature.

We typically employ periodic boundary conditions along the $x$ direction, but we have tested that our main results do not change when using outflow boundary conditions, similar to those described in \citet{Sironi2016}.  With the latter, it is possible to study the dynamical evolution of the reconnection system over multiple Alfv\'enic crossing times, whereas the evolution of a periodic simulation is limited to a few Alfv\'enic crossing times, before the periodic boundaries start affecting the reconnection physics.  We compare the results of simulations with outflow and periodic boundaries in Appendix \ref{sec:outvper}.

Fresh plasma, described by a Maxwell-J\"{u}ttner distribution, is introduced at two moving injectors. Each injector recedes from $y=0$ at the speed of light, and the simulation domain is enlarged when the injectors reach the boundaries, so that the injectors may continue receding in the $\pm \mathbf{\hat{y}}$ directions.  This strategy --- described in more detail in \citet{Sironi2011} --- ensures that the domain includes all causally connected regions throughout the evolution of the system, while making efficient use of the available memory and computing time.  Additional computational optimization is achieved by allowing the injectors to periodically ``jump'' backwards (toward $y=0$), removing all particles beyond the injectors and resetting the electromagnetic fields to their initial values \citep{Sironi2011}.

A hot, over-dense population of particles is initialized in the current sheet to balance the magnetic pressure from outside.
  These particles have temperature $k_{\rm B}T_{cs}/m_{\rm i} c^2=\sigma_{\rm i}/2 \eta,$ where $\eta$ is the over-density relative to the inflowing plasma; we use $\eta = 3$.  
Reconnection is triggered at the initial time by cooling by hand the over-dense population in the middle of the current sheet $(x,y) \approx (0,0)$.  This causes a local collapse of the layer, leading to the formation of an X-point, after which the system evolves self-consistently \citep{Sironi2016}.  

Adequate resolution of the electron skin depth $c/\omega_{\rm pe}$ is required for accuracy and stability of PIC codes.  We use 4 cells per electron skin depth, and fix $c=0.45$ cells/timestep, which is less than required by the Courant-Friedrichs-Lewy condition in 2D.  The time resolution of our simulations is then ${\Delta t \approx 0.1\,\omega_{\rm pe}^{-1}}$, which properly captures the physics at electron scales.  For two cases ($\betai = 0.0078$ and $\betai=2$, with the same $\sigma_{w}=0.1$ and $\teti=1$), we have tested for convergence by varying the spatial resolution (we have tested with $c/\omega_{\rm pe} = 2$ or $8$ cells), which has the effect of changing also the temporal resolution (we still fix $c=0.45$ cells/timestep). For both choices of $\betai$, our results are essentially the same (see Appendix \ref{sec:compconvergence}, where we study the convergence of our results with respect to the spatial resolution of the electron skin depth).  

For simulations with $\beta_{\rm i} = 2,$ we use $64$ particles per cell ($N_{\rm ppc}$), whereas $N_{\rm ppc}=16$ at lower $\beta_{\rm i}$.  We have found that these values of $N_{\rm ppc}$ are sufficient to keep numerical heating under control, even for $T_{\rm e}/T_{\rm i}\ll1$.
We have extensively tested the impact of numerical heating  in simulations with $\beta_{\rm i}=2$ for several values of $N_{\rm ppc},$ in some cases up to $N_{\rm ppc}=256$; see Appendix \ref{sec:ppc} for some discussion. 


\begin{table}\centering
\ra{1.3}
\begin{tabular}{@{}crrrrrr@{}}\toprule
\multicolumn{1}{c}{\textbf{ID}} & \multicolumn{1}{c}{\textbf{A[0]}} & \multicolumn{1}{c}{\textbf{A[1]}} & \multicolumn{1}{c}{\textbf{A[2]}} & \multicolumn{1}{c}{\textbf{A[3]}} & \multicolumn{1}{c}{\textbf{A[4]}} & \\
\midrule
$\beta_{\rm i}$			&0.0078	&0.031	&0.13	&0.50	& 2.0 &\\
$\beta_{\rm e}$			&0.00078	&0.0031	&0.013	&0.050	& 0.20 &\\
$\theta_{\rm i}$			&0.00041	&0.0016	&0.0066	& 0.028	& 0.16 &\\
$\theta_{\rm e}$			&0.0010	&0.0041	&0.017	& 0.070	& 0.39 &\\
$\upsilon_{\rm i}$	&0.00061      &0.0024 	& 0.010 	& 0.043 	& 0.27 &\\
$\upsilon_{\rm e}$     &0.0015       & 0.0062	& 0.025 	& 0.11  	& 0.78 &\\
$\sigma_{\rm i}$			& 0.10			& 0.10	& 0.10	& 0.11	& 0.15 &\\
$T_{\rm e}/T_{\rm i}$			& 0.10 			& 0.1	0	& 0.10	& 0.10	& 0.10 &\\
$N_{\rm ppc}$					& 16				&16	&16	&16	& 64 &\\
$c/ \omega_{\rm pi}$               & 20   &   20   &   20   &   19   &   16 &\\
$L_{x} [c/ \omega_{\rm pi}]$                &860   &   870  &   870    &   890   &   1100 &\\ \midrule
\multicolumn{1}{c}{\textbf{ID}} & \multicolumn{1}{c}{\textbf{B[0]}} & \multicolumn{1}{c}{\textbf{B[1]}} & \multicolumn{1}{c}{\textbf{B[2]}} & \multicolumn{1}{c}{\textbf{B[3]}} & \multicolumn{1}{c}{\textbf{B[4]}} & \\ \midrule
$\beta_{\rm i}$			&0.0078	&0.031	&0.13	&0.50	& 2.0 &\\
$\beta_{\rm e}$			&0.0023	&0.0094	&0.038	&0.15	& 0.60 &\\
$\theta_{\rm i}$			&0.00041	&0.0016	&0.0066	&0.029	& 0.18 &\\
$\theta_{\rm e}$			&0.0031	&0.012	&0.050	&0.21	& 1.3 &\\
$\upsilon_{\rm i}$		& 0.00061  	&0.0025 	&0.010 		&0.044  	& 0.32 &\\
$\upsilon_{\rm e}$  	& 0.0046 	&0.019 	&0.079 	&0.39  	& 3.3 &\\
$\sigma_{\rm i}$			&0.10	&0.10	&0.10	&0.11	& 0.17 &\\
$T_{\rm e}/T_{\rm i}$			&0.30	&0.30	&0.30	&0.30	& 0.30 &\\
$N_{\rm ppc}$					&16	&16	&16	&16	& 64 &\\
$c/ \omega_{\rm pi}$		&20   &   20  &   19   &   17  &    11 &\\ 
$L_{x} [c/ \omega_{\rm pi}]$                & 870   &   870   &   890   &   1000   &   1600 &\\ \midrule

\multicolumn{1}{c}{\textbf{ID}} & \multicolumn{1}{c}{\textbf{C[0]}} & \multicolumn{1}{c}{\textbf{C[1]}} & \multicolumn{1}{c}{\textbf{C[2]}} & \multicolumn{1}{c}{\textbf{C[3]}} & \multicolumn{1}{c}{\textbf{C[4]}} & \\
\midrule 
$\beta_{\rm i}$			&0.0078	&0.031	&0.13	&0.50	& 2.0 &\\
$\beta_{\rm i}$			&0.0078	&0.031	&0.13	&0.50	& 2.0 &\\
$\theta_{\rm i}$			&0.00041  &0.0016   &0.0067	&0.031	& 0.39 &\\
$\theta_{\rm e}$			&0.010	&0.041	&0.17     	&0.77       & 9.9  &\\
$\upsilon_{\rm i}$	&0.00061  &0.0024 	&0.010   	&0.048  	& 0.79 &\\
$\upsilon_{\rm e}$  	& 0.015  	&0.064 	&0.30 	&1.8          & 29 &\\
$\sigma_{\rm i}$			&0.10	&0.10	&0.10	&0.12	& 0.38 &\\
$T_{\rm e}/T_{\rm i}$			&1.0	        &1.0	         &1.0	         &1.0	         & 1.0 &\\
$N_{\rm ppc}$					&16	&16	&16	&16	& 64 &\\
$c/\omega_{\rm pi}$                &20  &    19  &    17   &   12  &    5.0 &\\ 
$L_{x} [c/ \omega_{\rm pi}]$                & 870  &    890 &     990  &    1500   &   3400 &\\ \bottomrule
\end{tabular}
\caption{\raggedright Initial parameters for the $m_{\rm i}/m_{\rm e}=25$ simulations with our fiducial $\sigma_w=0.1$.  The proton skin depth $c/\omega_{\rm pi}$, calculated according to Eq. \ref{eq:skindepth}, is expressed in number of cells. The definition of the various quantities is in  Section \ref{sec:setup}. Simulation sets \textbf{A}, \textbf{B}, and \textbf{C} differ by the initial temperature ratio, with $T_{\rm e}/T_{\rm i}=0.1,0.3,$ and $1$, respectively.  From left to right, $\beta_{\rm i}$ increases. We fix the mass ratio $m_{\rm i}/m_{\rm e}=25,$ magnetization $\sigma_{w}=0.1,$ electron skin depth $c/\omega_{\rm pe}=4$ cells, and domain size $L_{x} = 4318\,c/\omega_{\rm pe}.$ We also perform a number of additional simulations, up to the realistic mass ratio $m_{\rm i}/m_{\rm e}=1836$ and with higher magnetizations ($\sigma_w=0.3$, 1, 3, 10), as described in Section \ref{sec:setup}.}
\label{tab:params}
\end{table}

In our parameter scan (Tab.~\ref{tab:params}), we fix $\sigma_{w}$ and study the reconnection physics as a function of $\beta_{\rm i}$ and $T_{\rm e}/T_{\rm i}$.  We choose to fix $\sigma_{w}$ rather than $\sigma_{\rm i},$ given that the parameter space we probe involves relativistic particles whose thermal contribution to the inertia is non-negligible (see Eq. \ref{eq:sigmaw}). For a constant $\sigma_w$, the Alfv\'{e}n velocity
\begin{align}
\frac{v_{\rm A}}{c} &= \sqrt{\frac{\sigma_{w}}{1+\sigma_{w}}},
\end{align}
remains fixed across our simulations. The reconnection layer is evolved for $\sim 1$ Alfv\'{e}nic crossing time $(t_{\rm A}=L_x/ v_{\rm A})$, which for our reference magnetization of $\sigma_w=0.1$ and $L_x=4318\,c/\omega_{\rm pe}$ corresponds to $ t \approx 14000\,\omega_{\rm pe}^{-1}.$ 

The focus of our investigation is the so-called \textit{trans-relativistic} regime of reconnection, hence we select $\sigma_w = 0.1$ as our fiducial magnetization, and we vary $\beta_{\rm i}$ from $0.0078$ to $2.$  
Additionally, we study the effect of the initial electron-to-proton temperature ratio $T_{\rm e}/T_{\rm i}$ on the reconnection physics.  For each value of $\beta_{\rm i},$ we run three simulations with $T_{\rm e}/T_{\rm i} = 0.1, 0.3,$ and $1.$  Our choice of initial parameters, both physical ($\sigma_{w}, \beta_{\rm i}$, and $T_{\rm e}/T_{\rm i}$) and computational ($N_{\rm ppc}$, $c/\omega_{\rm pe}$), is summarized in Tab.~\ref{tab:params}. 
Other derived physical parameters in the inflow region, namely the electron plasma $\beta_{\rm e}=\beta_{\rm i}T_{\rm e}/T_{\rm i},$ the dimensionless proton and electron temperatures $\theta_{\rm i}=k_{\rm B}T_{\rm i}/m_{\rm i} c^2$ and $\theta_{\rm e}=k_{\rm B}T_{\rm e}/m_{\rm e} c^2,$ the dimensionless internal energy per particle for protons and electrons $\upsilon_{\rm i}\equiv u_{\rm i}/n_{0} m_{\rm i} c^2$ and $\upsilon_{\rm e} \equiv u_{\rm e}/n_{0} m_{\rm e} c^2$, and the ratio $\sigma_{\rm i}$ of magnetic pressure to rest mass energy density,  are also included.
 In addition to the simulations listed in the table, which employ mass ratio $m_{\rm i}/m_{\rm e}=25$, we also investigate mass ratios $m_{\rm i}/m_{\rm e}=10, 50,$ and $1836$ for $\beta_{\rm i}$ in the range $5 \times 10^{-4}-2$ (with fixed $\sigma_w=0.1$ and a fixed electron-to-proton temperature ratio $T_{\rm e}/T_{\rm i}=1$). With realistic mass ratios and $T_{\rm e}/T_{\rm i}=1$, we also explore the $\beta_{\rm i}$-dependence of the heating efficiency at higher values of the magnetization: $\sigma_w=0.3$, 1, 3 and 10.

\begin{figure*}[!tbh]
		\centering
		\includegraphics[width=\textwidth,trim={0 8.5cm 1.5cm 7.5cm},clip]{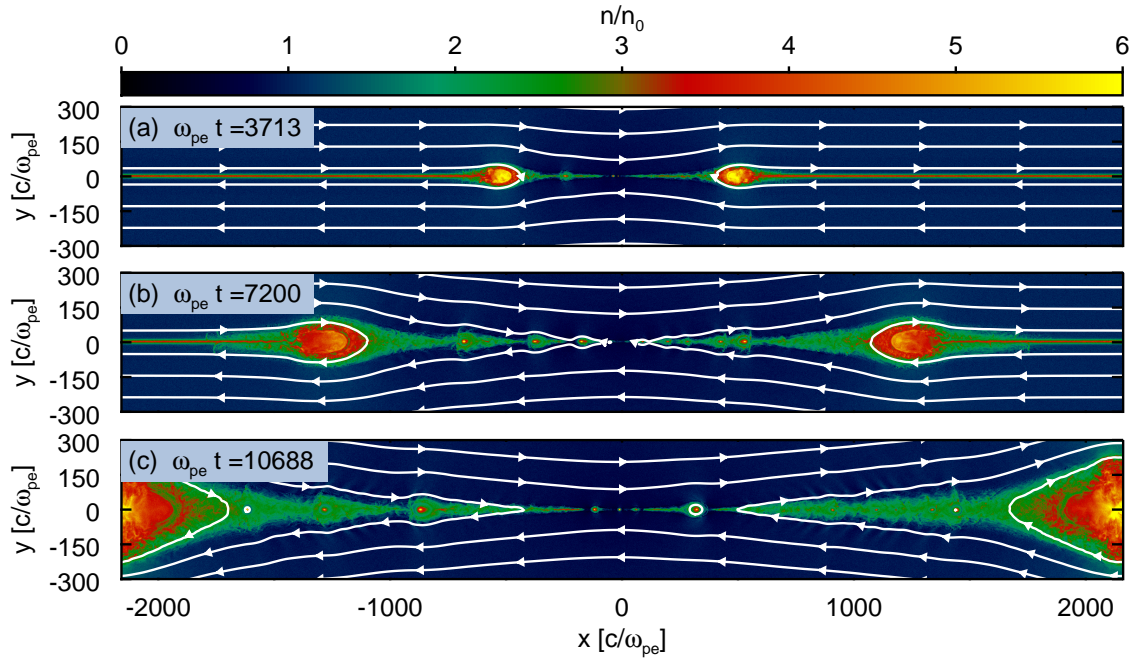} \\ %
			\caption{Time evolution of a representative low-$\beta_{\rm i}$ simulation (\textbf{A[0]} in Tab.~\ref{tab:params}), with $\beta_{\rm i}=0.0078$ and $T_{\rm e}/T_{\rm i}=0.1$.  The snapshots show number density of electrons in units of the initial density at (a): $ t = 3713\,\omega_{\rm pe}^{-1} \approx 0.25\,t_{\rm A}$; (b): $ t = 7200\,\omega_{\rm pe}^{-1} \approx 0.50\,t_{\rm A}$; (c): $ t = 10688\,\omega_{\rm pe}^{-1}\approx 0.75\,t_{\rm A}$.   We show the whole dimension of the box in $x$, and only a small portion close to the center in $y$.  A characteristic feature of this and other low-$\beta_{\rm i}$ simulations is the presence of {\it secondary} magnetic islands, i.e., structures like those at $x \approx 300\,c/\omega_{\rm pe}$ and $x \approx -900\,c/\omega_{\rm pe}$ (panel (c)).  These are to be distinguished from the large \textit{primary} island at $x\approx \pm 2200\,c/\omega_{\rm pe},$ whose properties depend on choices at initializiation.  As the primary island grows, it will eventually inhibit further accretion of magnetic flux and the reconnection process will terminate. \label{fig:lowbeta}} 
	\end{figure*}

\begin{figure*}
		\centering
		\includegraphics[clip, trim=0.5cm 11cm 1cm 11cm,width=1\textwidth]{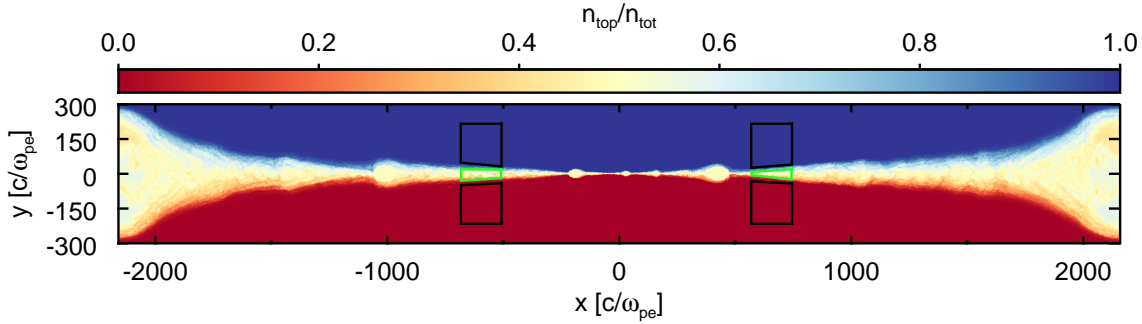} \\ 
			\caption{$\,$2D plot of the ratio of top-to-total particle density, $n_{\rm top}/n_{\rm tot},$ for a representative simulation with $\beta_{\rm i} = 0.0078$ and $T_{\rm e}/T_{\rm i}=0.1$ (\textbf{A[0]} in Tab.~\ref{tab:params}) at time $ t \approx 11000\, \omega_{\rm pe}^{-1}\approx0.8\,t_{A}$.  The green and black contours show the boundaries of the regions we use to calculate the downstream and upstream temperatures, respectively. The box edges at the interface between upstream and downstream change as the system evolves, and are calculated according to Eqs. \ref{eq:criterion1} and \ref{eq:criterion2}.
Particle mixing serves as a tracer for the downstream region.  Particles from the top ($y > 0$) of the domain are tagged;  as they enter the reconnection layer, they mix with particles from the bottom ($y<0$) of the domain.  The reconnection downstream is identified via the mixing fraction $n_{\rm top}/n_{\rm tot}$, and a choice of the threshold $r_{\rm down},$ as in  Eq.  \ref{eq:criterion1}.   \label{fig:mixing2d}} 
\end{figure*}

\begin{figure}
		\centering
		\includegraphics[width=0.45\textwidth]{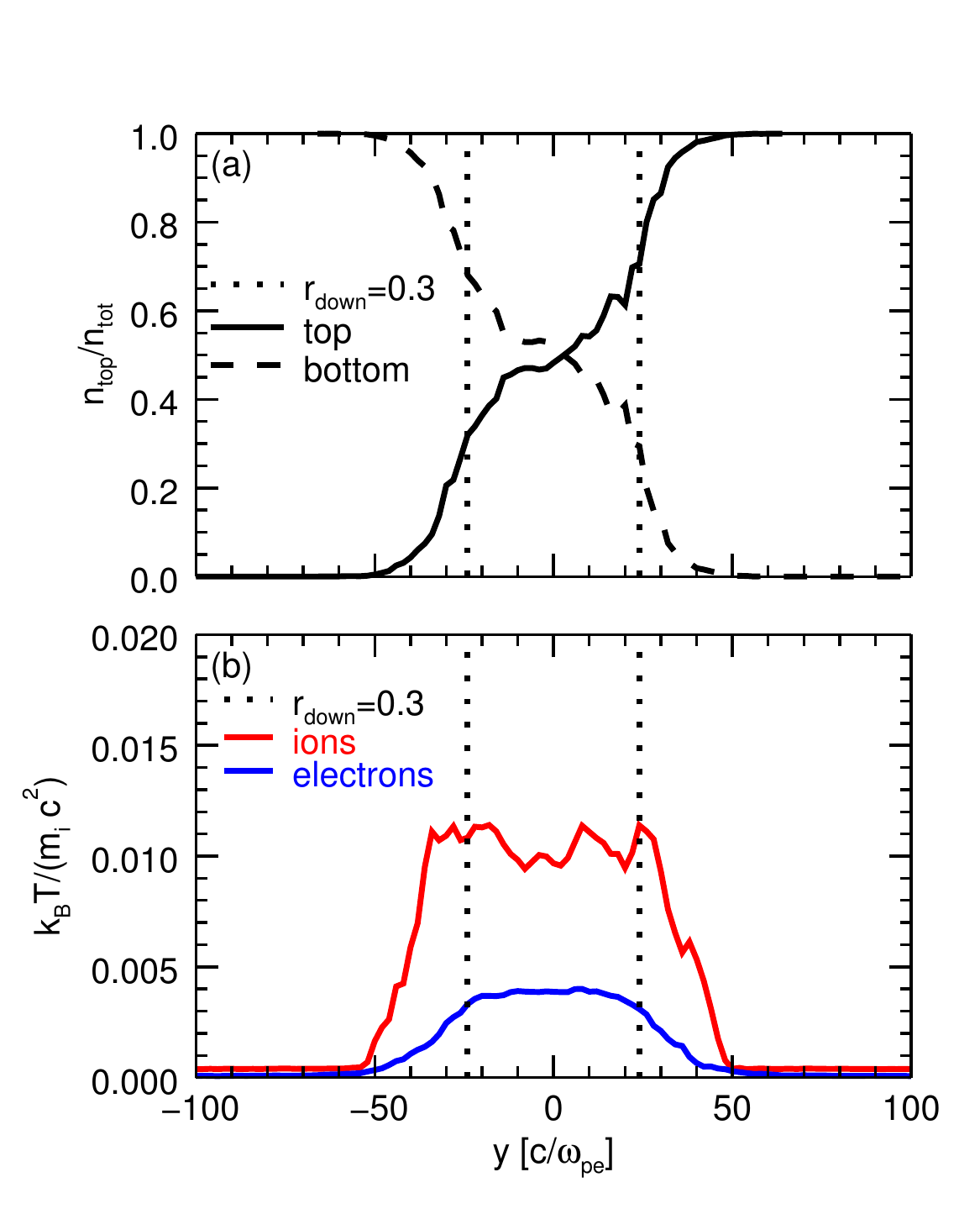} \\
			\caption{(a):$\,$ 1D profile along the $y$ direction of top-to-total particle density ratio (solid line) and bottom-to-total ratio (dashed line) in a slice at $x\approx1000\,c/\omega_{\rm pe}$, at time $t \approx 8400\,\omega_{\rm pe}^{-1}  \approx0.60\,t_{\rm A}.$  The profiles are from the same simulation we show in Fig.~\ref{fig:mixing2d} (with $\beta_{\rm i} = 0.0078$ and $T_{\rm e}/T_{\rm i}=0.1$).  Vertical dotted lines indicate the locations in $x$ where the top-to-total density ratio is between 0.3 and 0.7 (at $y\approx-25$ and $25\,c/\omega_{\rm pe},$ respectively).  Between the vertical dotted lines (i.e., in the region we define as the reconnection downstream), mixing has efficiently occurred. (b): Proton and electron temperature profiles in the same region.  In between the vertical dotted lines, the temperature profiles are nearly flat.  \label{fig:mixing}}
\end{figure}

\section{Technique for extracting the heating efficiency}\label{sec:technique}
In this section, we discuss our method of extracting the heating efficiency from PIC simulations.
First, in Section \ref{ssec:timeevol}, we discuss the time evolution of the reconnection layer for two representative cases at low and high $\beta_{\rm i}$.
Then, in Section \ref{ssec:tagged}, we describe the identification of inflow (upstream) and outflow (downstream) regions.
Lastly, in Section \ref{ssec:characterization}, we isolate the  irreversible heating, i.e., the part associated with a genuine increase in entropy, from the reversible heating induced by adiabatic compression.



\subsection{Time evolution of the reconnection layer}\label{ssec:timeevol}
To illustrate the time evolution of the reconnection layer, we show in Fig.~\ref{fig:lowbeta} a few snapshots of density from a representative simulation (\textbf{A[0]} in Tab.~\ref{tab:params}) with $\beta_{\rm i}=0.0078$ and $T_{\rm e}/T_{\rm i}=0.1$.  
We plot the 2D profile of the number density in units of the initial value, $n/n_{0}$.  In each panel, we show only a small fraction of the domain in the $y$ direction (we present only the region closest to the current sheet), and the full extent of the domain in $x$.  White lines with arrows show magnetic field lines.

Panels (a)--(c) show the time evolution of the system over $\sim1$ Alfv\'{e}nic crossing time.  By removing by hand the plasma pressure at the center of the current sheet ($x\sim 0$), we trigger a local collapse of the layer, forming an X-point. After the formation of the central X-point, two reconnection ``wavefronts'' are pulled outwards in the $\pm \mathbf{\hat{x}}$ directions by the magnetic tension of the field lines, and the fronts recede from the center at close to the Alfv\'{e}n speed. 
 In panels (a), (b), and (c), the wavefronts are located at $x\approx\pm 400,1100,$ and $1800\,c/\omega_{\rm pe},$ respectively, corresponding to the innermost (i.e., closest to $x=0$) locations of the large semi-circular red/yellow density blobs. 
 
The fronts carry away the hot particles initialized in the current sheet.  With periodic boundary conditions, this leads to the formation of a \textit{primary} island at the boundary of the simulation domain (in Fig.~\ref{fig:lowbeta}(c), located at $x\approx\pm 2200\,c/\omega_{\rm pe}).$  
The primary island continues to accrete plasma as the system evolves, but eventually it grows so large that further accretion of magnetic flux into the layer is inhibited, and reconnection stops.  

The primary island shows the hottest electron temperatures. Here, electron heating might be due in part to reconnection, but also in part to weak shocks at the interface between the reconnection outflow and the island. In addition, the plasma conditions in the island are sensitive to our arbitrary choice for the current sheet initialization. For these reasons, we choose not to focus on the heating physics in the primary island. 

In this paper, we focus exclusively on the outflow (i.e., before the the plasma reaches the primary island; see also \citet{Shay2014},  in the context of non-relativistic reconnection), shown by the green region between the two wavefronts in Fig. \ref{fig:lowbeta}.  
In Section \ref{ssec:tagged}, we detail the steps we take to avoid contamination of our temperature measurements by the primary island.  

As the two reconnection fronts recede from the center, plasma flows into the reconnection layer and particles are heated and accelerated as a bulk, flowing along $\pm \mathbf{\hat{x}}$ toward the domain boundaries.  The dense (green) region in between the two wavefronts is the reconnection outflow.  A key feature of low-$\beta_{\rm i}$ simulations is the formation in the reconnection exhausts of \textit{secondary} islands due to the secondary tearing instability, e.g., Fig.~\ref{fig:lowbeta}(c) at $x\approx 300\,c/\omega_{\rm pe}$ and $x\approx -900\,c/\omega_{\rm pe}$ \citep{Daughton2007,Uzdensky2010}.  Between each pair of secondary islands, there is a secondary X-point, e.g., at $x\approx-1000\,c/\omega_{\rm pe}$.  We discuss the structure of the reconnection layer as a function of $\beta_{\rm i}$ in Section \ref{ssec:examples}.

\subsection{Upstream and downstream identification} 
\label{ssec:tagged}
\label{sec:updownid}
We now describe how we determine which computational cells in the simulation domain belong to the upstream (or, inflow) and downstream (or, outflow) regions.
We identify downstream cells by using a particle mixing criterion between the two sides of the current sheet.  
Particles that originate above $y=0$ (top of the domain) are tagged, to distinguish them from particles originating below $y=0$ (bottom of the domain).  

In Fig.~\ref{fig:mixing2d}, we show the ratio of top-to-total number density.  Away from the current sheet, i.e., in the blue and red regions, there is no mixing between the two populations.  Particles from the two sides of the current sheet get mixed as they enter the reconnection layer; the region with the greatest amount of mixing is shown in white/light-yellow.
We compute the ratio of top-particle density $n_{\rm top}$ to total-particle density $n_{\rm tot}=n$ (including particles from both top and bottom) in each cell.  If this ratio in a given cell exceeds a chosen threshold $r_{\rm down}$ and is below the complementary threshold, i.e.,
\begin{align}
\label{eq:criterion1}
r_{\rm down} < \frac{n_{\rm top}}{n_{\rm tot}} < 1-r_{\rm down},
\end{align}
then the cell is counted as one where plasma has reconnected (i.e., the cell belongs to the reconnection downstream). This technique is similar to that used in \citet{Daughton2014}.  In our analysis, we choose $r_{\rm down}=0.3,$ but we have verified that the identification of the reconnection region, and therefore the heating efficiencies that we extract, do not significantly depend on this choice.  For $r_{\rm down}$ in the range $0.1$ -- $0.3$, the heating efficiencies typically differ only by $\sim 15\%.$  
The choice $r_{\rm down}=0.3$ is restrictive enough to exclude contamination by the upstream region.  This is especially important for high $\beta_{\rm i}$, where, even if the electron gyrocenter is located in a cell that is safely part of the downstream, if the cell is close to the interface between downstream and upstream, the particle gyro-motion may temporarily lead this ``downstream'' electron to the upstream side. If $r_{\rm down}$ were to be too small, the region where the electron motion extends into the upstream might be incorrectly counted as part of the downstream, biasing our temperature estimates toward lower values. Our choice of $r_{\rm down}$ is to some extent arbitrary, but we have found that a relatively large value like $r_{\rm down}=0.3$ is  suitable for identifying the genuine reconnection downstream. 

In Fig.~\ref{fig:mixing}, we show 1D plots of the density fraction of tagged particles and the temperature profiles along the $y$ direction, in a slice located at $x\approx1000\,c/\omega_{\rm pe}.$ In panel (a), we show the profiles of the ratio of top- and bottom-density to total density, denoted by solid and dashed lines, respectively, at time $t \approx 8400\, \omega_{\rm pe}^{-1} \approx 0.60\,t_{\rm A}$.  Between the two vertical dotted lines, the ratio of top-to-total density ranges between 0.3 and 0.7, as required to satisfy our mixing criterion. As shown in panel (b), both the electron (blue) and the proton (red) temperature in the region between the vertical lines are remarkably uniform, proving that our mixing criterion can confidently capture the reconnection downstream.



The upstream region is identified via
\begin{align}
\label{eq:criterion2}
\left(  \frac{n_{\rm top}}{n_{\rm tot}}< r_{\rm up} \right) \;{\rm or}\; \left( \frac{n_{\rm top}}{n_{\rm tot}} > 1-r_{\rm up} \right),
\end{align}
and we choose $r_{\rm up}=3\times 10^{-5}$.  As before, this definition avoids contamination of the upstream region by any ``downstream'' particles that leak out of the current sheet.  In practice, a buffer zone with a width on the order of a few tens of  $c/\omega_{\rm pe}$ is established between the regions we identify as upstream and downstream.

While Eq. \ref{eq:criterion1} (Eq. \ref{eq:criterion2}, respectively) identifies the whole reconnection outflow (inflow, respectively), we enforce an additional constraint on the downstream and upstream regions that we employ to extract our heating efficiencies. We select downstream regions far enough from the central X-point that the electron and proton outflow bulk velocities have saturated, and also that the electron and proton temperatures have reached their asymptotic values.  At the same time, we select these regions to be far enough from the boundaries to avoid contamination from the material inside the primary island, and only capture the genuine reconnection outflow.  
 The downstream region that satisfies these constraints (identified by the green contours in Fig.~\ref{fig:mixing2d}) varies for different simulations: for $\beta_{\rm i}<2$ it is located at a distance of $\sim 630\,c/\omega_{\rm pe}$  from the center, whereas for $\beta_{\rm i}=2$ it is at $\sim 350\,c/\omega_{\rm pe}$ from the center (as we show below, the primary island tends to be larger at higher  $\beta_{\rm i}$).  The extent of the downstream region across the layer (i.e., along $y$) is determined by the mixing criterion in Eq. \ref{eq:criterion1}, while the length along the layer is fixed at $\sim170\,c/\omega_{\rm pe} $ (see the green contours in Fig.~\ref{fig:mixing2d}). The corresponding upstream values are measured at the same distance from the center of the layer, within the black contours in Fig.~\ref{fig:mixing2d}. Their exent along the $y$ direction does not significantly affect our results.




\subsection{Characterization of heating} \label{ssec:characterization}
In this section, we describe our assessment of particle heating.  
First, in Section \ref{sssec:tcalc}, we describe our calculation of rest-frame internal energy and temperature.
Next, in Section \ref{sssec:efficiency}, we define ratios that characterize the total amount of heating. 
Finally, in Section \ref{sssec:compnoncomp}, we provide a more detailed analysis of the heating physics by isolating the effect of a genuine entropy increase (which we call ``irreversible heating'') from the contribution of adiabatic compression (giving ``adiabatic heating'').

\subsubsection{Temperature calculation}\label{sssec:tcalc}
We measure the total particle energy density in the simulation frame, then extract the corresponding fluid-frame internal energy and temperature, by employing the perfect, isotropic fluid approximation, i.e.
\begin{align}\label{eq:pf}
	T^{\mu \nu} = \left( {e} + {p} \right) U^{\mu} U^{\nu} - {p} g^{\mu \nu},
\end{align}
where $T^{\mu \nu}$ is the stress-energy tensor of the fluid, ${e}$ is the rest-frame energy density,  
${p}$ is the pressure, $U^{\mu}$ is the fluid dimensionless four-velocity, and $g^{\mu \nu}$ is the flat-space Minkowski metric.  
The rest-frame energy density  is the sum of rest-mass and internal energy densities, i.e.
\begin{align}
{e} &= \xoverline{n} m c^{2} + {u} \\
&= \xoverline{n} m c^{2} + \frac{{p}}{{\hat{\gamma}} - 1},
\end{align}
where $\xoverline{n}$ is the rest-frame particle number density, ${u}$ is the internal energy density, and ${\hat{\gamma}}$ is the adiabatic index.
The dimensionless internal energy per particle in the fluid rest frame ${\upsilon}_{s}$ may be expressed as
\begin{align} \label{eq:approxequation}
	{\upsilon}_{s} &= \frac{(T^{00}_{s}/n_{s} m_{s} c^2 - \Gamma_{s}) \Gamma_{s}}{1 + {\hat{\gamma}}_{s} (\Gamma_{s}^2 - 1)},
\end{align}
where $T^{00}_{s}$ is the total energy density in the simulation frame, $n_{s}$ is the lab-frame particle number density, $\Gamma_{s}$ is the Lorentz factor corresponding to the local fluid velocity, ${\hat{\gamma}}_{s}$ is the adiabatic index, and the subscript $s=\rm e, i$ refers to the particle species.  

To make use of Eq. \ref{eq:approxequation}, we need to express the adiabatic index $\hat{\gamma}_{s}$ as a function of the internal energy per particle, so that the equation may be solved iteratively.  For a plasma described by a Maxwell-J\"{u}ttner distribution with dimensionless temperature $\theta_s$, 
\begin{align}
f_{\rm MJ}(\gamma, \theta_{s}) \propto \gamma \sqrt{\gamma^{2} - 1} \exp \left( -\gamma / \theta_{s} \right),
\end{align}
where $\gamma$ denotes the particle Lorentz factor, the dimensionless internal energy is given by
\begin{align}
	\upsilon_{s} = \frac{\int_{1}^{\infty} \gamma f_{\rm MJ}(\gamma, \theta_{s}) d \gamma}{\int_{1}^{\infty}  f_{\rm MJ}(\gamma, \theta_{s}) d \gamma} - 1.
\end{align}
We have numerically evaluated the integral on the right hand side for a range of temperatures and thereby produced interpolating tables for $\hat{\gamma}_{s}(\upsilon_{s})$ and $\theta_{s}(\upsilon_{s})$, to be used for finding $\upsilon_{s}$ in Eq. \ref{eq:approxequation}.

Eqs. \ref{eq:pf} and \ref{eq:approxequation} assume that the stress-energy tensor is diagonal and isotropic in the fluid frame.  We have explicitly tested this assumption by measuring all the components of the stress-energy tensor in  our computational domain. By boosting into the local fluid frame, we can calculate all the components of the pressure tensor. We find that the off-diagonal components are generally negligible. As regard to the diagonal components, we quantify the degree of anisotropy with the temperature ratios  $T_{xx}/T_{\rm tot}, \,T_{yy}/T_{\rm tot},$ and $T_{zz}/T_{\rm tot},$ where $T_{\rm tot}=(T_{xx}+T_{yy}+T_{zz})/3$. For an isotropic fluid, $T_{xx}/T_{\rm tot}=T_{yy}/T_{\rm tot}=T_{zz}/T_{\rm tot}=1$.
 For electrons in the reconnection downstream, we find that these ratios typically lie in the range $T_{yy}/T_{\rm tot} \approx T_{zz}/T_{\rm tot} \approx 0.9$ -- $0.95$  and $T_{xx}/T_{\rm tot} \approx 1.2$ -- $1.1$  (see Appendix \ref{sec:aniso} for further discussion, including the dependence of the anisotropy on $\beta_{\rm i}$ and $T_{\rm e}/T_{\rm i}$).  We find greater anisotropy along the outflow direction $\mathbf{\hat{x}}$ than either $\mathbf{\hat{y}}$ or $\mathbf{\hat{z}}$. This is in  qualitative agreement  with the findings of \citet{Shay2014}, who demonstrated that the electron pressure tensor in the immediate reconnection exhausts is anisotropic, with the  component parallel to the local magnetic field larger than the perpendicular component.

As an additional test, we have also measured the temperature and internal energy via an explicit boost of the stress-energy tensor into the fluid rest frame, and compared the results to those computed by employing the perfect-fluid approximation as described above.  We find that the disagreement between the two methods is only of order $\sim 1\%,$  providing \textit{a posteriori} a justification for the perfect-fluid assumption.

\subsubsection{Total heating} 
\label{sssec:efficiency}
The main focus of our investigation is particle heating by reconnection, and how the heating efficiency depends on the upstream parameters.  From each simulation, we extract a dimensionless ratio $M_{u\rm e,tot},$ which we define as  
\begin{align}
M_{u\rm e,tot} &\equiv \frac{\upsilon_{\rm e,down}-\upsilon_{\rm e,up}}{\sigma_{\rm i} m_{\rm i} / m_{\rm e}} \label{eq:mue}.
\end{align}
The numerator is the difference in dimensionless internal energy per electron between downstream and upstream, while the denominator represents (apart from a factor of two) the available magnetic energy per electron in the upstream, in units of the electron rest mass energy ($=B_0^2/4 \pi n_0 m_{\rm e} c^2$).  The ratio $M_{u\rm e,tot}$ is then a measure of the efficiency of reconnection in converting available magnetic energy to electron heating.  Alternatively, the efficiency parameter may be phrased in terms of the dimensionless temperature, 
\begin{align}
M_{T\rm e,tot} &\equiv \frac{\theta_{\rm e,down}-\theta_{\rm e,up}}{\sigma_{\rm i} m_{\rm i}/m_{\rm e}} \label{eq:mte},
\end{align}
as in \citet{Shay2014}.  We define analogous ratios for protons as 
\begin{align}
M_{u\rm i,tot} &\equiv \frac{\upsilon_{\rm i,down}-\upsilon_{\rm i,up}}{\sigma_{\rm i}} \label{eq:mui},
\end{align}
and
\begin{align}
M_{T\rm i,tot} &\equiv \frac{\theta_{\rm i,down}-\theta_{\rm i,up}}{\sigma_{\rm i}} \label{eq:mti}.
\end{align}
For the results presented below, we average the dimensionless internal energy and temperature appearing in the above equations over time, starting at $\sim0.3$ Alfv\'{e}nic crossing times (or equivalently, $ \sim 4500\;\omega_{\rm pe}^{-1}$), when the two reconnection wavefronts --- and with them, the particles initialized in the current sheet --- have moved beyond the region that we use for our computations (green and black boxes in Fig.~\ref{fig:mixing2d}). We typically time-average our results  over an interval of $\sim0.3$ Alfv\'{e}nic crossing times.



\subsubsection{Adiabatic and irreversible heating} 
\label{sssec:compnoncomp}
When gas is adiabatically compressed, its internal energy increases while its entropy remains constant.  The reconnecting plasma may experience such adiabatic heating, since the downstream region is denser than the upstream (see Fig.~\ref{fig:lowbeta}). However, adiabatic heating is not a genuine signature of the conversion of field energy into particle energy.  We isolate the irreversible heating generated by magnetic field dissipation by subtracting out the adiabatic heating from the total particle heating. 

The predicted internal energy per particle in the downstream resulting from adiabatic compression alone (which we shall call $\upsilon^{\rm ad}_{s, \rm down}$ for species $s$) is calculated from the upstream internal energy per particle $\upsilon_{s, \rm up}$, the upstream rest-frame number density $\bar{n}_{s, \rm up}$ and the downstream rest-frame number density $\bar{n}_{s, \rm down}$ using the second law of thermodynamics for constant entropy,
\begin{align} \label{eq:igl}
dU_s &= -p_s dV
\end{align}
From the ideal gas equation of state, the pressure is  $p_{s}=\bar{n}_{s} k_{\rm B} T_{s}= (\hat{\gamma}_{s}-1)u_{s}$. Using the relation $U_s/V=u_s=\upsilon_s \bar{n}_s m_s c^2$, we can integrate Eq.~\ref{eq:igl} to obtain
\begin{align}
\int_{\upsilon_{s,\rm up}}^{\upsilon^{\rm ad}_{s,\rm down}} \frac{1}{(\hat{\gamma}(\upsilon_{s}) - 1) \upsilon_{s}} d\upsilon_{s} - \log \left(\frac{\bar{n}_{s,\rm down}}{\bar{n}_{s,\rm up}} \right) &= 0 \label{eq:log}.
\end{align}
We compute the argument of the logarithm in Eq. \ref{eq:log} as the ratio of downstream to upstream rest-frame density, spatially averaged over the regions marked in Fig.~\ref{fig:mixing2d}. 
The lower bound of the integral $\upsilon_{s,\rm up}$ is computed as a density-weighted spatial average in the selected upstream region.  The adiabatic index $\hat{\gamma}_s(\upsilon_s)$ is tabulated as discussed above.  We numerically solve Eq.~\ref{eq:log} for the predicted downstream  dimensionless internal energy per particle $\upsilon^{\rm ad}_{s,\rm down}$ resulting from adiabatic compression.  We refer to the corresponding dimensionless temperature as $\theta^{\rm ad}_{s,\rm down}$.
We call the difference between the initial and the predicted dimensionless temperature or internal energy per particle due to adiabatic compression, i.e., $\Delta \theta_{s,\rm ad} \equiv \theta_{s,\rm down}^{\rm ad}-\theta_{s,\rm up}$  and $\Delta \upsilon_{s,\rm ad} = \upsilon_{s,\rm down}^{\rm ad}-\upsilon_{s,\rm up}$, as the ``adiabatic'' component of heating.

The irreversible heating, which is associated with a genuine increase in entropy, is the residual between the total heating and the adiabatic heating:
\begin{align}
	\Delta \theta_{s,\rm irr} &= (\theta_{s,\rm down} - \theta_{s,\rm up}) - \Delta \theta_{s,\rm ad}, \\
	\Delta \upsilon_{s,\rm irr} &= (\upsilon_{s,\rm down} - \upsilon_{s,\rm up}) - \Delta \upsilon_{s,\rm ad}. 
\end{align} 

As in Section \ref{sssec:efficiency}, we introduce efficiency ratios to characterize the irreversible and adiabatic heating of electrons,
\begin{alignat}{3} \label{eq:mtencmtec}
M_{T\rm e,\rm irr} &\equiv \frac{\Delta \theta_{\rm e,irr}}{\sigma_{\rm i} m_{\rm i}/m_{\rm e}}, \qquad M_{T\rm e,ad} && \equiv \frac{\Delta \theta_{\rm e,ad}}{\sigma_{\rm i} m_{\rm i}/m_{\rm e}},  \\
\label{eq:muencmuec}
M_{u\rm e,irr} &\equiv \frac{\Delta \upsilon_{\rm e,irr}}{\sigma_{\rm i} m_{\rm i}/m_{\rm e}}, \,\,\,\qquad M_{u\rm e,ad} &&\equiv \frac{\Delta \upsilon_{\rm e,ad}}{\sigma_{\rm i} m_{\rm i}/m_{\rm e}}, 
\end{alignat}
and define analogous ratios for protons
\begin{alignat}{3} \label{eq:mtincmtic}
M_{T\rm i,irr} &\equiv \frac{\Delta \theta_{\rm i,irr}}{\sigma_{\rm i}}, \qquad M_{T\rm i,ad} && \equiv \frac{\Delta \theta_{\rm i,ad}}{\sigma_{\rm i}}, \\
M_{u\rm i,irr} &\equiv \frac{\Delta \upsilon_{\rm i,irr}}{\sigma_{\rm i} }, \qquad M_{u\rm i,ad} && \equiv \frac{\Delta \upsilon_{\rm i,ad}}{\sigma_{\rm i}}.
\end{alignat}

\begin{figure*}[!h]
		\centering
		\includegraphics[width=\textwidth,trim={0 7.5cm 0 5cm},clip]{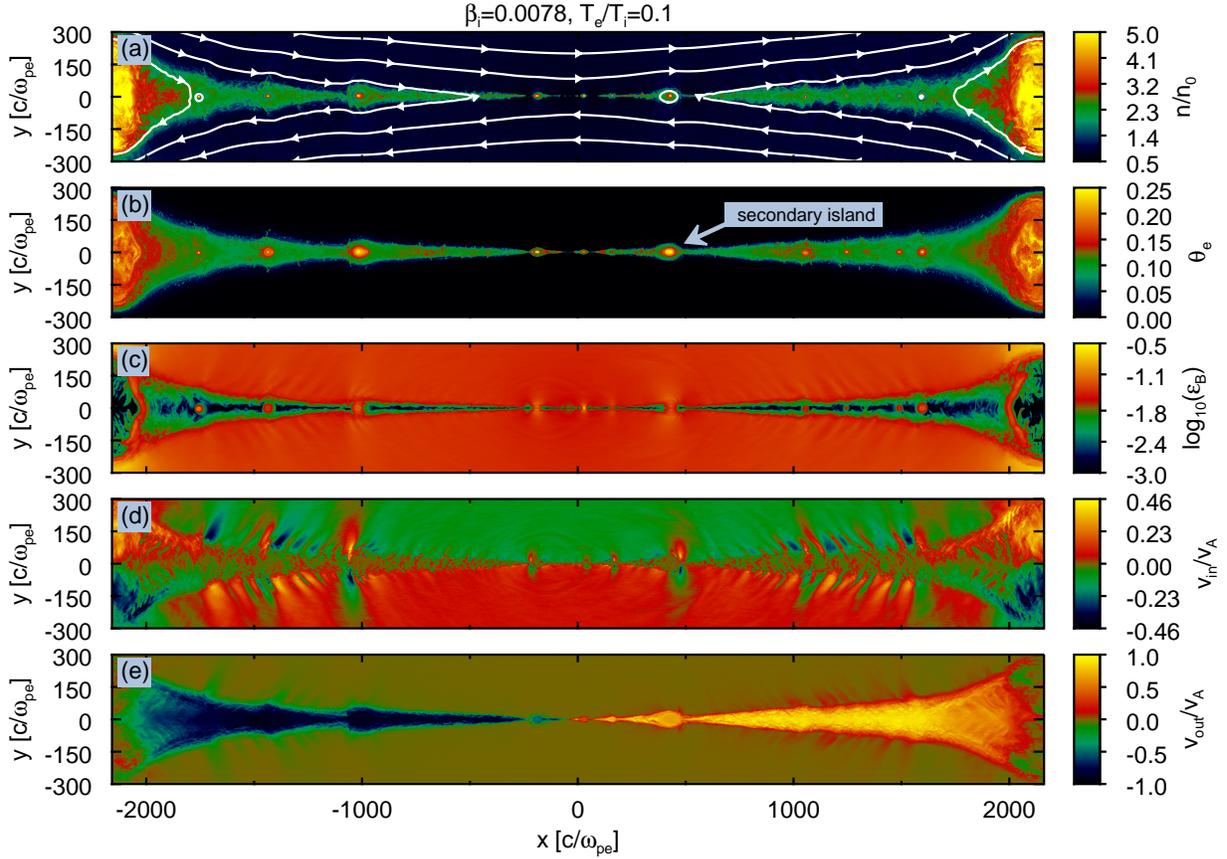} \\
			\caption{$\,$2D structure at $t = 11250\,\omega_{\rm pe}^{-1} \approx 0.75 \,t_{\rm A}$ from a representative low-$\betai$ simulation (\textbf{A[0]} in Tab.  \ref{tab:params}) with $\betai=0.0078$, $\sigma_w=0.1$, $\teti=0.1$ and $\mime=25$. We present  2D plots of (a): particle density in units of the  upstream  initial value, $n/n_{0}$, with overplotted magnetic field lines; (b): dimensionless electron temperature, $\theta_{\rm e}$; (c): logarithm of the magnetic energy fraction, $\varepsilon_{B}=B_{0}^{2}/8 \pi n_0 m_{\rm i} c^{2}$; (d): inflow velocity, in units of Alfv\'{e}n speed $v_{\rm in}/v_{\rm A}=\mathbf{v} \cdot \mathbf{\hat{y}}/v_{\rm A};$ (e): outflow velocity, in units of Alfv\'{e}n speed $v_{\rm out}/v_{\rm A}=\mathbf{v} \cdot \mathbf{\hat{x}}/v_{\rm A}$. We show the full extent of the domain in the $x$ direction ($L_{x}=4318\,c/\omega_{\rm pe}$), and only a small fraction of the box close to the current sheet in the $y$ direction.		
The primary island, which contains the particles initialized in the current sheet, can be seen at the boundaries ($x=\pm2200\,c/\omega_{\rm pe}$).
As shown in panel (a), the density reaches $n/n_{0}\approx 2.3$ in the bulk of the outflow, with sharp increases up to $n/n_{0}\approx 5$ in the core of secondary islands (e.g., at $x=-1000\,c/\omega_{\rm pe}$ and  $x=300\,c/\omega_{\rm pe}$).  The primary island has a high density throughout its interior, $n/n_{0}\approx 5.$
Similarly, the temperature (panel (b)) is uniform $\theta_{\rm e} \approx 0.1$ in the bulk of the outflow, with spikes up to $\theta_{\rm e} \approx 0.25$ at the center of secondary islands.  The primary island has a temperature $\theta_{\rm e} \approx 0.15$ throughout its interior.
In panel (c), we show that the magnetic energy fraction $\varepsilon_{B}$ is extremely small in the outflow, $\varepsilon_{B} \lesssim 0.01$.
The inflow velocity in panel (d) is a fraction of the Alfv\'{e}n limit $|v_{\rm in}|/v_{\rm A}\approx0.08$, and  the outflow velocity in panel (e) approaches the Alfv\'{e}n limit, $|v_{\rm out}|/v_{\rm A} \approx 1.$
\label{fig:lowbeta2d}} 
\end{figure*}
\begin{figure*}
		\centering
		\includegraphics[width=\textwidth,trim={0 8cm 0 5cm},clip]{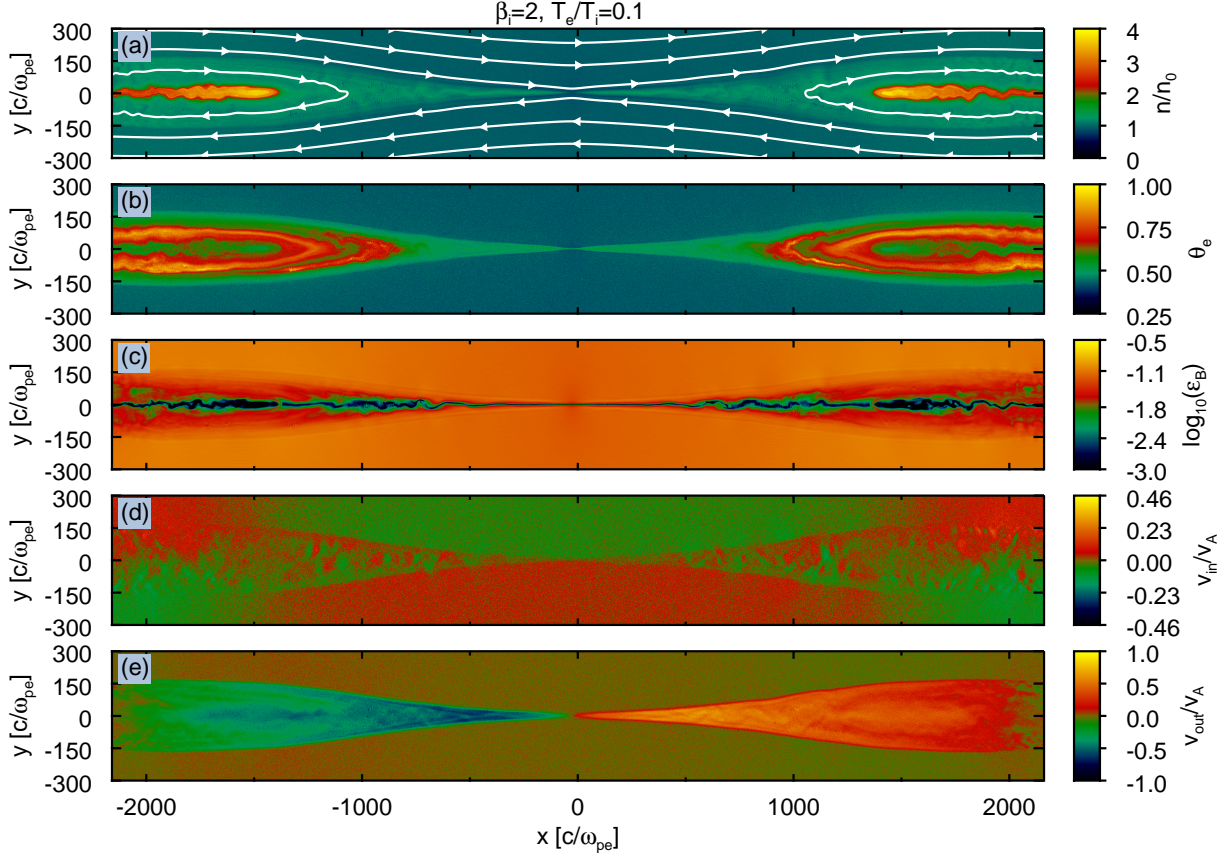} \\
			\caption{2D structure at $t = 11250\,\omega_{\rm pe}^{-1} \approx 0.75 \,t_{\rm A}$ from a representative high-$\betai$ simulation (\textbf{A[4]} in Tab.  \ref{tab:params}) with $\betai=2$, $\sigma_w=0.1$, $\teti=0.1$ and $\mime=25$ (i.e., apart from $\betai$, with the same parameters as in Fig.~\ref{fig:lowbeta2d}). The panels show the same quantities as in Fig.~\ref{fig:lowbeta2d}.   
As shown in panel (a), the density is roughly $n/n_{0}\approx 1.2$ in the bulk of the outflow, which is only slightly larger than the upstream density.  In the primary island, the density reaches $n/n_{0}\approx 4.$
The electron temperature (panel (b)) is nearly uniform in the reconnection exhausts (i.e., within a distance of $\approx700\,c/\omega_{\rm pe}$ from the central X-point), with $\theta_{\rm e} \approx 0.6$. Within the primary island, the temperature reaches $\theta_{\rm e} \approx 0.8.$
In panel (c), we present the logarithm of magnetic energy fraction $\varepsilon_{B}$, showing that the reconnection layer is weakly magnetized ($\varepsilon_{B} \lesssim 0.01$).
Panel (d) shows that the inflow velocity is nearly uniform in the upstream, with a typical value $|v_{\rm in}|/v_{\rm A}\approx0.04.$
Panel (e) shows that the outflow velocity in the reconnection exhausts is $|v_{\rm out}|/v_{\rm A}\approx 0.6$.  At the center of the primary island, $x\approx\pm2200\,c/\omega_{\rm pe},$ the plasma from the reconnection outflows comes to rest, $|v_{\rm out}|/v_{\rm A}\approx 0.$  
			\label{fig:highbeta2d}} 
	\end{figure*}

\begin{figure*}
		\centering
		\includegraphics[width=\textwidth,trim={0 7.5cm 0 7.5cm},clip]{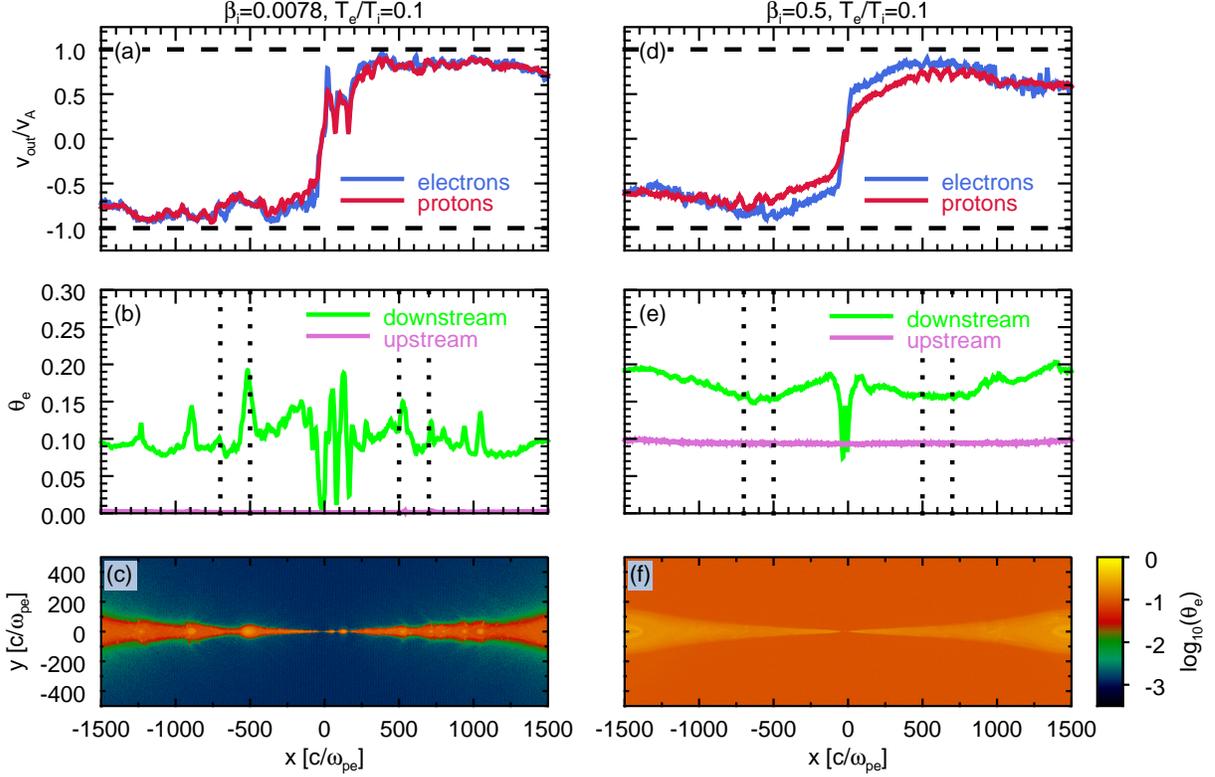} \\
			\caption{Comparison between a low-$\betai$ (left column, with $\betai=0.0078$, \textbf{A[0]} in Tab.  \ref{tab:params}) and  a high-$\betai$ (right column, with $\betai=0.5$, \textbf{A[3]} in Tab.~\ref{tab:params}) simulation, at time $ t = 9225\,\omega_{\rm pe}^{-1} \approx 0.65\,t_{\rm A}$. In both cases, $\sigma_w=0.1$, $\teti=0.1$ and $\mime=25$.  (a),(d): 1D profiles along $x$ (averaged along $y$ within the reconnection downstream, as identified by Eq. \ref{eq:criterion1}) of proton (red) and electron (blue) outflow velocity in units of the Alfv\'{e}n speed, $v_{\rm out}/v_{\rm A}$; (b),(e): 1D profiles along $x$ of the upstream (magenta) and downstream (green) dimensionless electron temperature, $\theta_{\rm e}$ (the two slabs in between the vertical dotted lines indicate the regions we use to calculate the downstream and upstream temperatures); (c),(f): 2D plots of $\log(\theta_{\rm e})$.  In both the low- and high-$\beta_{\rm i}$ cases, the spatial profiles of outflow velocity and electron temperature show that the downstream region reaches a quasi-steady state.   \label{fig:saturation}} 
	\end{figure*}

\begin{figure}
		\centering
		\includegraphics[width=0.45\textwidth]{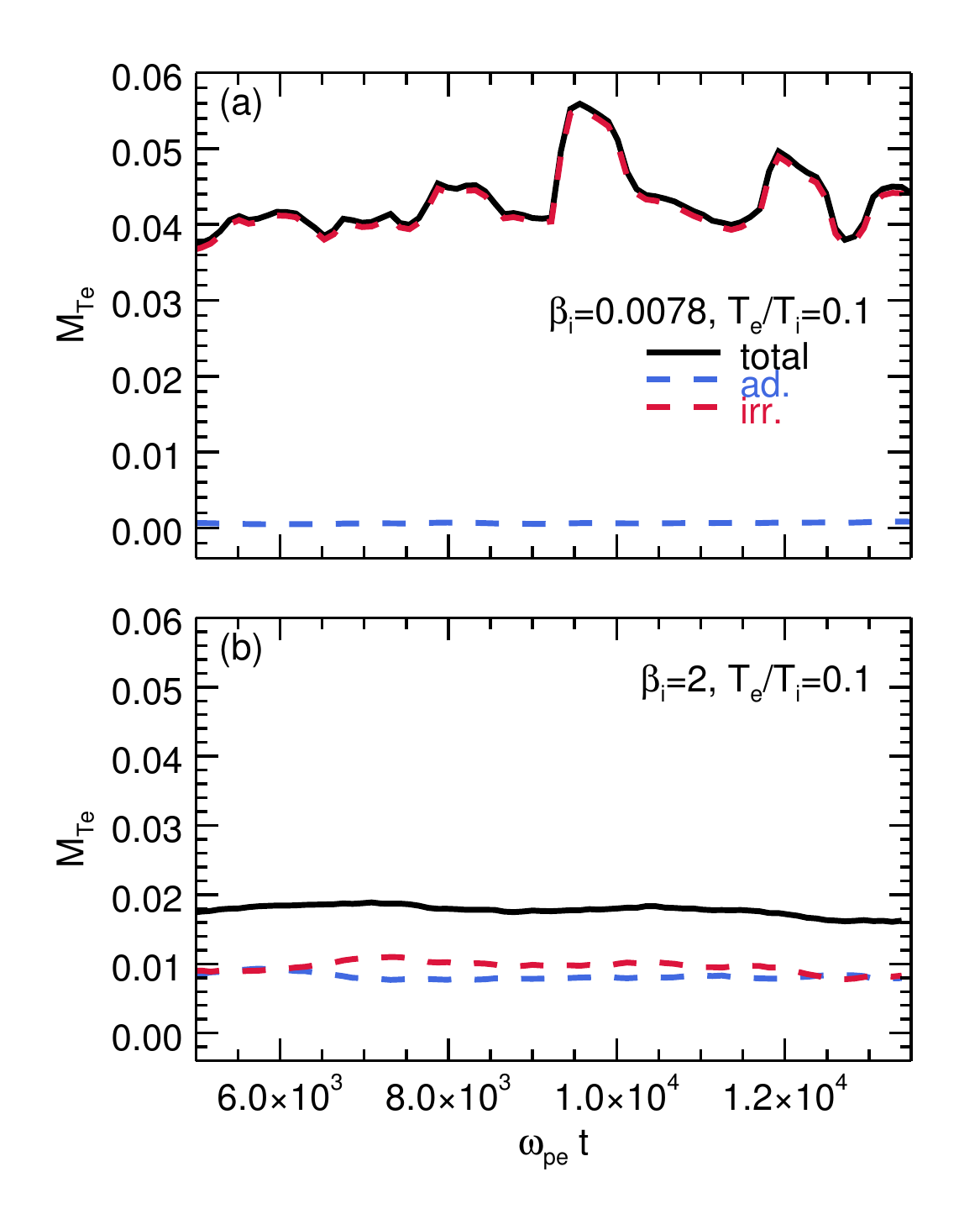}
			\caption{Time evolution of total ($M_{T\rm e,tot}$; black solid), irreversible ($M_{T\rm e,irr}$; red dashed), and adiabatic ($M_{T\rm e,ad}$; blue dashed) heating efficiency, for a low-$\betai$ simulation (top panel, with $\beta_{\rm i}=0.0078$) and a high-$\betai$ case (bottom panel, with $\beta_{\rm i}=2$). In both cases, $\sigma_w=0.1$, $\teti=0.1$ and $\mime=25$. The heating efficiencies are measured starting at $t \approx 5000\,\omega_{\rm pe}^{-1},$ at which point the two reconnection wavefronts recede past the location of the downstream region used for our computations (shown in Fig.~\ref{fig:mixing2d} with the green contours).  For the low-$\beta_{\rm i}$ case, the total heating efficiency oscillates around $M_{T\rm e} \approx 0.04,$ and it is dominated by genuine/irreversible heating (panel (a)).  For high $\beta_{\rm i},$ the total heating efficiency saturates at a smaller value, $M_{T\rm e}\approx0.016.$ Here, adiabatic and irreversible heating equally contribute (panel (b)).
			\label{fig:MTetimeevol}} 
	\end{figure}

\section{Results} \label{sec:results}
In this section, we describe our main results, focusing on the dependence of the heating efficiency on the plasma conditions.
First, in Section \ref{ssec:examples}, we present the dynamics of the reconnection layer, and describe the main differences between low-$\beta_{\rm i}$ and high-$\beta_{\rm i}$ cases, for our fiducial magnetization $\sigma_w=0.1$ and mass ratio $m_{\rm i}/m_{\rm e}=25$.
Next, in Section \ref{ssec:inflowoutflow}, we discuss the inflow and outflow rates as a function of $\beta_{\rm i}$ and $T_{\rm e}/T_{\rm i}.$
Then, in Section \ref{ssec:moneyplots}, we show the dependence of electron and proton heating on $\beta_{\rm i}$ and $T_{\rm e}/T_{\rm i}, $ still for our fiducial magnetization $\sigma_w=0.1$ and mass ratio $m_{\rm i}/m_{\rm e}=25$.
In Section \ref{ssec:massratio}, we extend our results for $T_{\rm e}/T_{\rm i}=1$ and $\sigma_w=0.1$ up to the physical mass ratio  $m_{\rm i}/m_{\rm e}=1836,$ emphasizing the $\beta_{\rm i}$-dependence of the particle heating efficiencies.
Finally, in Section \ref{ssec:sigdep}, we show how the heating physics changes when the magnetization $\sigma_{w}$ extends above unity (i.e., in the regime of ultra-relativistic reconnection), for mass ratio $m_{\rm i}/m_{\rm e}=1836$ and temperature ratio  $\teti=1$.

\subsection{Reconnection physics as a function of \texorpdfstring{$\beta_{\MakeLowercase{i}}$}{betai}} \label{ssec:examples}
The physics of reconnection shows a marked difference between low- and high-$\beta_{\rm i}$ regimes.  In Figs. \ref{fig:lowbeta2d} and \ref{fig:highbeta2d}, we present various fluid quantities for representative low- and high-$\beta_{\rm i}$ simulations, respectively ($\beta_{\rm i}=0.0078$ in Fig.~\ref{fig:lowbeta2d} and $\beta_{\rm i}=2$ in Fig.~\ref{fig:highbeta2d}). In both cases, $\sigma_w=0.1$, $T_{\rm e}/T_{\rm i}=0.1$ and $m_{\rm i}/m_{\rm e}=25$. 
At $ t = 11250\,\omega_{\rm pe}^{-1} \approx 0.75\,t_{\rm A}$, we show 2D plots of: (a) the total density in the simulation frame in units of the initial density, $n/n_{0}$; (b) the dimensionless electron temperature $\theta_{\rm e}$; (c) the magnetic energy fraction $\varepsilon_{B}=B^{2}/8 \pi n_0 m_{\rm i} c^{2}$; (d) the inflow velocity $v_{\rm in}/v_{\rm A}=\mathbf{v} \cdot \mathbf{\hat{y}} / v_{\rm A}$ ($v_{\rm A}$ is the Alfv\'{e}n speed), and (e) the outflow velocity $v_{\rm out}/v_{\rm A}=\mathbf{v} \cdot \mathbf{\hat{x}} / v_{\rm A}$.

A striking difference between the simulations shown in Figs.~\ref{fig:lowbeta2d} and \ref{fig:highbeta2d} is that, while the reconnection outflow at high $\beta_{\rm i}$ is nearly homogeneous, a number of secondary magnetic islands appear at low $\beta_{\rm i}$ (see Fig.~\ref{fig:lowbeta2d}(a)). The secondary islands are over-dense, and at their center they can reach temperatures a few times larger than  the bulk of the outflow (Fig.~\ref{fig:lowbeta2d}(b)). They also correspond to peaks in magnetic energy (Fig.~\ref{fig:lowbeta2d}(c)).

The difference in electron temperature between inflow and outflow regions  is more pronounced in the low- than in the high-$\beta_{\rm i}$ case (compare Figs. \ref{fig:lowbeta2d}(b) and \ref{fig:highbeta2d}(b)).  However, as we demonstrate in Section \ref{ssec:moneyplots}, the fraction of available magnetic energy converted into total electron heating is roughly comparable between the two cases.

The inflow velocity $v_{\rm in}/v_{\rm A}=\mathbf{v} \cdot \mathbf{\hat{y}}/v_{\rm A}$ is shown in panel (d).  
For low-$\beta_{\rm i}$, the inflow velocity is $|v_{\rm in}|/v_{\rm A} \approx 0.08.$ It is nearly uniform in the upstream, with the exception of the regions ahead of the secondary islands, where the velocity reverses its sign relative to the ambient inflow (see, e.g., Fig.~\ref{fig:lowbeta2d}(d) at $x\approx-1100\,c/\omega_{\rm pe}).$  This reversal occurs as the secondary island moves along the outflow direction, pushing aside the inflowing plasma.  For high-$\beta_{\rm i},$ the plasma inflow is remarkably uniform, with  $|v_{\rm in}|/v_{\rm A}\approx0.04,$ which is half the value of the low-$\beta_{\rm i}$ case.  The inflow velocity at high $\betai$ shows no reversals near the reconnection exhausts, as there are no secondary islands.

The outflow velocity $v_{\rm out}/v_{\rm A}=\mathbf{v} \cdot \mathbf{\hat{x}}/v_{\rm A}$ is shown in panel (e).  For low-$\beta_{\rm i},$ the outflow speed
nearly reaches the Alfv\'{e}n limit, $|v_{\rm out}|/v_{\rm A} \approx 1$, whereas for high-$\beta_{\rm i}$ it approaches a smaller value, $|v_{\rm out}|/v_{\rm A} \approx 0.6.$  For both low and high $\beta_{\rm i},$ the outflow velocity is nearly uniform in the reconnection exhausts, but it drops close to the periodic boundaries at $x\approx\pm2200,$ as the outflowing plasma accretes onto the primary island. 

We show in Fig.~\ref{fig:saturation} a direct comparison between one low-$\betai$ and one high-$\betai$ simulation. The left column in Fig.~\ref{fig:saturation} refers to $\betai=0.0078$ (the same as in Fig. \ref{fig:lowbeta2d}), whereas $\betai=0.5$ for the right column. In both cases, $\sigma_w=0.1$, $\teti=0.1$   and $\mime=25$.
In the top row, we show the profile along $x$ of the outflow velocity, for protons (red) and electrons (blue). We find that electrons move slightly faster than protons in the vicinity of the central X-point, but at larger distances the  speeds of the two species are the same, and they saturate at a fixed fraction of the Alfv\'{e}n limit.
 We show in the middle row of panels the $x$-profile of the dimensionless electron temperature $\theta_{\rm e}$, in the upstream (magenta) and downstream (green). The secondary magnetic islands present in the low-$\beta_{\rm i}$ simulation (panel (c)) are correlated with spikes in the downstream electron temperature (see the peak at $x\approx -500\,c/\omega_{\rm pe}$ in Fig.~\ref{fig:saturation}(b)). Aside from the temperature spikes at low $\beta_{\rm i}$, the two panels in the middle row of Fig.~\ref{fig:saturation} demonstrate that, far enough from  the central X-point, the  electron temperature is nearly uniform.  
 
To estimate the reconnection heating efficiency, we measure the downstream temperature in the two slabs delimited by the vertical dotted lines in Fig.~\ref{fig:saturation}(b) and (e) (more precisely, within the green contours in Fig.~\ref{fig:mixing2d}). The time evolution of the total electron heating efficiency $M_{T\rm e, tot}$, of the adiabatic contribution $M_{T\rm e,ad}$ and of the irreversible component $M_{T\rm e,irr}$ are shown in  Fig.~\ref{fig:MTetimeevol} with black, dashed blue and dashed red lines, respectively. The top panel refers to a low-$\betai$ simulation with $\betai=0.0078$, whereas the bottom panel refers to the high-$\betai$ case $\betai=2$. In both cases, $\sigma_w=0.1$, $\teti=0.1$   and $\mime=25$.
The horizontal axis in the figure starts from $t = 5000\,\omega_{\rm pe}^{-1}$, which corresponds to the time when the two reconnection wavefronts pass beyond the region that we employ for calculating the downstream quantities (as discussed above, after this time the measurements are no longer affected by our choice of initialization of the current sheet).\footnote{This time is typically in the range $t \approx 4000-5000\,\omega_{\rm pe} ^{-1}$, with marginal dependence on $\beta_{\rm i}$ and on the initial sheet thickness $\Delta$.} While the heating efficiencies are nearly constant in time for high $\beta_{\rm i}$ (bottom panel), the temporal profiles at low $\beta_{\rm i}$ (top panel) present quasi-periodic modulations. They mark the passage of secondary islands --- whose temperature is typically hotter than the bulk outflow --- through the region used for our computations. To minimize the temperature variations associated with secondary islands, we average the heating efficiencies over time, as described in Section \ref{sssec:efficiency}. In doing so, the results we obtain are a reliable assessment of the steady-state heating physics in reconnection.
 
  
  Panels (a) and (b) in Fig.~\ref{fig:MTetimeevol} also demonstrate that the fractional contributions of adiabatic and irreversible heating to the total electron heating significantly depend  on $\beta_{\rm i}$, as we further discuss in Section \ref{ssec:moneyplots}. In the low-$\beta_{\rm i}$ regime, adiabatic heating is unimportant  as compared to the irreversible part, whereas the two components are comparable at high $\beta_{\rm i}$.


\begin{figure}
		\centering
		\includegraphics[width=0.43\textwidth,trim={0.75cm 3cm 0 0cm},clip]{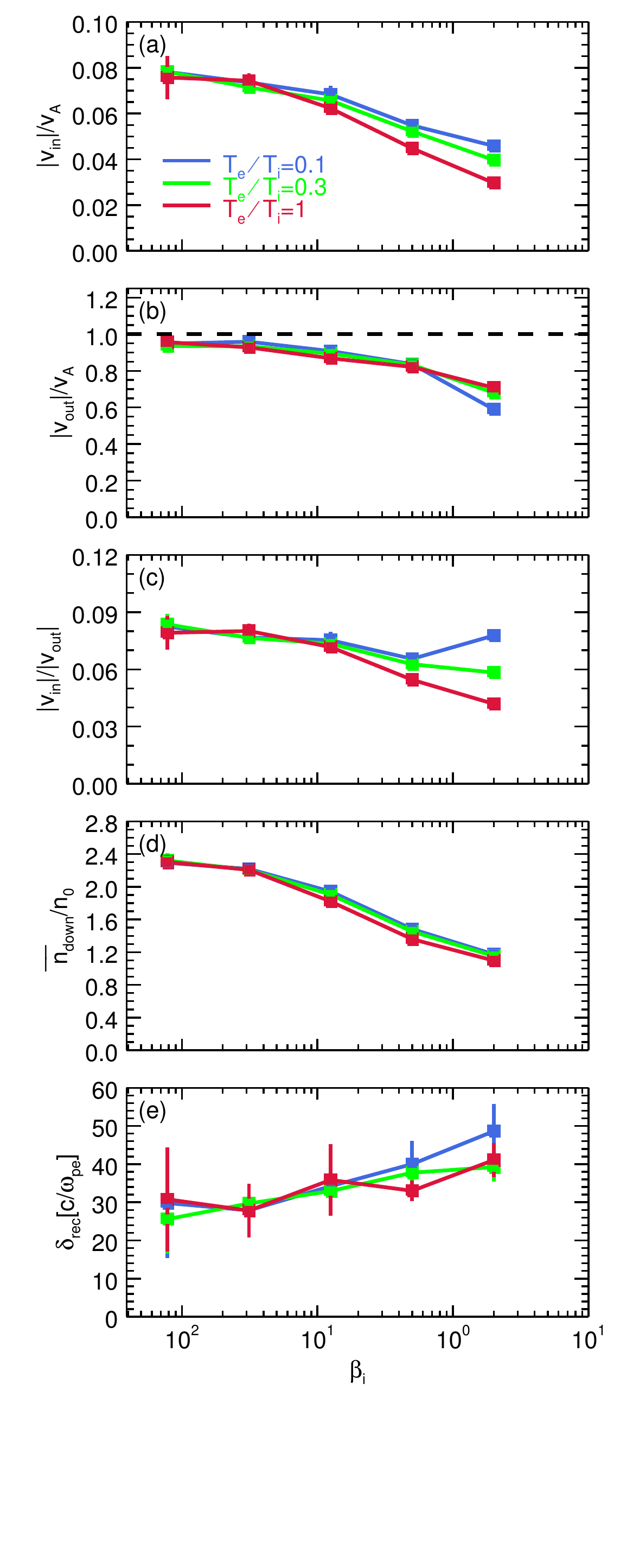}
			\caption{For temperature ratios $T_{\rm e}/T_{\rm i} = 0.1$ (blue), $0.3$ (green), and 1 (red), $\beta_{\rm i}$-dependence of (a): inflow velocity $|v_{\rm in}|/v_{\rm A}$; (b): outflow velocity $|v_{\rm out}|/v_{\rm A};$ (c): reconnection rate $|v_{\rm in}|/|v_{\rm out}|$; (d): downstream density in units of initial density in the upstream $\xoverline{n}_{\rm down}/n_{0}$; (e): width of reconnection layer $\delta_{\rm rec}$.  Error bars represent one standard deviation from the mean.  
The inflow velocity is averaged over a region of length $L_{x}/3\approx 1440\,c/\omega_{\rm pe}$ in $x$ and width $20\,c/\omega_{\rm pe}$ in $y$, located $|y|\sim 100\,c/\omega_{\rm pe}$ upstream of the central X-point.  
We have checked that the saturation value is insensitive to the choice of  averaging region. 
 The outflow velocity is computed as an average over the 20 cells with the largest $|\mathbf{v} \cdot \mathbf{\hat{x}}|$ located along the central region of the outflow ($|y|\lesssim4\,c/\omega_{\rm pe}$).  We have tested that the resulting outflow velocity is nearly insensitive to our averaging procedure.
 The regions used for measuring density in the upstream and downstream are described in Section \ref{sec:updownid}.  The width of the reconnection layer is measured at a distance $\sim430\,c/\omega_{\rm pe}$ downstream of the central X-point.  All quantities are time-averaged over $\sim0.3\,t_{A} \approx 4500\,\omega_{\rm pe}^{-1}$.  Both inflow and outflow velocities tend to decrease with $\beta_{\rm i}$, with weak dependence on $\teti$ (noticeable only at high $\beta_{\rm i}$).  
 The density compression decreases with $\beta_{\rm i}.$ The width $\delta_{\rm rec}$ of the layer increases with $\beta_{\rm i}$, yet with large error bars.}  \label{fig:betainout}
\end{figure}

\vspace{0.1in}
\subsection{Dependence of inflow and outflow velocity on \texorpdfstring{$\beta_{\MakeLowercase{i}}$}{betai} and \texorpdfstring{$T_{\MakeLowercase{e}}/T_{\MakeLowercase{i}}$}{teti}}	\label{ssec:inflowoutflow}
In Fig.~\ref{fig:betainout}, we show the dependence on $\beta_{\rm i}$ and $\teti$ of various fluid quantities, from a suite of simulations with fixed $\sigma_w=0.1$ and $\mime=25$. We present the (a) inflow velocity normalized to the Alfv\'en speed $|v_{\rm in}|/v_{\rm A}$; (b) outflow velocity normalized to the Alfv\'en speed $|v_{\rm out}|/v_{\rm A}$; (c) ratio of inflow to outflow velocity $|v_{\rm in}|/|v_{\rm out}|$; (d) downstream rest-frame density in units of the initial density in the upstream $\bar{n}_{\rm down}/n_{0}$; (e) width of the reconnection region at a distance of $430\,c/\omega_{\rm pe}$ from the center of the layer. Blue, green, and red points denote simulations with upstream temperature ratios $T_{\rm e}/T_{\rm i} = 0.1, 0.3,$ and $1, $ respectively.

As described in Section~\ref{sssec:efficiency}, the quantities we extract are time-averaged, typically over 0.3 Alfv\'{e}nic crossing times, corresponding to $\sim 4500 \,\omega_{\rm pe}^{-1}.$  The points in Fig.~\ref{fig:betainout} represent the time averages, with vertical error bars indicating one standard deviation.  
At low $\beta_{\rm i},$ the inflow velocity is $|v_{\rm in}|/v_{\rm A} \approx 0.08,$ independent of the upstream temperature ratio (panel (a)). 
In the high-$\beta_{\rm i}$ case, the inflow speed is smaller, $|v_{\rm in}|/v_{\rm A}\approx0.04$, and shows a weak dependence on the temperature ratio, with higher temperature ratios attaining lower values of $|v_{\rm in}|/v_{\rm A}$.

The outflow velocity (panel (b)) nearly saturates the Alfv\'{e}n limit at low $\beta_{\rm i}$ (the Alfv\'{e}n limit is indicated with the horizontal dashed black line), whereas for high $\beta_{\rm i}$ it is sub-Alfv\'{e}nic, $|v_{\rm out}|/v_{\rm A} \approx 0.75$.
For low values of $\beta_{\rm i}$, i.e.,  $\beta_{\rm i} \lesssim 0.1,$ the outflow velocity is nearly independent of the temperature ratio, whereas at high $\beta_{\rm i}$ it shows a weak dependence on $\teti$, with higher temperature ratios corresponding to greater outflow speeds.  

The dependence of the reconnection rate $|v_{\rm in}|/|v_{\rm out}|$ on $\betai$ and $\teti$ (panel (c)) follows from the variations in inflow speed and outflow velocity that we have just discussed.  At low $\beta_{\rm i},$ the reconnection rate is  $|v_{\rm in}|/|v_{\rm out}| \approx 0.08$ regardless of the temperature ratio, whereas at high $\beta_{\rm i}$, and specifically for $\beta_{\rm i} = 2,$ the reconnection rate at $T_{\rm e}/T_{\rm i} =1$ is $|v_{\rm in}|/|v_{\rm out}| \approx 0.04,$ only half that of the $T_{\rm e}/T_{\rm i} =0.1$ case. So, in the high-$\beta_{\rm i}$ regime reconnection proceeds slower for hotter upstream electrons.

As $\beta_{\rm i}$ increases, the plasma is less prone to be compressed during the reconnection process.  As shown in Fig.~\ref{fig:betainout} (d), the downstream to upstream density ratio decreases as $\beta_{\rm i}$ increases.  The value of $\xoverline{n}_{\rm down}/n_{0}$ is nearly independent of the upstream temperature ratio.
  Though the ratio $\xoverline{n}_{\rm down}/n_{0}$ approaches  unity at high $\beta_{\rm i},$ this does not necessarily imply that the fractional contribution of adiabatic heating to total heating is negligible at high $\beta_{\rm i}$ (we demonstrate this in Section~\ref{ssec:moneyplots}).
  
Lastly, in panel (e) we show the $\beta_{\rm i}$-dependence of the reconnection layer width $\delta_{\rm rec},$ in units of the electron skin depth $c/\omega_{\rm pe}$.  We measure the width across the reconnection layer, as identified by Eq.~\ref{eq:criterion1}, at a distance $\sim430\,c/\omega_{\rm pe}$ downstream of the central X-point.  The width shows strong variability in time at low $\beta_{\rm i}$, as secondary islands pass through the  region employed for our measurements (note the large error bars). Despite the uncertainty in the measurement, panel (e) shows a consistent trend of increasing reconnection layer width  $\delta_{\rm rec}$ with $\beta_{\rm i}$.  The measured values of $\delta_{\rm rec}$ lie in the range $25$ -- $50\,c/\omega_{\rm pe},$ which is apparently close to the chosen current sheet thickness at initialization, $\Delta = 40\,c/\omega_{\rm pe}.$  However, we demonstrate in Appendix \ref{sec:recl_conv} that the measured reconnection layer width is independent of our choice of the initial sheet thickness. It follows that our measurement leads to a reliable assessment of the opening angle of the reconnection outflow, $\sim \delta_{\rm rec}/(430\,c/\omega_{\rm pe})\sim 0.1$.

\begin{figure}
		\centering
		\includegraphics[width=0.45\textwidth,trim={0.2cm 1cm 0 1cm},clip]{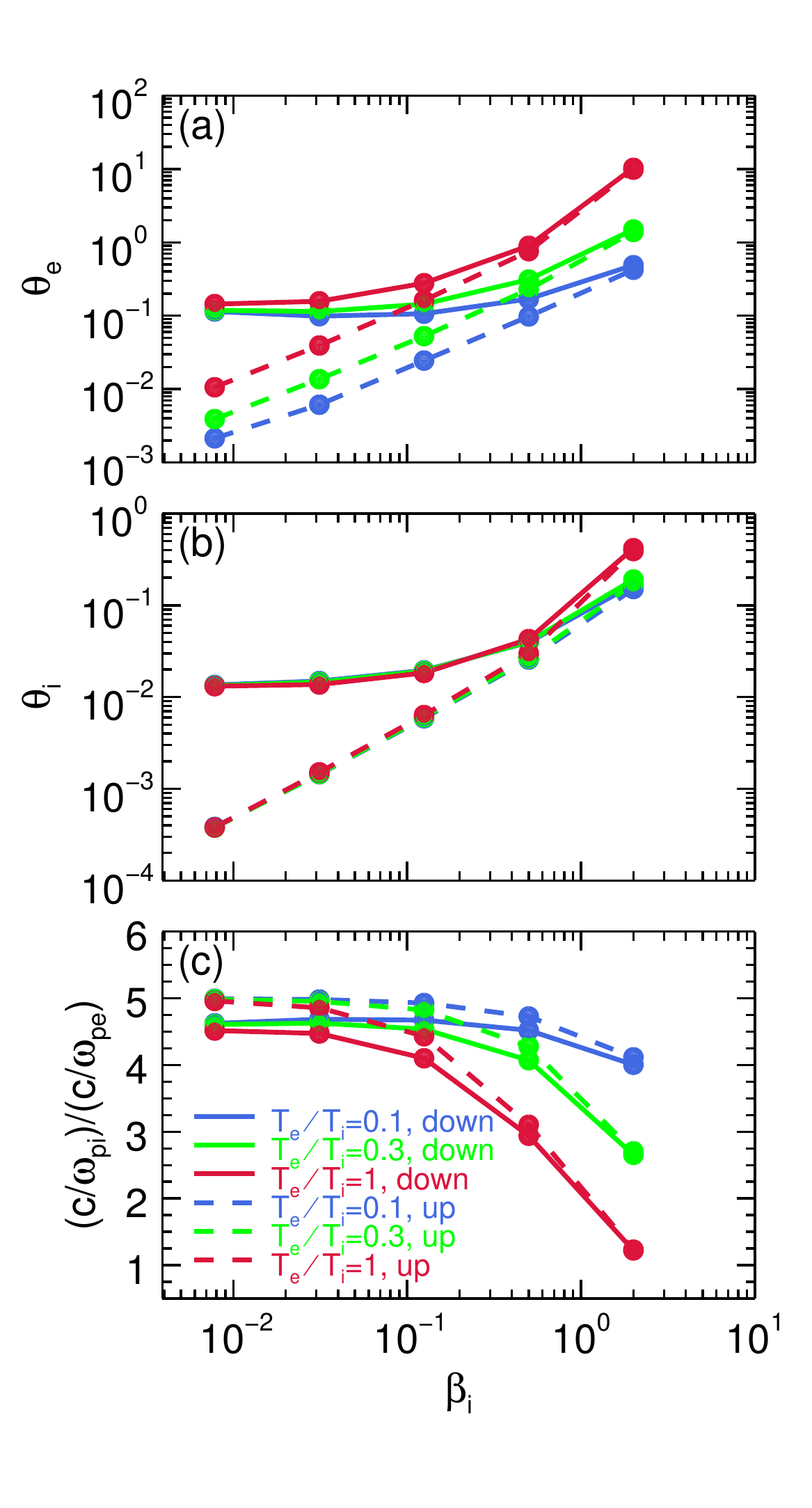} \\
			\caption{For upstream temperature ratios $T_{\rm e}/T_{\rm i} =0.1$ (blue), 0.3 (green), and 1 (red), we present the $\beta_{\rm i}$-dependence of various upstream (dashed) and downstream (solid) quantities; (a): electron dimensionless temperature, $\theta_{\rm e}$; (b): proton dimensionless temperature, $\theta_{\rm i}$; (c): proton-to-electron skin depth ratio, $(c/\omega_{\rm pi})/(c/\omega_{\rm pe}).$ The simulations shown here use a mass ratio $m_{\rm i}/m_{\rm e}=25$ and magnetization $\sigma_{w}=0.1$. 
			 \label{fig:theta_updown}} 
	\end{figure}

\begin{figure*}
		\centering
		\includegraphics[width=\textwidth]{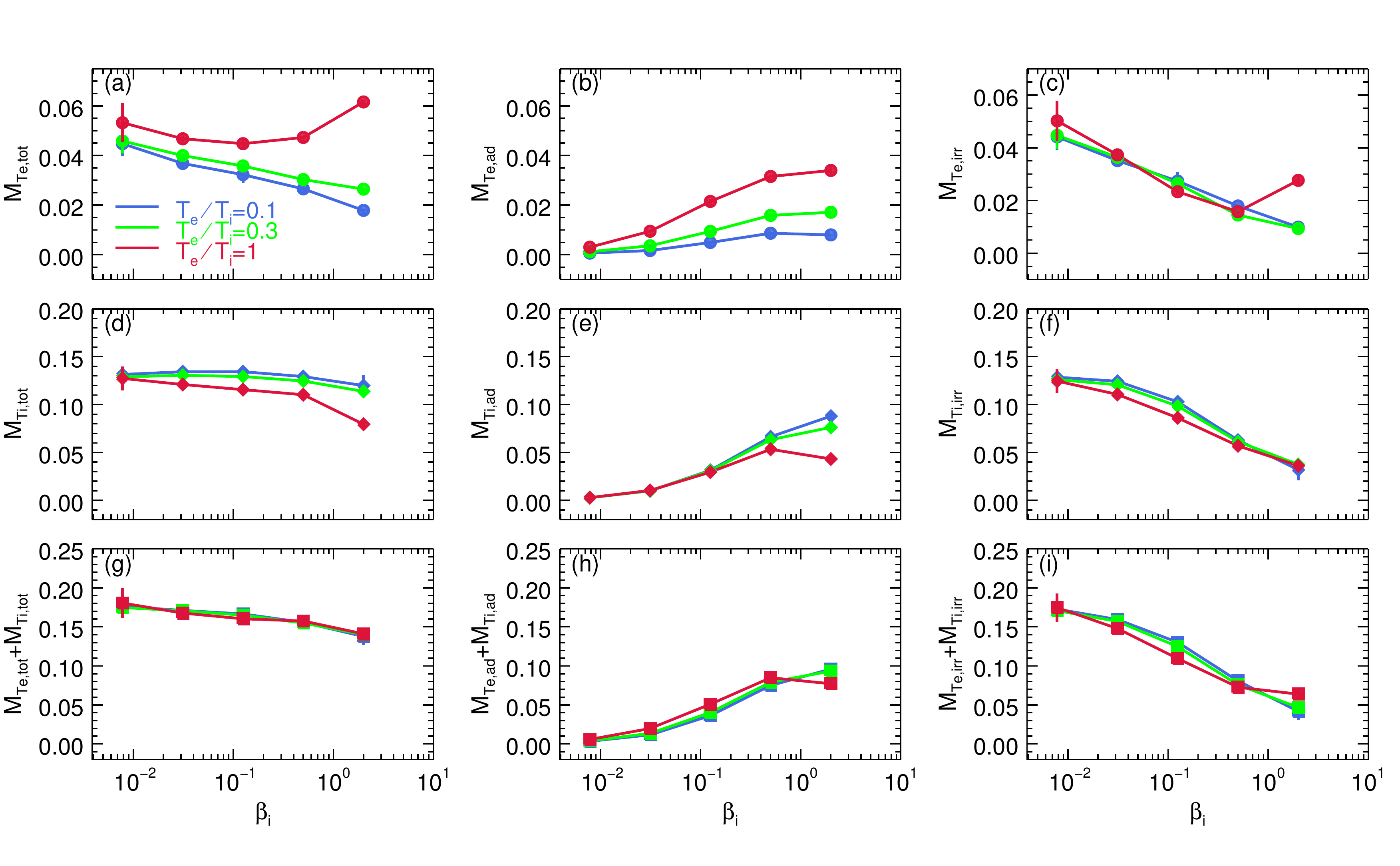} \\
			\caption{For upstream temperature ratios $T_{\rm e}/T_{\rm i} =0.1$ (blue), 0.3 (green), and 1 (red), $\beta_{\rm i}$-dependence of heating efficiencies; (a): electron total, $M_{T\rm e, tot}$; (b): electron adiabatic, $M_{T\rm e,ad}$; (c): electron irreversible, $M_{T\rm e,irr}$; (d): proton total, $M_{T\rm i, tot}$; (e): proton adiabatic, $M_{T\rm i,ad}$; (f): proton irreversible, $M_{T\rm i,irr}$; (g): electron and proton total, $M_{T\rm e, tot}+M_{T\rm i, tot}$; (h): electron and proton adiabatic, $M_{T\rm e,ad}+M_{T\rm i,ad}$; (i): electron and proton irreversible, $M_{T\rm e,irr}+M_{T\rm i,irr}$.  The simulations shown here use a mass ratio $m_{\rm i}/m_{\rm e}=25$ and magnetization $\sigma_{w}=0.1$.  Error bars, mostly smaller than the plotted symbols, represent one standard deviation from the mean.  The decomposition of total heating into irreversible and adiabatic components shows that electron and proton heating at low $\beta_{\rm i}$ is accompanied by an increase in entropy, while heating in the high-$\beta_{\rm i}$ regime tends to be dominated by adiabatic compression. 
			 \label{fig:mttt}} 
	\end{figure*}

\subsection{Dependence of particle heating on \texorpdfstring{$\beta_{\MakeLowercase{i}}$}{betai} and \texorpdfstring{$T_{\MakeLowercase{e}}/T_{\MakeLowercase{i}}$}{teti}}	\label{ssec:moneyplots}	
In Fig.~\ref{fig:theta_updown}, we show the $\beta_{\rm i}$ and $T_{\rm e}/T_{\rm i}$ dependence of electron (panel (a)) and proton (panel (b)) dimensionless temperature, and the ratio of proton-to-electron skin depth (panel (c); see Eq.~\ref{eq:skindepth}).  In each panel, solid and dashed lines indicate downstream and upstream quantities, respectively.
As in Fig.~\ref{fig:betainout}, blue, green, and red points refer to electron-to-proton temperature ratios $T_{\rm e}/T_{\rm i}=0.1, 0.3$, and $1$, respectively.  The upstream electron dimensionless temperatures lie in the range $10^{-3}$ to $10$, as in Table~\ref{tab:params}; for protons, the dimensionless temperature in the upstream ranges from $4\times10^{-4}$ to $0.4$.  

The range of temperatures in the downstream is smaller than in the upstream (compare the solid and dashed lines in Fig.~\ref{fig:theta_updown}(a) and (b)).  At low $\beta_{\rm i}$, the available magnetic energy is large compared to the particle thermal content in the upstream, so dissipation of the magnetic field leads to electron and proton temperatures in the downstream that are nearly independent of $\beta_{\rm i}$.  At high $\beta_{\rm i}$, the energy transferred from the fields to the particles is much smaller than the initial particle thermal content, giving a minor increase of temperature from upstream to downstream. 
Even if the fractional increase in temperature is extremely small at high $\beta_{\rm i}$, the fraction of available magnetic energy being converted into particle heating might still be as large as at low $\beta_{\rm i}$. The rest of the section addresses this question.

We show the plasma-$\beta_{\rm i}$ and temperature ratio $T_{\rm e}/T_{\rm i}$ dependence of electron and proton heating in Fig.~\ref{fig:mttt}.  The simulations presented here are those referenced in Tab.~\ref{tab:params}, for which $m_{\rm i}/m_{\rm e}=25$ and $\sigma_{w}=0.1$.  We indicate the total, adiabatic, and irreversible heating by $M_{T\rm e,tot}$ (Eq. \ref{eq:mte}), $M_{T\rm e,ad}$ and $M_{T\rm e,irr}$ (Eqs. \ref{eq:mtencmtec}) for electrons, and by $M_{T\rm i,tot}$ (Eq. \ref{eq:mti}), $M_{T\rm i,ad}$ and $M_{T\rm i,irr}$ (Eqs. \ref{eq:mtincmtic}) for protons.  Blue, green, and red points indicate simulations with upstream electron-to-proton temperature ratios of $0.1, 0.3,$ and $1$, respectively.  
As in Section \ref{ssec:inflowoutflow}, filled points are the time-averaged results of our simulations, and vertical error bars indicate one standard deviation from the mean.  The top, middle, and bottom rows show heating fractions of electrons,  protons, and of the overall fluid, respectively, which we now discuss in turn.

In Fig.~\ref{fig:mttt}(a), we show the dependence of the total electron heating efficiency $M_{T\rm e,tot}$ on $\beta_{\rm i}$ and $\teti$.  
Although the initial plasma $\beta_{\rm i}$ spans more than two orders of magnitude, and the initial temperature ratio an order of magnitude, even the most extreme values of $M_{T\rm e,tot}$ differ by no more than a factor of $\sim4$.  The value of $M_{T\rm e,tot}$ in our $\beta_{\rm i} \lesssim 0.5$ simulations, for which electrons stay non-relativistic both in the upstream and in the downstream, is $\sim 0.04,$ which is consistent with the results of non-relativistic reconnection by \citet{Shay2014} for mass ratio $m_{\rm i}/m_{\rm e}=25$.\footnote{In \citet{Shay2014}, the magnetization was $\sigma_w\approx0.004-0.1$. However, as long as $\sigma_w\ll1$ and all the species stay at non-relativistic temperatures, we expect the reconnection physics to be independent of the flow magnetization.}  As shown by \citet{Shay2014}, the electron heating efficiency in non-relativistic reconnection is expected to decrease with increasing mass ratio;  in Sections \ref{ssec:massratio} and \ref{ssec:tratio1836}, we present the dependence of the electron and proton heating fractions in trans-relativistic reconnection on $\mime$, up to the physical value.

The total electron heating fraction $M_{T\rm e,tot}$ is decomposed into adiabatic and irreversible components in panels (b) and (c).  By comparing the two panels, we see that most of the heating in the low-$\beta_{\rm i}$ regime comes from irreversible processes, i.e., it is accompanied by a genuine increase in entropy, while heating at high $\beta_{\rm i}$ mostly results from adiabatic compression.

The electron adiabatic heating efficiency increases with the inflow temperature ratio $T_{\rm e}/T_{\rm i}$ (Fig.~\ref{fig:mttt}(b)).  The dependence is most apparent at high $\beta_{\rm i},$ where adiabatic heating represents a significant contribution to the total heating.  The dependence of adiabatic heating on temperature ratio can be simply understood through the adiabatic law $T/\xoverline{n}^{\hat{\gamma}-1}=\text{const.}$  As electrons get compressed from upstream to downstream, the adiabatic heating fraction can be written as\footnote{Eq. \ref{eq:compapprox} assumes that the adiabatic index is constant as electrons pass from upstream to downstream, which is a good approximation when electrons are ultra-relativistic in both regions (so, for high $\beta_{\rm i}$ and large $\teti$); still, in all the simulations used in Fig.~\ref{fig:mttt}, we find that the electron adiabatic index   changes by no more than $\hat{\gamma}_{\rm e,up} - \hat{\gamma}_{\rm e,down} \approx0.1$.  In any case, Eq. \ref{eq:compapprox} is presented only for illustrative purposes, and we properly account for the effect of a variable adiabatic index in our calculation of the heating fractions.}
\begin{align} \label{eq:compapprox}
M_{T\rm e,ad} &= \frac{1}{2}\betai \frac{T_{\rm e}}{T_{\rm i}}\left[ \left(\frac{\xoverline{n}_{\rm down}}{n_{0}} \right)^{\hat{\gamma}_{\rm e}-1} - 1\right].
\end{align}
As shown in Fig.~\ref{fig:betainout}(d), the ratio of downstream to initial upstream density $\xoverline{n}_{\rm down}/n_{0}$ is nearly independent of the upstream temperature ratio, so that $M_{T\rm e,ad}\propto\teti$ at fixed $\betai$.
Eq. \ref{eq:compapprox} also provides insight into the $\beta_{\rm i}$-dependence of  adiabatic heating.  It shows that, for a given temperature ratio, the adiabatic heating efficiency would scale linearly with $\beta_{\rm i}$, if the compression ratio 
$\xoverline{n}_{\rm down} /n_{0}$ were to be fixed. As shown in Fig. \ref{fig:betainout}(d), the downstream to upstream density ratio decreases with $\betai$, approaching unity at high $\betai$. However, the decrease of $\xoverline{n}_{\rm down} /n_{0}$ with $\betai$ is quite shallow, and insufficient to counteract the linear dependence on $\beta_{\rm i}$ in Eq. \ref{eq:compapprox}. It follows that at low $\beta_{\rm i}$ the effect of adiabatic heating is negligible, while at high $\beta_{\rm i}$ the role of adiabatic heating can be more important than that of irreversible heating.

This statement can be further justified by considering electron energization in the diffusion region as the main source of irreversible electron heating, following \citet{Le2016}. In the diffusion region, the electron energy will increase by $e E_{\rm rec} \ell_{\rm e}$, where $E_{\rm rec}\sim 0.1 (v_{\rm A}/c) B_0$ is the reconnection electric field (assuming a reconnection inflow rate of $\sim 0.1 \,v_{\rm A}/c$, see Fig. \ref{fig:betainout}(a)) and $\ell_{\rm e}$ is the distance traveled by electrons along the electric field (along $z$, in our geometry). For the sake of simplicity, let us now assume that $\betai$ is sufficiently small that $w\sim n_0 m_{\rm i} c^2$ and so $\sigma_w\sim \sigma_{\rm i}$ (this is the case for $\betai\lesssim0.1$, at our reference magnetization $\sigma_w=0.1$). The corresponding irreversible heating efficiency can be written in the case $\sigma_w\sim \sigma_{\rm i}\lesssim1$ as 
\begin{equation}\label{eq:dwetot}
M_{T\rm e,irr}\sim 0.1\frac{\ell_{\rm e}}{c/\omega_{\rm pi}}
\end{equation}
which does not depend explicitly on $\beta_{\rm i}$. It follows that, as long as $\ell_{\rm e}$ is a weak function of $\beta_{\rm i}$, the adiabatic heating efficiency in Equation \ref{eq:compapprox}, which scales linearly with $\betai$, will be unimportant at low $\betai$, whereas it will dominate over irreversible heating at high $\betai$. 

We remark that Equation \ref{eq:dwetot} does not capture a number of important effects. First, by tracking individual particle orbits, we have found  that the in-plane components of the electric field, that we have neglected above, can provide a significant contribution to the total electron energization (a comprehensive discussion of the physics of electron heating will be presented elsewhere). Second, we have neglected the $\betai$-dependence of the reconnection rate. Third, we have assumed $w\sim n_0 m_{\rm i} c^2$, which is incorrect at high $\betai$. Fourth, we do not have a direct measure of $\ell_{\rm e}$, which would assess its dependence on the flow conditions. For these reasons, it is likely that the electron irreversible heating will be dependent on $\betai$.

In fact, the irreversible electron heating efficiency (shown in Fig.~\ref{fig:mttt}(c)) systematically decreases with $\beta_{\rm i}$, and the trend is largely independent of the initial temperature ratio, apart from the case with $\beta_{\rm i}=2$ (rightmost points in  Fig.~\ref{fig:mttt}(c)).
Here, the irreversible heating fraction reaches $M_{T\rm e,irr} \approx0.03$ for $T_{\rm e}/T_{\rm i}=1$, a factor of $\sim3$ larger than for the $\beta_{\rm i}=2$ cases with lower temperature ratios, $T_{\rm e}/T_{\rm i}=0.1$ and 0.3.\footnote{We have extensively checked this result, finding that it holds regardless of the simulation boundary conditions (periodic or outflow in the $x$ direction, or double periodic; see Appendix~\ref{sec:outvper}) and the number of computational particles per cell.}   We attribute the peculiar behavior of this case to the fact that, among the $m_{\rm i}/m_{\rm e}=25$ simulations presented in Fig.~\ref{fig:mttt}, the $\beta_{\rm i}=2,\, T_{\rm e}/T_{\rm i}=1$ case is the only one for which the scale separation $(c/\omega_{\rm pi})/(c/\omega_{\rm pe})$ between protons and electrons approaches unity (see Fig.~\ref{fig:theta_updown}(c)). For the case $\beta_{\rm i}=2, T_{\rm e}/T_{\rm i}=1$ in Fig.~\ref{fig:mttt}, this statement holds true for both the {\it upstream} and the {\it downstream} scale separation, since the reconnection process at high $\betai$ does not appreciably change the plasma thermal content. 
However, as we further discuss in the next two subsections, where we investigate the dependence of our results on the mass ratio and the magnetization, we find that the necessary and sufficient condition for the electron and proton heating efficiencies to be comparable is that the {\it downstream} scale separation approaches unity. In retrospect, this is not surprising, since if $(c/\omega_{\rm pi})/(c/\omega_{\rm pe})\rightarrow 1$ in the downstream, the fluid effectively behaves like an electron-positron plasma.

In Fig.~\ref{fig:mttt} (second row of panels), we also explore the $\beta_{\rm i}$-dependence of (d) total, (e) adiabatic, and (f) irreversible proton heating.  As before, blue, green, and red points correspond to simulations with upstream $T_{\rm e}/T_{\rm i}$ of $0.1, 0.3,$ and $1,$ respectively (we change the temperature ratio  by varying the electron temperature, while the proton temperature at a given $\betai$ is kept fixed). While the initial dimensionless electron temperature in our simulations ranges from  non-relativistic to  ultra-relativistic values, protons always stay at non-relativistic or trans-relativistic energies, $\theta_{\rm i} \approx 0.0004$ -- $0.5$ (this is true in both upstream and downstream).
At low $\betai$, protons are heated more efficiently than electrons,  typically by a factor of 2 -- 3 at mass ratio $m_{\rm i}/m_{\rm e}=25$ (compare panels (a) and (d), $M_{T\rm e, tot} \approx 0.05$ while $M_{T\rm i, tot} \approx 0.13$). At larger values of $m_{\rm i}/m_{\rm e},$ the ratio of proton to electron heating is even larger, as we discuss in Sections~\ref{ssec:massratio} and \ref{ssec:tratio1836}. Once again, the notable exception is the high-$\betai$ case with $\beta_{\rm i}=2 $ and $T_{\rm e}/T_{\rm i}=1,$ where the electron and proton heating fractions are comparable, $M_{T\rm e, tot} \approx 0.06$ and $M_{T\rm i, tot} \approx 0.08$. 
 Similar to electrons, the irreversible component of proton heating decreases with $\beta_{\rm i},$ and shows only weak dependence on the upstream temperature ratio $T_{\rm e}/T_{\rm i}$ (panel (f)).  As shown in panel (e), the fractional contribution of adiabatic heating to the total proton heating increases with $\beta_{\rm i}$, as for electrons.  

Finally, we show the total particle (i.e., sum of electron and proton) heating, as well as the corresponding adiabatic and irreversible components, in Fig.~\ref{fig:mttt}(g)-(i). Given that protons are heated more efficiently than electrons, the trends in the bottom row of Fig.~\ref{fig:mttt} are primarily controlled by protons (again, with the exception of the case $\beta_{\rm i}=2, $ $T_{\rm e}/T_{\rm i}=1$). Panel (g) shows that the total particle heating efficiency is ${\sim0.15}$ across all simulations, with a weakly declining trend with increasing $\beta_{\rm i}$. Panels (h) and (i) show that, as discussed for electrons and protons individually, heating in the low-$\beta_{\rm i}$ regime is associated with an increase in entropy, while   at high $\beta_{\rm i}$ it is dominated by adiabatic compression. 

While we cast the heating fractions in Fig.~\ref{fig:mttt} in terms of temperature differences between upstream and downstream, they may be expressed, alternatively, via differences in internal energy per particle; see Appendix \ref{sec:appmu}.

\begin{figure*}
	\centering
		\includegraphics[width=\textwidth,trim={0cm 0cm 0 0cm},clip]{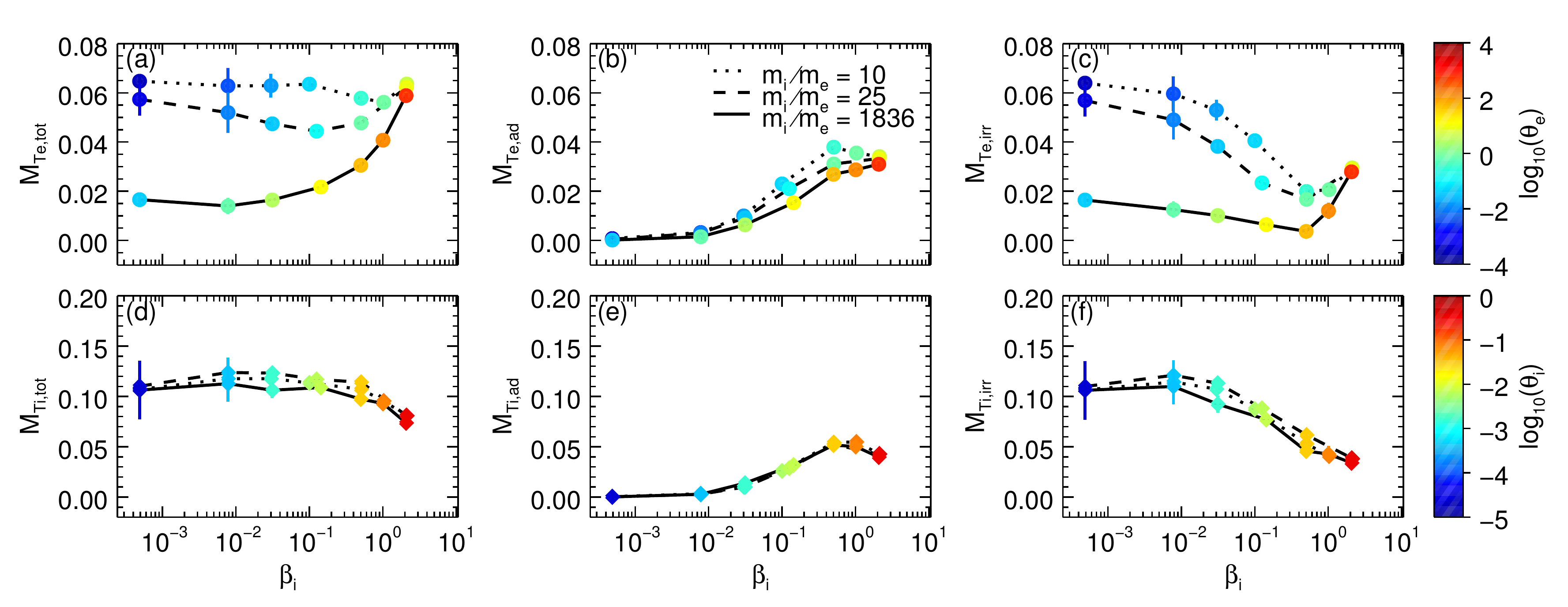} \\
		\caption{Mass ratio $m_{\rm i}/m_{\rm e}$ dependence of heating efficiencies; (a): electron total, $M_{T\rm e,tot}$; (b): electron adiabatic, $M_{T\rm e,ad}$; (c): electron irreversible,  $M_{T\rm e,irr}$; (d): proton total, $M_{T\rm i,tot}$; (e): proton adiabatic, $M_{T\rm i,ad}$; (f): proton irreversible,  $M_{T\rm i,irr}$, for $T_{\rm e}/T_{\rm i}=1$ simulations with mass ratios $m_{\rm i}/m_{\rm e}=10$ (dotted), $25$ (dashed), and $1836$ (solid); the legend is located in the upper part of panel (b).  Points in panels (a)--(c) are colored according to electron dimensionless temperature in the upstream (color bar is to the right of (c)), and points in (d)--(f) according to proton dimensionless temperature in the upstream (color bar is to the right of (f)).  The irreversible heating is remarkably independent of mass ratio at high $\beta_{\rm i}(= 2),$  while at low $\beta_{\rm i}$, the irreversible electron heating efficiency decreases with increasing mass ratio.  \label{fig:weirdpoint}}   
\end{figure*}

\begin{figure*}
	\centering
	\includegraphics[width=0.9\textwidth,trim={0.5cm 0cm 0cm 0cm},clip]{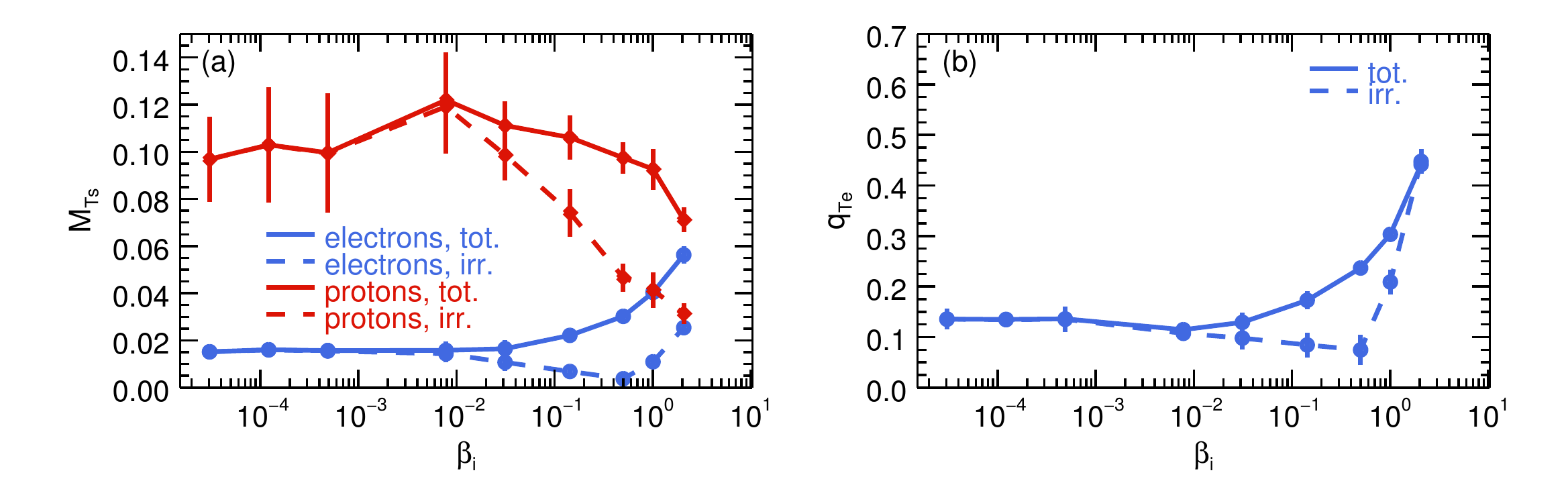} \\
		\caption{(a): $\beta_{\rm i}$-dependence of electron total heating  $M_{T\rm e,tot}$ (solid blue), electron irreversible heating  $M_{T\rm e,irr}$ (dashed blue), proton total heating  $M_{T\rm i,tot}$ (solid red), and proton irreversible heating  $M_{T\rm i,irr}$ (dashed red); (b): $\beta_{\rm i}$-dependence of electron-to-overall total heating ratio $q_{T\rm e,tot}$ (solid blue) and electron-to-overall irreversible heating ratio $q_{T\rm e,irr}$ (dashed blue), as defined in Eqs. \ref{eq:qte1}, \ref{eq:qte2}.  
		Here, $\sigma_{w} = 0.1,$ $T_{\rm e}/T_{\rm i}=1,$ and $m_{\rm i}/m_{\rm e}=1836.$ \label{fig:qte}}   
\end{figure*}

\subsection{Dependence of particle heating on \texorpdfstring{$m_{\MakeLowercase{i}}/m_{\MakeLowercase{e}}$}{mime}}	\label{ssec:massratio}	
We have extended our results up to the physical mass ratio $m_{\rm i}/m_{\rm e}=1836$, and in this section we focus on the case with $T_{\rm e}/T_{\rm i} = 1$ (runs with $\sigma_w=0.1$ and unequal temperature ratios are presented in Section \ref{ssec:tratio1836}). The separation between the electron scale $c/\omega_{\rm pe}$ and the proton scale $c/\omega_{\rm pi}$ is regulated by Eq. \ref{eq:skindepth}. For non-relativistic particles, the ratio of proton to electron skin depth is $\sqrt{m_{\rm i}/m_{\rm e}}\sim 40$, so that a large simulation domain is required to properly capture the proton physics. However, in the trans-relativistic regime of our simulations, the particles can approach (or exceed, in the case of electrons) relativistic temperatures. Here, the effective increase in electron inertia can bring the ratio of proton to electron skin depth close to unity (see Eq. \ref{eq:skindepth}). This condition holds, for example, in simulations \textbf{C[3], C[4],} and \textbf{B[4]}, when the mass ratio is increased to $m_{\rm i}/m_{\rm e}=1836$ at fixed $\sigma_w$ and $\betai$.

We show in Fig.~\ref{fig:weirdpoint} the dependence of total (a), adiabatic (b), and irreversible (c) electron heating on $\beta_{\rm i},$ for mass ratios $m_{\rm i}/m_{\rm e}=10,25,$ and $1836.$ We fix the magnetization $\sigma_{w}=0.1,$ and the temperature ratio $T_{\rm e}/T_{\rm i}=1;$ the legend is shown in the upper part of panel (b).  The points are colored according to the dimensionless temperature of upstream electrons (the corresponding colorbar is to the right of panel (c)), ranging from non-relativistic ($\theta_{\rm e} \sim 10^{-4}$) to ultra-relativistic ($\theta_{\rm e} \sim 10^{3}$) values. In agreement with earlier studies of non-relativistic reconnection by \citet{Dahlin2014} and \citet{Le2016}, we find that the total electron heating efficiency at low $\betai$ is a decreasing function of mass ratio. For the realistic mass ratio,  at low $\beta_{\rm i}$ the total heating fraction $M_{T\rm e,tot} \approx 0.016$ is in good agreement with the observed value in the magnetopause, $M_{T\rm e,tot}=0.017$ \citep{Phan2013}.
At $\beta_{\rm i} = 2,$ the  electron heating efficiency is remarkably insensitive to the mass ratio, with $M_{T\rm e,tot} \approx 0.06$. As we have anticipated above, in this case the upstream and downstream skin depths of protons and electrons are comparable (once we account for the effects of relativistic inertia), so the physics should resemble that of an electron-positron plasma, regardless of the mass ratio. The adiabatic heating efficiency (panel (b)) shows only a weak dependence on  mass ratio, in agreement with Eq. \ref{eq:compapprox}. For realistic mass ratios, electron heating is governed by irreversible processes at low $\beta_{\rm i},$ adiabatic heating dominates at intermediate $\beta_{\rm i}\sim 0.1-1$, while the two components equally contribute  at  high $\beta_{\rm i}\sim 2$. 

We show the $\beta_{\rm i}$-dependence of the proton heating fractions $M_{T\rm i,tot}, M_{T\rm i,ad},$ and $M_{T\rm i,irr}$ in panels (d), (e), and (f).  The points are colored according to the upstream dimensionless proton temperature, $\theta_{\rm i}$ (the scale is to the right of panel (f)).  The upstream proton temperatures are non-relativistic or trans-relativistic, with $\theta_{\rm i} \lesssim 0.5$. At fixed $\sigma_w$ and $\beta_{\rm i}$, the initial proton temperature stays the same, when we vary the mass ratio (as opposed to the electron temperature, which increases with mass ratio). So, the proton heating efficiencies are expected to remain unchanged, as long as the box size $L_{x}$ is sufficiently large (in units of the proton skin depth $c/\omega_{\rm pi}$) to capture the physics of proton heating.
In the bottom row of Fig.~\ref{fig:weirdpoint}, the proton heating fractions $M_{T\rm i,tot}, M_{T\rm i,ad},$ and $M_{T\rm i,irr}$ are nearly independent of the mass ratio, which demonstrates that even for the realistic mass ratio, the box used here is sufficiently large to capture the physics of proton heating (and even more so, of electron heating).  The results  discussed in Section \ref{ssec:moneyplots} for $m_{\rm i}/m_{\rm e}=25$ and $\teti=1$ are therefore still valid here:  proton heating is dominated by irreversible processes at low $\beta_{\rm i},$ whereas irreversible and adiabatic components equally contribute  at high $\beta_{\rm i};$ the irreversible heating efficiency of protons is a decreasing function of $\beta_{\rm i}$; protons are heated more efficiently than electrons (although the total proton-to-electron heating ratio for $m_{\rm i}/m_{\rm e} =1836$ is $\sim 7$ at low $\beta_{\rm i}$, larger than the value measured for $m_{\rm i} / m_{\rm e}=25$, since the electron heating efficiency decreases with mass ratio);  the heating fractions of the two species approach comparable values at $\beta_{\rm i} = 2,$ with $M_{T\rm i, tot} \approx 0.08$ and  $M_{T\rm e, tot} \approx 0.06$.

In Fig.~\ref{fig:qte}(a), we directly compare the $\beta_{\rm i}$-dependence of electron and proton heating fractions $M_{T\rm e,tot}$ (solid blue), $M_{T\rm e,irr}$ (dashed blue), $M_{T\rm i,tot}$ (solid red), and $M_{T\rm i,irr}$ (dashed red)  for $m_{\rm i}/m_{\rm e}=1836, \sigma_{w}=0.1,$ and $T_{\rm e}/T_{\rm i}=1$.\footnote{The error bars in Fig.~\ref{fig:qte}(a) are larger for protons than electrons (for electrons, they are smaller than the size of the plot symbols), but the fractional error is the same. 
Additionally, the error bars are larger at low $\beta_{\rm i}$.  As described in \ref{ssec:inflowoutflow}, this results from the frequent formation of secondary islands at low $\beta_{\rm i}.$}  As anticipated above, the proton and electron total and irreversible heating fractions differ roughly by a factor of $\sim 7$ at low $\beta_{\rm i},$ but they approach a similar value at $\beta_{\rm i}=2$ ($\approx0.03$ for the irreversible component and $\approx0.06$ for the total). 
In Fig.~\ref{fig:qte}(b), we show the $\beta_{\rm i}$-dependence of the ratio of electron-to-overall total heating ratio (solid blue),
\begin{align} \label{eq:qte1}
q_{T\rm e,tot} = \frac{M_{T\rm e,tot}}{M_{T\rm e,tot} + M_{T\rm i,tot}},
\end{align}
and similarly, the ratio of electron-to-overall irreversible heating ratio (dashed blue),
\begin{align} \label{eq:qte2}
q_{T\rm e,irr} = \frac{M_{T\rm e,irr}}{M_{T\rm e,irr} + M_{T\rm i,irr}}.
\end{align}
At low $\beta_{\rm i},$ the electron-to-overall total heating ratio is $q_{T\rm e,tot} \approx 0.14$, and it increases with $\beta_{\rm i}$ up to $q_{T\rm e,tot} \approx 0.45$ at $\beta_{\rm i}=2.$ The corresponding ratio of the irreversible components $q_{T\rm e, irr}$ is comparable to $q_{T\rm e, tot}$ at both low $\beta_{\rm i}$ (where adiabatic heating is negligible) and $\beta_{\rm i}=2$ (where adiabatic and irreversible contributions are similar), but for intermediate $\beta_{\rm i}$ we find that $q_{T\rm e, irr}$ can be as low as $0.07$, smaller than $q_{T\rm e, tot}$ by up to a factor of $\approx 3$.

\begin{figure*}
	\centering
	\includegraphics[width=\textwidth,trim={0cm 0cm 0cm 0cm},clip]{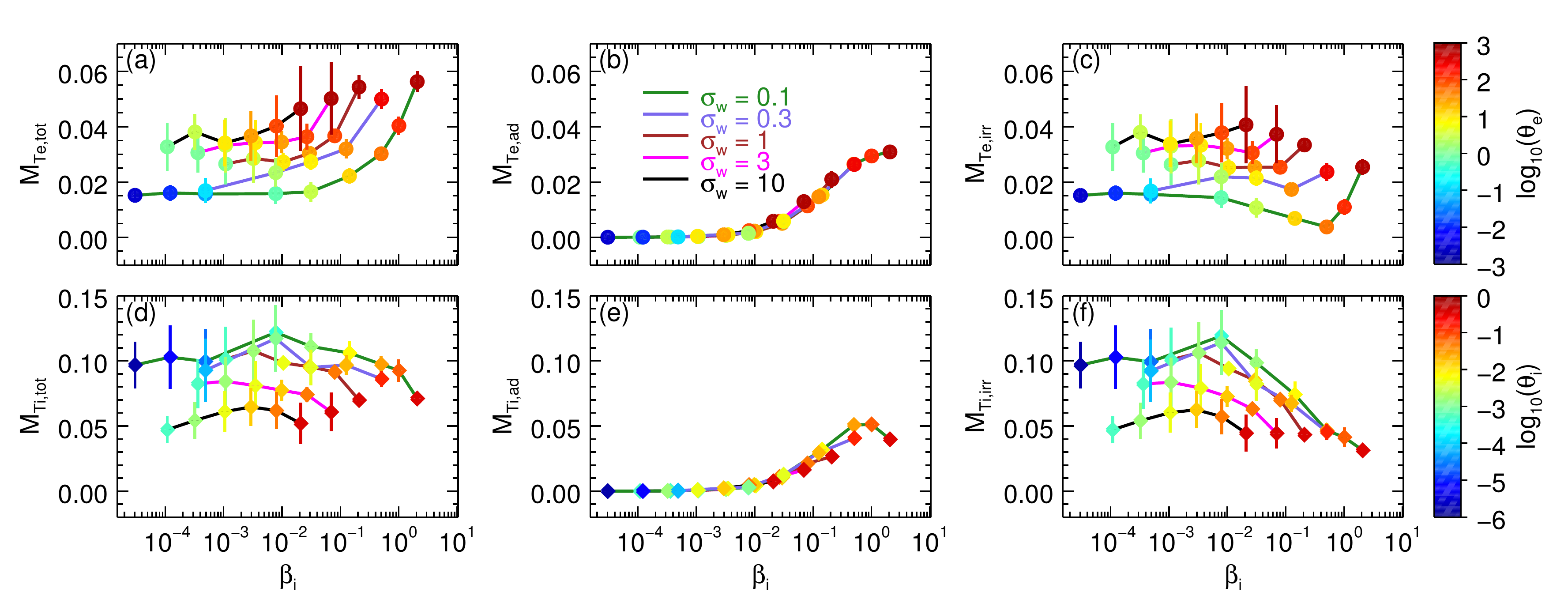} \\
		\caption{Dependence of the heating efficiencies on the magnetization $\sigma_w$ (normalized to the enthalpy density), with a layout similar to that of Fig.~\ref{fig:weirdpoint}; (a): electron total, $M_{T\rm e,tot}$; (b): electron adiabatic, $M_{T\rm e,ad}$; (c): electron irreversible,  $M_{T\rm e,irr}$; (d): proton total, $M_{T\rm i,tot}$; (e): proton adiabatic, $M_{T\rm i,ad}$; (f): proton irreversible,  $M_{T\rm i,irr}$. We fix  $T_{\rm e}/T_{\rm i}=1$ and $\mime=1836$, and vary the magnetization $\sigma_{w}=0.1$ (green), $0.3$ (purple), $1$ (brown), $3$ (magenta), $10$ (black); the legend is located in the upper part of panel (b).  Points in panels (a)--(c) are colored according to $\theta_{\rm e, up}$ (color bar is to the right of (c)), and points in (d)--(f) according to $\theta_{\rm i, up}$ (color bar is to the right of (f)).  
		\label{fig:mime1836_sig}}   
\end{figure*}

\begin{figure}
	\centering
	\includegraphics[width=0.45\textwidth,trim={0 0 0 0},clip]{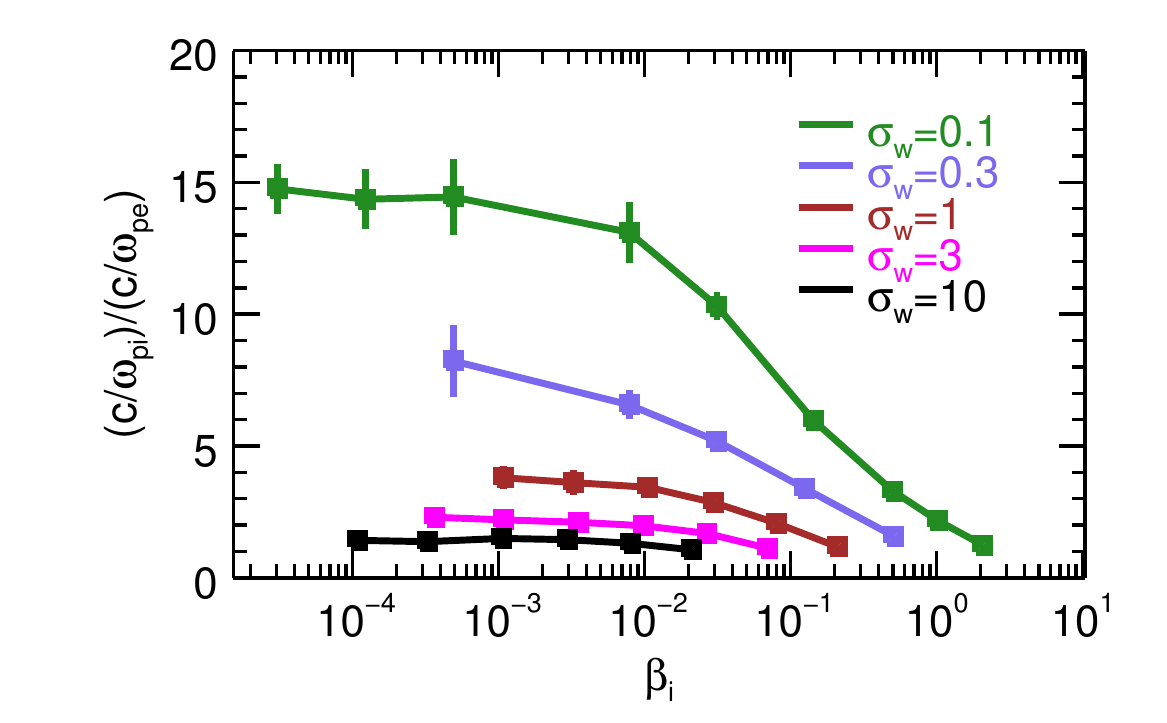} \\
		\caption{$\beta_{\rm i}$-dependence of downstream proton-to-electron skin depth ratio, $(c/\omega_{\rm pi})/(c/\omega_{\rm pe})$ (see Eq.~\ref{eq:skindepth}), for magnetizations $\sigma_{w}=0.1$ (green), 0.3 (purple), 1 (brown), 3 (magenta), and 10 (black).  
		For these simulations, the upstream electron-to-proton temperature ratio is $T_{\rm e}/T_{\rm i}=1,$ and $m_{\rm i}/m_{\rm e}=1836.$ \label{fig:scalesep_sig}}   
\end{figure}

\begin{figure*}
	\centering
	\includegraphics[width=0.9\textwidth,trim={0.5cm 0cm 0cm 0cm},clip]{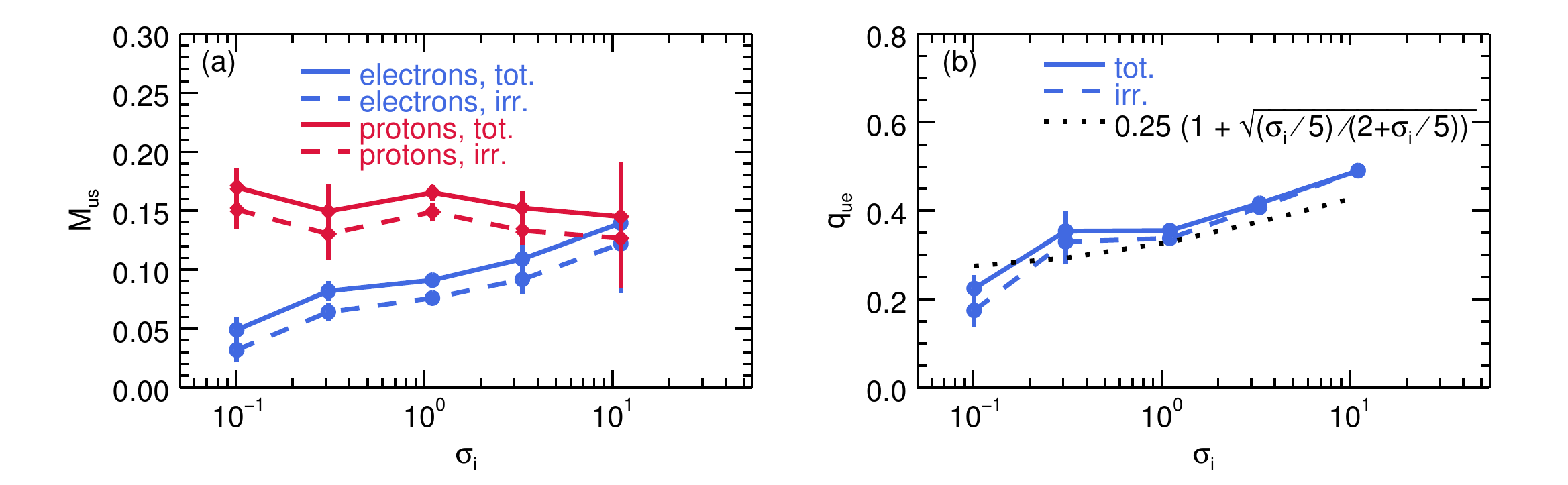} \\
		\caption{(a): Dependence on the magnetization $\sigma_{\rm i}$ (normalized to rest mass energy density) of  electron total heating efficiency $M_{u\rm e,tot}$ (solid blue), irreversible heating efficiency $M_{u\rm e,irr}$ (dashed blue), proton total heating efficiency $M_{u\rm i,tot}$ (solid red), and proton irreversible heating efficiency $M_{u\rm i,irr}$ (dashed red). (b): Dependence on the magnetization $\sigma_{\rm i}$ of the electron-to-overall heating ratio $q_{u\rm e,tot}$ (solid blue) as in Eq. \ref{eq:quetot}, electron-to-overall irreversible heating ratio $q_{u\rm e,irr}$ (dashed blue) as in Eq. \ref{eq:que}, and empirical formula Eq. \ref{eq:que_emp} (dotted black) obtained by \citet{Werner2016} in the case $\beta_{\rm i}=0.01$.  Here, $\beta_{\rm i} \approx 0.03,$ $T_{\rm e}/T_{\rm i}=1,$ and $m_{\rm i}/m_{\rm e}=1836.$ \label{fig:que}}   
\end{figure*}
\begin{figure}
	\centering
	\includegraphics[width=0.45\textwidth,trim={0cm 0cm 0cm 0cm},clip]{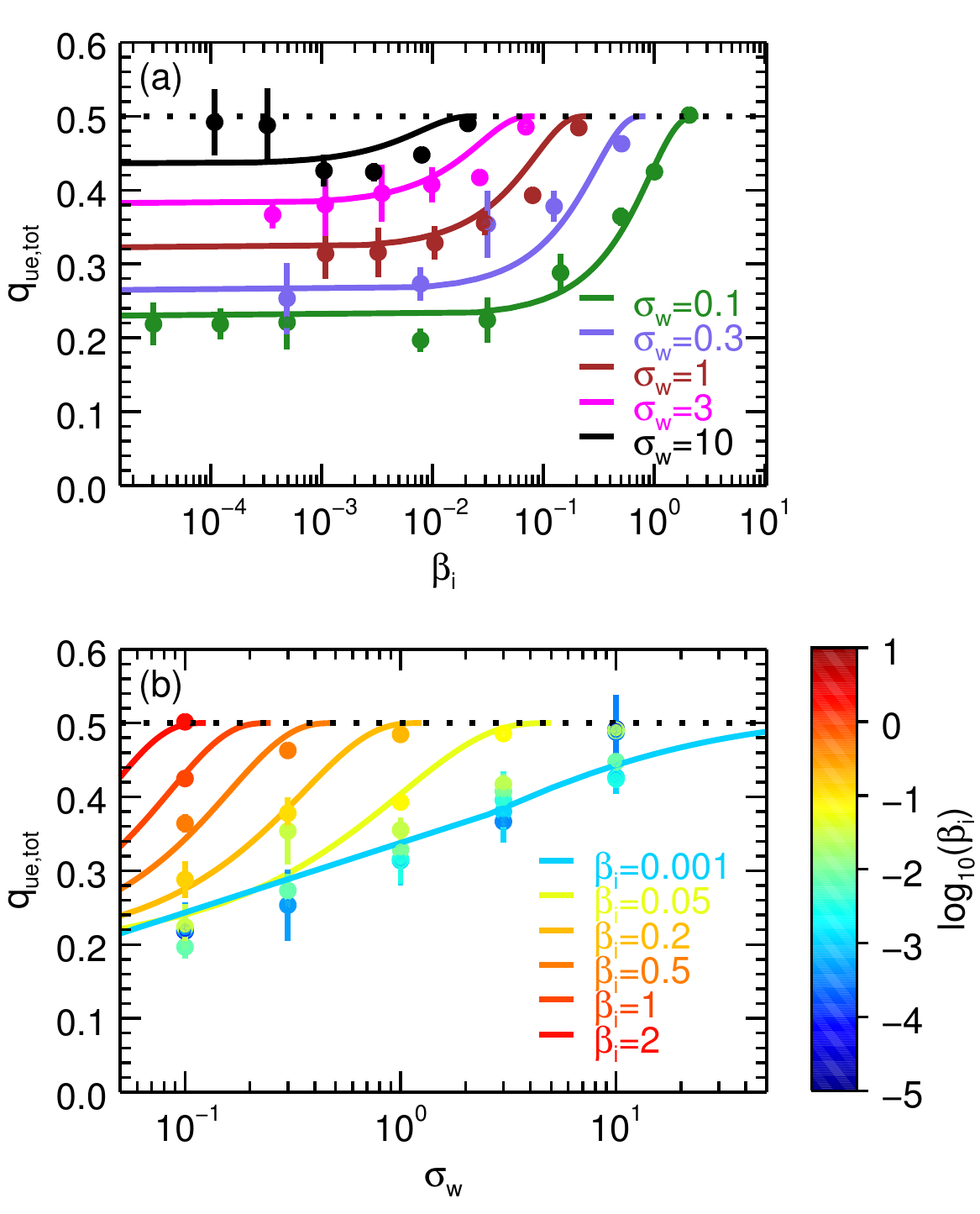} \\
		\caption{Comparison of the electron-to-overall heating ratio $q_{u\rm e,tot}$ between our simulations with $\mime=1836$ and $\teti=1$ (filled circles with error bars) and the best fitting formula in Eq.~\ref{eq:fit} (solid curves). We show the dependence on (a) plasma-$\beta_{\rm i}$ and (b) magnetization $\sigma_w$. In panel (a), the different colors represent magnetizations $\sigma_{w}=0.1$ (green), $0.3$ (purple), $1$ (brown), $3$ (magenta), and $10$ (black). In panel  (b), the color coding of the curves is indicated in the legend (from cyan to red for increasing $\betai$), while the color of the filled points refers to the colorbar on the right.
		In both panels, the black dotted line at $q_{u\rm e,tot}=0.5$ shows the limit of comparable heating efficiencies between electrons and protons, expected when $\beta_{\rm i} \rightarrow \beta_{\rm i, max}$ (regardless of $\sigma_{w}$) or $\sigma_{w} \gg1$ (independently of $\beta_{\rm i}$).  \label{fig:fitfunc}}   
\end{figure}

\subsection{Dependence of particle heating on magnetization}	\label{ssec:sigdep}
In the previous sections, we have focused on the case $\sigma_{w} = 0.1;$  in
Fig.~\ref{fig:mime1836_sig}, we show the $\beta_{\rm i}$-dependence of the heating efficiencies for a suite of simulations with $\sigma_{w} =0.1, 0.3, 1, 3,$ and $10$.\footnote{At high $\sigma_{w}$, the rate of secondary island production is enhanced \citep{Sironi2016}.
In the simulations with $\sigma_{w} = 1, 3, 10,$ we employ outflow boundary conditions in order to evolve the system to longer times. This allows us to average the downstream quantities in the reconnection exhausts over a longer timespan,  and obtain more reliable estimates.}  We fix the temperature ratio $T_{\rm e}/T_{\rm i}=1$ and the mass ratio $m_{\rm i}/m_{\rm e} = 1836.$
The panels are similar to those in Fig.~\ref{fig:weirdpoint}: (a), (b), and (c) show the electron heating fractions $M_{T\rm e, tot}, M_{T\rm e,ad},$ and $M_{T\rm e,irr}$;
(d), (e), and (f) show the proton heating fractions $M_{T\rm i, tot}, M_{T\rm i,ad},$ and $M_{T\rm i,irr}.$  The legend is in panel (b): green, purple, brown, magenta, and black curves connect the points having $\sigma_{w} = 0.1, 0.3, 1, 3,$ and $10$, respectively, to guide the eye.
The points of panels (a)--(c) are colored according to the upstream dimensionless electron temperature $\theta_{\rm e}$, as indicated by the color bar to the right of panel (c).  
Similarly, in panels (d)--(f) the points are colored according to the upstream dimensionless proton temperature $\theta_{\rm i}$, as indicated by the color bar to the right of panel (f).  For fixed $\beta_{\rm i}, T_{\rm e}/T_{\rm i},$ and $m_{\rm i}/m_{\rm e},$ an increase in magnetization leads to an increase in the upstream dimensionless temperature of both electrons and protons, which can be seen by comparing the colors of data points in panel (a) or (d) at fixed $\beta_{\rm i}$.  

We note that the data points in Fig.~\ref{fig:mime1836_sig} extend up to a maximum value of $\betai$ that depends on $\sigma_w$. For our choice of defining the magnetization using the enthalpy density, rather than the rest-mass energy density,  the ion $\betai$ cannot exceed $\beta_{\rm i,max}\sim 1/4\sigma_w$. For each value of $\sigma_{w},$ the points with the highest value of $\betai$ are also those for which the proton-to-electron scale separation ratio $(c/\omega_{\rm pi})/(c/\omega_{\rm pe})$ is the smallest {(see Fig.~\ref{fig:scalesep_sig})}.
We find that in the limit $\betai\rightarrow \beta_{\rm i,max}$, the total electron heating efficiency shows a characteristic upturn (panel (a)), with a typical value $M_{T\rm e,tot}\approx 0.05$ that is nearly independent of $\sigma_w$. In the low-$\betai$ regime, the electron total heating efficiency approaches a $\sigma_w$-dependent plateau, with higher $\sigma_w$ yielding larger electron efficiencies (panel (a)). The opposite holds for protons: higher magnetizations give smaller proton heating efficiencies (panel (d)). Indeed, for $\sigma_w=10$ the electron and proton efficiencies are comparable in the whole range of $\betai$ we have explored, in agreement with the results by \citet{Sironi2015c}. 

As anticipated in Section \ref{ssec:moneyplots},	we find that the necessary and sufficient condition for having comparable electron and proton heating efficiencies is that the separation between the electron and proton scales in the downstream be of order unity  (or equivalently, that the two species be relativistically hot, with comparable temperatures). As shown in {Fig.~\ref{fig:scalesep_sig}}, this can be achieved in two ways: ({\it i}) at high $\sigma_w$, regardless of $\betai$, the reconnection process transfers so much magnetic energy to the particles that both species become relativistically hot, with comparable temperatures; ({\it ii}) at low $\sigma_w$ and in the limit $\betai\rightarrow \beta_{\rm i,max}$, both electrons and protons already start relativistically hot in the upstream region (and more so, will be relativistically hot in the downstream). 

Most of the $\sigma_w$-dependences that we have now presented for the total heating efficiencies $M_{T\rm e, tot}$ and $M_{T\rm i, tot}$ also apply to the irreversible components $M_{T\rm e, irr}$ and $M_{T\rm i, irr}$, since the adiabatic contribution is independent of the magnetization, at fixed $\betai$ (see Eq.~\ref{eq:compapprox}). However, 
since the magnetization affects the efficiency of irreversible heating at fixed $\beta_{\rm i}$, while the adiabatic component remains the same, this can lead to a significant change in the relative contributions of irreversible and adiabatic heating.  This can be seen, for example, at $\beta_{\rm i}\approx0.5$.  For $\sigma_{w}=0.1$, $M_{T\rm e,irr}/M_{T\rm e,tot}\approx 0.1,$ whereas at $\sigma_{w}=0.3$, we find $M_{T\rm e,irr}/M_{T\rm e,tot} \approx 0.5.$

To connect with the recent work of \citet{Werner2016}, we show in Fig.~\ref{fig:que} the dependence of electron and proton heating on the magnetization $\sigma_{\rm i}$, defined with the rest-mass energy density (see Eq. \ref{eq:sigmai}). We fix temperature ratio $T_{\rm e}/T_{\rm i}=1$, mass ratio $m_{\rm i}/m_{\rm e}=1836,$ and $\beta_{\rm i} \approx 0.03$ (which is close to the upstream plasma $\beta_{\rm i}$ employed in \citet{Werner2016}, $\beta_{\rm i}=0.01$). In panel (a), we show the $\sigma_{\rm i}$-dependence of the electron total (solid blue), electron irreversible (dashed blue), proton total (solid red), and proton irreversible (dashed red) heating fractions, phrased in terms of internal energy as in \citet{Werner2016}, $M_{u\rm e,tot}$, $M_{u\rm e,irr}$, $M_{u\rm i,tot}$, and $M_{u\rm i,irr}$ (see Eqs.~\ref{eq:mue},~\ref{eq:mui}). As $\sigma_{\rm i}$ increases, the downstream scale separation between protons and electrons gets reduced {(see Fig.~\ref{fig:scalesep_sig})}, and the two species approach comparable heating efficiencies (whereas the two differ by a factor of $\sim3$ at low magnetization). This holds for both the total efficiencies  $M_{u\rm e,tot} $ and $M_{u\rm i,tot} $ and the irreversible components $M_{u\rm e,irr} $ and $M_{u\rm i,irr} $, since the amount of adiabatic heating at fixed $\betai$ does not depend on $\sigma_w$.

This is further illustrated in Fig.~\ref{fig:que}(b), where we show the $\sigma_{\rm i}$-dependence of the electron-to-overall total heating fraction, phrased in terms of internal energy (solid blue), 
\begin{align} \label{eq:quetot}
q_{u\rm e,tot} = \frac{M_{u\rm e,tot}}{M_{u\rm e,tot}+ M_{u\rm i,tot}},
\end{align}
and the electron-to-overall irreversible heating ratio (dashed blue),
\begin{align} \label{eq:que}
q_{u\rm e,irr} = \frac{M_{u\rm e,irr}}{M_{u\rm e,irr}+ M_{u\rm i,irr}}.
\end{align}
Blue circles show the results of our simulations, and the black dotted line indicates the empirical formula suggested by \citet{Werner2016},
\begin{align} \label{eq:que_emp}
q_{u\rm e,emp} = \frac{1}{4} \left( 1 + \sqrt{\frac{\sigma_{\rm i}/5}{2 + \sigma_{\rm i}/5}} \right)
\end{align}
We find reasonable agreement between this empirical formula and our simulations, for $\betai\approx0.03$.  
For low values of the magnetization, $q_{u\rm e,tot} \approx q_{u\rm e,irr}\approx0.25$, but as $\sigma_{\rm i}$ increases toward the ultra-relativistic limit, $q_{u\rm e,tot}$ and $q_{u\rm e,irr}$ approach $\approx 0.5$, i.e., electrons and protons are heated with comparable efficiencies. However, Fig.~\ref{fig:mime1836_sig} shows that, at fixed magnetization, the heating efficiencies depend on $\betai$, a trend which cannot be properly captured by the empirical formula of \citet{Werner2016}.

We then propose the following formula, which captures the dependence of the electron-to-overall heating ratio $q_{u\rm e,tot}$  on both magnetization $\sigma_w$ and proton $\beta_{\rm i}$:
\begin{align}
q_{u\rm e,fit}=\frac{1}{2} \exp \left[ \frac{-(1-\beta_{\rm i}/\beta_{\rm i,max})^{3.3}}{1 + 1.2\, \sigma_{w}^{0.7}} \right], \label{eq:fit}
\end{align}
where $\betai\leq\beta_{\rm i,max}=1/4\sigma_{w}.$  The formula in Eq.~\ref{eq:fit} has two desirable, and physically motivated, features.  First, for $\beta_{\rm i} \rightarrow \beta_{\rm i,max},$ the electron-to-overall heating ratio approaches $0.5,$ independently of the magnetization.  Second, for $\sigma_{w} \gg1,\;q_{u\rm e, tot}=0.5$, regardless of $\betai$.  In both these limits, the scale separation between electrons and protons in the downstream will be of order unity (as we have discussed above), which we have demonstrated is a necessary and sufficient condition for comparable heating efficiencies between electrons and protons. 

In Fig.~\ref{fig:fitfunc}, we compare Eq.~\ref{eq:fit} to the results of simulations with $m_{\rm i}/m_{\rm e}=1836$ and $ T_{\rm e}/T_{\rm i}=1$ (this is the same set of simulations presented earlier in this section, as well as in Section~\ref{ssec:massratio}).  In Fig.~\ref{fig:fitfunc}(a), we show the $\beta_{\rm i}$-dependence of the electron-to-overall heating ratio $q_{u\rm e,tot}$ for a range of $\sigma_{w}$ (see the legend).  The simulation results are shown by solid filled circles, while solid lines are based on Eq.~\ref{eq:fit}.  The curves are plotted up to to the maximum allowed value of $\beta_{\rm i}$,  namely $\beta_{\rm i,max}=1/4\sigma_w$.  The black dotted line at $q_{u\rm e, tot}=0.5$ shows the limit of comparable heating efficiencies for electrons and protons, which will be reached as $\beta_{\rm i} \rightarrow \beta_{\rm i,max}$, independently of $\sigma_w$. We find that both the simulation data and the fitting formula in Eq.~\ref{eq:fit} asymptote to a constant value for $\betai\ll\beta_{\rm i,max}$, with smaller heating ratios at lower $\sigma_w$. In the non-relativistic limit $\sigma_w\ll1$, our formula prescribes that $q_{u\rm e,fit}\rightarrow0.18$, not very different from the value $q_{u\rm e,fit}\approx 0.22$ obtained for $\sigma_w=0.1$. This is consistent with the expectation that in the non-relativistic regime $\sigma_w\ll1$, the heating efficiencies will be independent from the magnetization.

In Fig.~\ref{fig:fitfunc}(b), we show the dependence of the electron-to-overall heating ratio on the magnetization $\sigma_w$, for a range of $\beta_{\rm i}$.  The simulations results are shown by filled solid circles, which are colored according to the value of $\beta_{\rm i}$ in the upstream (the color scale is located to the right of Fig.~\ref{fig:fitfunc}(b)). We select a few representative values of $\betai$ and plot the corresponding predictions based on Eq.~\ref{eq:fit} with the solid curves (see the legend in the plot).
 The curves are plotted up to $\sigma_{w,\rm max},$ which for a fixed $\beta_{\rm i}$ is given by $\sigma_{w,\rm max} \sim 1/4 \beta_{\rm i}$.
In summary, Figs.~\ref{fig:fitfunc}(a) and (b) show that our proposed formula (Eq.~\ref{eq:fit}) properly captures the magnetization and plasma $\beta_{\rm i}$ dependence of the electron-to-overall heating ratio over the whole range of $\sigma_{w}$ and $\beta_{\rm i}$ explored in this work.

\begin{figure*}
		\centering
		\includegraphics[width=\textwidth]{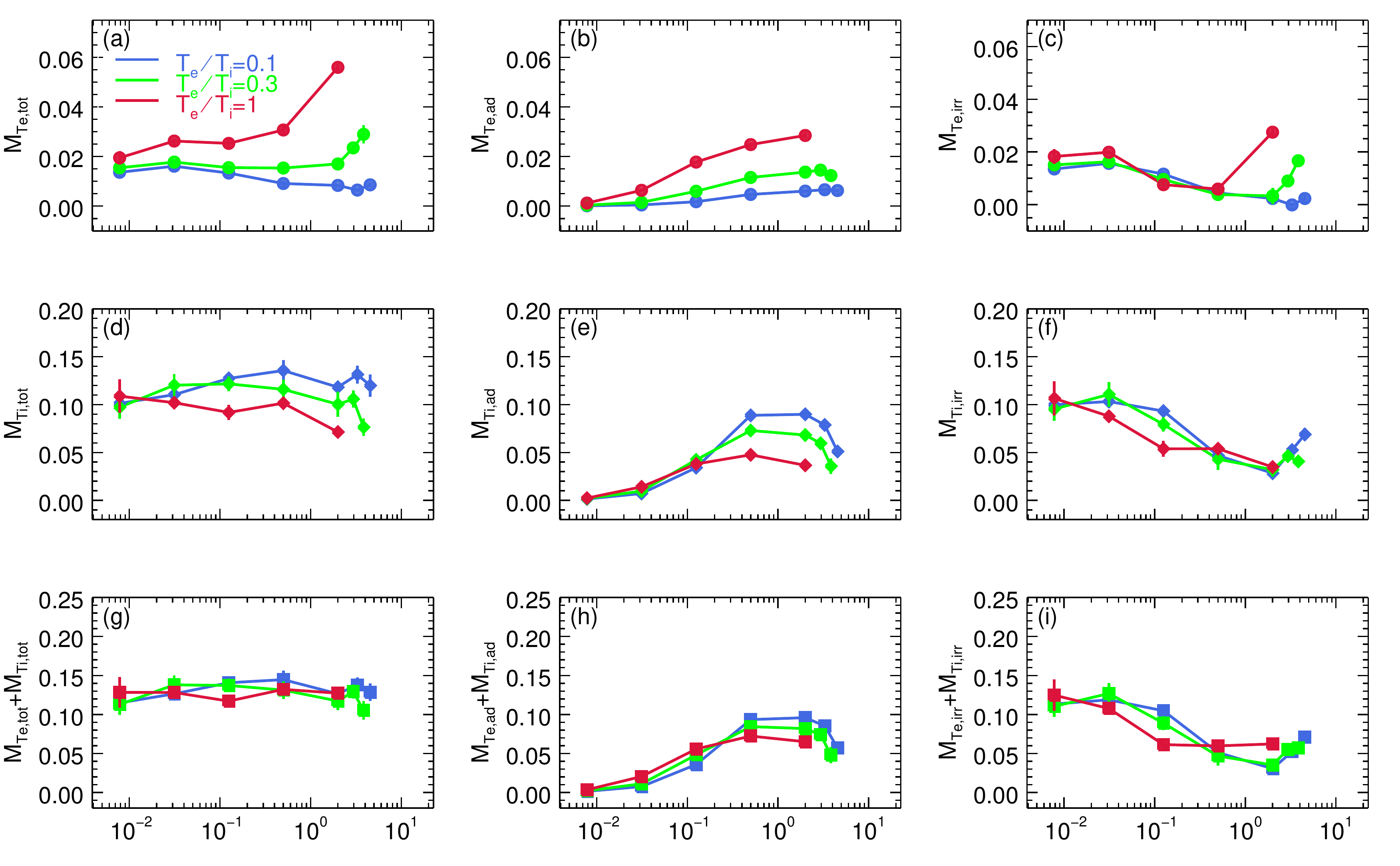} \\
			\caption{For mass ratio $m_{\rm i}/m_{\rm e} = 1836$, magnetization $\sigma_w=0.1$ and upstream temperature ratios $T_{\rm e}/T_{\rm i} =0.1$ (blue), 0.3 (green), and 1 (red), we present the $\beta_{\rm i}$-dependence of heating efficiencies; (a): electron total, $M_{T\rm e, tot}$; (b): electron adiabatic, $M_{T\rm e,ad}$; (c): electron irreversible, $M_{T\rm e,irr}$; (d): proton total, $M_{T\rm i, tot}$; (e): proton adiabatic, $M_{T\rm i,ad}$; (f): proton irreversible, $M_{T\rm i,irr}$; (g): electron and proton total, $M_{T\rm e, tot}+M_{T\rm i, tot}$; (h): electron and proton adiabatic, $M_{T\rm e,ad}+M_{T\rm i,ad}$; (i): electron and proton irreversible, $M_{T\rm e,irr}+M_{T\rm i,irr}$.}
			 \label{fig:mttt1836} 
	\end{figure*}

\subsection{Dependence of particle heating on \texorpdfstring{$T_{\MakeLowercase{e}}/T_{\MakeLowercase{i}}$}{teti} for \texorpdfstring{$m_{\MakeLowercase{i}}/m_{\MakeLowercase{e}} = 1836$}{mime}} \label{ssec:tratio1836}
In Fig. \ref{fig:mttt1836}, we present the  dependence of electron and proton heating efficiencies on the proton beta $\beta_{\rm i}$ and the temperature ratio $T_{\rm e}/T_{\rm i}$ for the realistic mass ratio $m_{\rm i}/m_{\rm e} = 1836$ (the figure layout is the same as in Fig. \ref{fig:mttt}, where we had employed a reduced mass ratio $\mime=25$).  We fix $\sigma_w=0.1$. Even at the realistic mass ratio, the conclusions drawn in the reduced mass ratio case $m_{\rm i}/m_{\rm e}=25$ (see Section \ref{ssec:moneyplots}) still hold: electron and proton heating at low $\beta_{\rm i}$ is dominated by irreversible processes, while heating in the high-$\beta_{\rm i}$ regime is mostly a result of adiabatic compression; the irreversible component of electron heating is independent of $T_{\rm e}/T_{\rm i}$ at $\beta_{\rm i}\lesssim 1$ (Fig. \ref{fig:mttt1836} (c)); the proton irreversible heating shows only a weak dependence on temperature ratio (Fig. \ref{fig:mttt1836} (f));  protons are heated more efficiently than electrons (compare the top and middle rows).

For both electrons and protons, the adiabatic heating efficiencies for $\mime=1836$ (Figs. \ref{fig:mttt1836}(b) and (e)) are similar to those of the reduced mass ratio case.  In fact, according to Eq. \ref{eq:compapprox}, the adiabatic heating efficiency is independent of mass ratio.\footnote{While Eq. \ref{eq:compapprox} is written for electrons, an analogous equation holds for the adiabatic heating of protons, if we replace $\beta_{\rm i}\teti\rightarrow \beta_{\rm i}$ and $\hat{\gamma}_{\rm e}\rightarrow\hat{\gamma}_{\rm i}$.} 
For protons, the adiabatic heating efficiency decreases at $\betai \gtrsim 2$; this is largely an effect of the decrease in the adiabatic index, as the protons transition from non-relativistic to relativistic temperatures.

For $\mime=1836$, the irreversible heating of protons at low $\betai$ is a factor of $\sim 5-7$ greater than that of electrons; in the $\mime=25$ case, the ratio of proton-to-electron irreversible heating was smaller, $\sim 2 - 3$.  As in the reduced mass ratio case, the simulation with $\betai=2$ and $\teti=1$ shows a sharp increase in irreversible electron heating as compared to the decreasing trend observed at lower $\betai$ (Fig. \ref{fig:mttt1836} (c)), and the heating efficiencies of the two species become comparable. As we argued in Section \ref{ssec:sigdep}, the electron and proton heating efficiencies are about equal if and only if the downstream scale separation is of order unity. Even for the highest values of $\betai$ that we can explore  ($\approx 3.9$ for $\teti=0.1,$ and $\approx 4.6$ for $\teti=0.3.$), this condition is not realized for smaller temperature ratios ($(c/\omega_{\rm pi})/(c/\omega_{\rm pe}) \gtrsim 3.2$ for $\teti=0.1,$ and $(c/\omega_{\rm pi})/(c/\omega_{\rm pe}) \gtrsim 1.8$ for $\teti=0.3$), which explains why --- despite the upturn in electron heating efficiency at high $\betai$ (Fig. \ref{fig:mttt1836} (c)) --- the ratio of irreversible proton to electron heating for $\teti=0.1$ and 0.3 remains larger than unity.
  

\section{Summary and discussion} \label{sec:conclusion}
In this work, we have presented the results of a series of 2D fully-kinetic PIC simulations to explore electron and proton heating by magnetic reconnection in the trans-relativistic regime. Here, protons are typically non-relativistic, yet electrons can be moderately relativistic or even ultra-relativistic. We vary the flow magnetization $\sigma_w$, the proton $\betai$ and the electron-to-proton temperature ratio $\teti$, extending our results up to the physical mass ratio $\mime=1836$.

We show that heating in the high-$\beta_{\rm i}$ regime is primarily dominated by adiabatic compression, while for low $\beta_{\rm i}$ the heating is genuine, in the sense that it is associated with an increase in entropy. At our fiducial $\sigma_{ w}= 0.1$, we find that for $\beta_{\rm i}\lesssim 1$ the irreversible heating efficiency is independent of $T_{\rm e}/T_{\rm i}$ (which we vary from $0.1$ up to $1$), for both electrons and protons. For $\teti=1$, the fraction of inflowing magnetic energy converted to electron irreversible heating at realistic mass ratios decreases from $\sim 1.6\%$ down to $\sim 0.2\%$ as $\beta_{\rm i}$ ranges from $\beta_{\rm i}\sim 10^{-2}$ up to $\beta_{\rm i}\sim 0.5$, but then it increases up to $\sim 3\%$ as $\beta_{\rm i}$ approaches $\sim2$. Protons are heated much more efficiently than electrons at low and moderate $\beta_{\rm i}$  (by a factor of $\sim7$), whereas the electron and proton heating efficiencies become comparable at $\beta_{\rm i}\sim 2$ if $T_{\rm e}/T_{\rm i}=1$. We find that comparable heating efficiencies between electrons and protons are achieved when the scale separation between the two species in the reconnection exhaust approaches unity, so that the electron-proton plasma effectively resembles an electron-positron fluid. This occurs at high $\betai$ for low magnetizations, or regardless of $\betai$ at high magnetizations (i.e., in the regime $\sigma_w\gg1$ of ultra-relativistic reconnection). We propose a fitting formula (Eq.~\ref{eq:fit}) that captures the magnetization and plasma-$\beta_{\rm i}$ dependence of the electron-to-overall heating ratio over the whole range of $\sigma_{w}$ and $\beta_{\rm i}$ explored in this work.


The low- and high-$\beta_{\rm i}$ cases differ with respect to secondary island formation.  The formation of secondary islands is suppressed at high $\beta_{\rm i}$, which leads to a homogeneous reconnection outflow.  Secondary islands occur frequently at low $\beta_{\rm i}$ and high magnetizations.

We also measure the inflow speed for our fiducial magnetization $\sigma_w=0.1$, finding that it decreases from $v_{\rm in}/v_{\rm A} \approx 0.08$ down to $0.04$ as $\beta_{\rm i}$ ranges from $\beta_{\rm i}\sim 10^{-2}$ up to $\beta_{\rm i}\sim 2$ (here, $v_{\rm A}$ is the Alfv\'en speed). Similarly, the outflow speed saturates at the Alfv\'{e}n velocity for low $\beta_{\rm i}$, but it decreases with increasing $\beta_{\rm i}$ down to $v_{\rm out}/v_{\rm A}\approx 0.7$ at $\beta_{\rm i}\sim2.$
The inflow (outflow, respectively) speed is independent of $T_{\rm e}/T_{\rm i}$ at low $\beta_{\rm i}$, with only a minor tendency for lower (higher, respectively) speeds at  larger $T_{\rm e}/T_{\rm i}$ in the high-$\beta_{\rm i}$ regime.

This investigation provides important insights into the physics of low-luminosity accretion flows, such as the accretion disk of Sgr A$^{*}.$  Collisionless accretion flows are often assumed to be two-temperature, and our results indeed show that in the trans-relativistic regime relevant to hot accretion flows and accretion disk coronae, magnetic reconnection preferentially heats protons more than electrons.
Our results --- and in particular, our fitting formula in Eq.~\ref{eq:fit} --- can be used to provide general relativistic MHD simulations of accretion flows with the sub-grid physics of energy partition between electrons and protons \citep{Ressler2015,Ressler2017,Sadowski2017}.   This ingredient is of fundamental importance in producing emission models that can be compared with the forthcoming observations by the Event Horizon Telescope  \citep{Doeleman2008}.

To conclude, we note a few lines of investigation that have not been considered in the current work.  First, we limited our focus to the case of symmetric, anti-parallel reconnection.  The more general case of guide-field reconnection will be a topic of future investigation.  Second, while we have provided a quantitative description of energy partition between electrons and protons, we have not addressed the question of the underlying heating mechanism.  A detailed study of the heating mechanism is deferred to future work.  Lastly, we have focused on thermal heating, as opposed to nonthermal acceleration.  The dependence of nonthermal acceleration efficiency on magnetization is the focus of \citet{Werner2016}, though the dependence on $\beta_{\rm i}$ and $T_{\rm e}/T_{\rm i}$ remains unexplored.

\section*{Acknowledgements}
This work is supported in part by NASA via the TCAN award grant NNX14AB47G and by the Black Hole Initiative at Harvard University, which is supported by a grant from the Templeton Foundation.  
LS acknowledges support from DoE DE-SC0016542, NASA Fermi NNX16AR75G, NSF ACI-1657507, and NSF AST-1716567.
The simulations were performed 
on Habanero at Columbia, 
on the BHI cluster at the Black Hole Initiative,
and on NASA High-End Computing (HEC) resources.
The authors acknowledge computational support from NSF via XSEDE resources (grants TG-AST80026N and TG-AST120010).

\clearpage
\appendix

\section{Appendix A:  Convergence with respect to domain size}\label{sec:lxconvergence} 
For most of the simulations presented in the main body of this work, we employ a domain size of $L_{x}\approx4000\,c/\omega_{\rm pe}$.
However, as we demonstrate in this appendix, the heating efficiencies are insensitive to the domain size.  While we have extensively checked for convergence with boxes ranging in size from $L_{x} \approx 500 \,c/\omega_{\rm pe}$ up to $L_{x} \approx 5000 \,c/\omega_{\rm pe},$ we focus here on a low-$\beta_{\rm i}$ case and a high-$\beta_{\rm i}$ case, and compare domains of size $L_{x} \approx 2000 \,c/\omega_{\rm pe}$ and $L_{x} \approx 4000 \,c/\omega_{\rm pe}$. 

\begin{figure*}[h] 
		\centering
		\includegraphics[width=\textwidth]{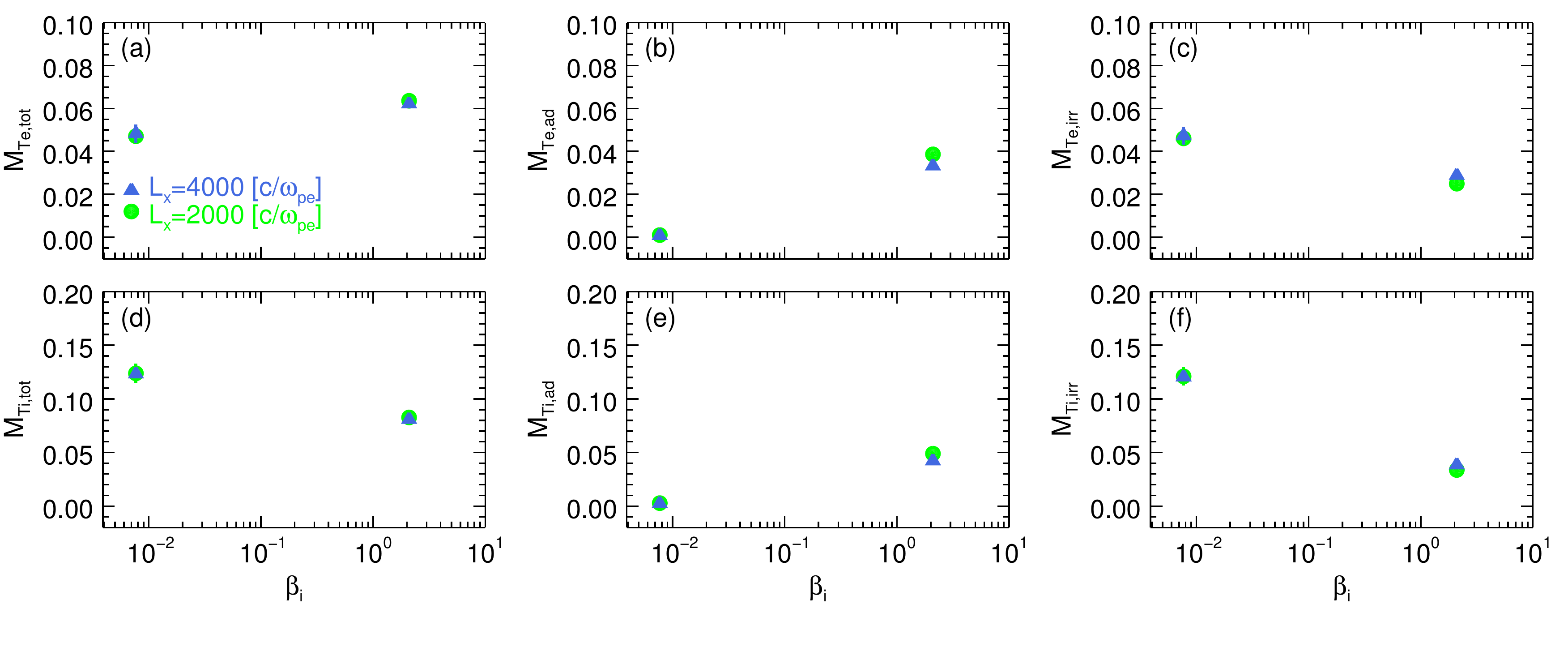} \\ 
			\caption{Comparison between domain sizes  $L_{x}\approx2000\,c/\omega_{\rm pe}$ (green circles) and $L_{x}\approx4000\,c/\omega_{\rm pe}$ (blue triangles) of the following heating fractions; (a): electron total, $M_{T\rm e,tot}$; (b): electron adiabatic, $M_{T\rm e,ad}$; (c): electron irreversible, $M_{T\rm e,irr}$; (d): proton total, $M_{T\rm i,tot}$; (e): proton adiabatic, $M_{T\rm i,ad}$; (f): proton irreversible, $M_{T\rm i,irr}$. We present a low-$\betai$ case with $\beta_{\rm i}=0.0078,\,T_{\rm e}/T_{\rm i}=0.1$, and a high-$\betai$ case with $\beta_{\rm i}=2,\, T_{\rm e}/T_{\rm i}=1;$ in both cases, the mass ratio is $m_{\rm i}/m_{\rm e}=25$ and $\sigma_{w}=0.1.$ \label{fig:lxbeta}} 
	\end{figure*}

We show in Fig.~\ref{fig:lxbeta} the electron heating fractions $M_{T\rm e,tot}, \,M_{T\rm e,ad}, \,M_{T\rm e,irr}$ (panels (a), (b), and (c)) and proton heating fractions $M_{T\rm i,tot}, \,M_{T\rm i,ad},\,M_{T\rm i,irr}$ (panels (d), (e), and (f)).  Green circles indicate simulations with $L_{x}\approx 2000\,c/\omega_{\rm pe}$, and blue triangles $L_{x}\approx4000\,c/\omega_{\rm pe}.$  The comparison is performed for two cases: $\beta_{\rm i}=0.0078, \,T_{\rm e}/T_{\rm i}=0.1$ and $\beta_{\rm i}=2, \,T_{\rm e}/T_{\rm i}=1$. For both the low- and high-$\beta_{\rm i}$ simulations, $\sigma_{w}=0.1$ and $m_{\rm i}/m_{\rm e}=25.$  For each pair of simulations (at low and high $\beta_{\rm i}$), the downstream and upstream dimensionless temperatures that enter into the heating fractions are measured at the same physical distance (in units of the electron skin depth) downstream of the central X-point.  The electron and proton heating fractions show minimal dependence on the box size.

\begin{figure*}[h] 
		\centering
		\includegraphics[width=0.9\textwidth,trim={0cm 0cm 0cm 0cm},clip]{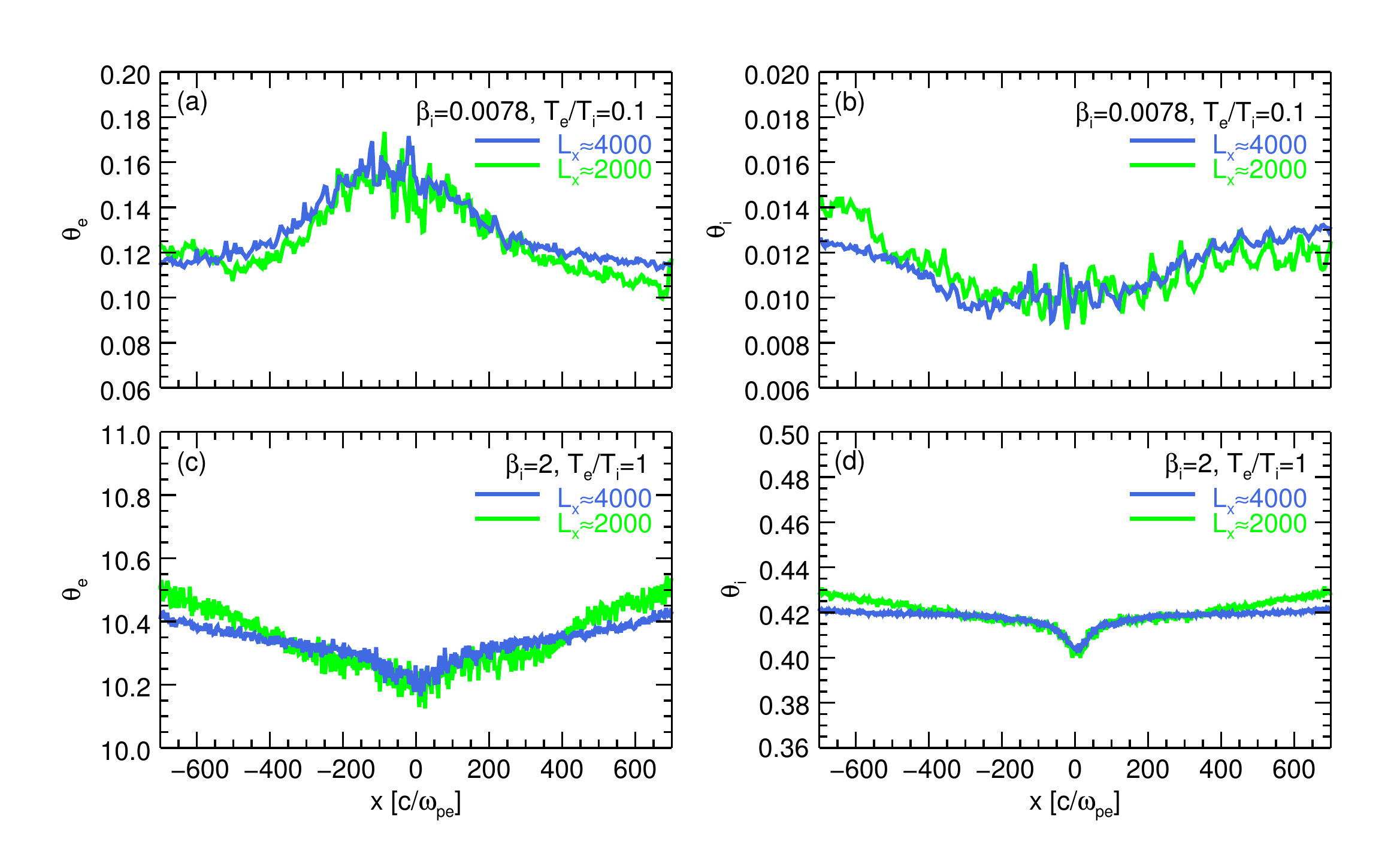} \\ 
			\caption{Spatial profiles along the reconnection outflow of 
(a): dimensionless electron temperature $\theta_{\rm e}$ for $\beta_{\rm i} = 0.0078, \,T_{\rm e}/T_{\rm i}=0.1;$ 
		(b): dimensionless proton temperature $\theta_{\rm i}$ for $\beta_{\rm i} = 0.0078,\, T_{\rm e}/T_{\rm i}=0.1;$
	(c): dimensionless electron temperature $\theta_{\rm e}$ for $\beta_{\rm i} = 2,\, T_{\rm e}/T_{\rm i}=1;$ 
	(d): dimensionless proton temperature $\theta_{\rm i}$ for $\beta_{\rm i} = 2, \,T_{\rm e}/T_{\rm i}=1$.
			The mass ratio is $m_{\rm i}/m_{\rm e}=25$ and $\sigma_{w}=0.1.$ 
			The spatial profiles are extracted from simulations with domain size $L_{x}\approx 2000\,c/\omega_{\rm pe}$ (green), and $L_{x}\approx 4000\,c/\omega_{\rm pe}$ (blue).
			These spatial profiles are from the same simulations shown in Fig.~\ref{fig:lxbeta}, at time $t \approx 1 t_{\rm A}$.
 \label{fig:lxtime}} 
	\end{figure*}

In Fig.~\ref{fig:lxtime}, we show --- for box sizes $L_{x}\approx2000\,c/\omega_{\rm pe}$ (green) and $L_{x}\approx4000\,c/\omega_{\rm pe}$ (blue) --- the spatial profiles along the outflow direction (i.e., along $x$, and averaged along $y$ in the cells identified by Eq. \ref{eq:criterion1} as belonging to the reconnection downstream) of:
	(a) dimensionless electron temperature $\theta_{\rm e}$ for $\beta_{\rm i} = 0.0078, \,T_{\rm e}/T_{\rm i}=0.1;$ 
		(b) dimensionless proton temperature $\theta_{\rm i}$ for $\beta_{\rm i} = 0.0078,\, T_{\rm e}/T_{\rm i}=0.1;$
	(c) dimensionless electron temperature $\theta_{\rm e}$ for $\beta_{\rm i} = 2,\, T_{\rm e}/T_{\rm i}=1;$ 
	(d) dimensionless proton temperature $\theta_{\rm i}$ for $\beta_{\rm i} = 2, \,T_{\rm e}/T_{\rm i}=1$.  The simulations shown in Fig.~\ref{fig:lxtime} correspond to the same simulations presented in Fig.~\ref{fig:lxbeta}. 	 
	The dimensionless temperature profiles are shown at $t\approx 1\,t_{\rm A}$;  this corresponds to $t \approx 6900\,\omega_{\rm pe}^{-1}$ for $L_{x}\approx2000\,c/\omega_{\rm pe},$ and to $t \approx 14000\,\omega_{\rm pe}^{-1} $ for $L_{x}\approx4000\,c/\omega_{\rm pe}.$ The horizontal axes range from $x\approx -700\, c/\omega_{\rm pe}$ to  $+700\, c/\omega_{\rm pe}$, which accounts for most of the smaller box, but only a fraction of the larger one.  For low $\beta_{\rm i},$ the region used for our measurements  is located at $x \approx \pm 630\,c/\omega_{\rm pe},$ whereas it is  at $x \approx \pm 350\,c/\omega_{\rm pe}$ for high $\beta_{\rm i}$; in each case, the chosen distance is far enough from the central X-point that the temperature profiles attain a quasi-uniform value, and far enough from the domain boundaries to be unaffected by the primary island (Section \ref{sec:technique}).  
	
	In Figs.~\ref{fig:lxtime}(a) and (b), which correspond to the low-$\beta_{\rm i}$ case, the dimensionless temperature profiles show similar spatial dependence within $x \approx \pm 630\,c/\omega_{\rm pe},$ and for the high-$\beta_{\rm i}$ profiles shown in (c) and (d), the temperatures agree within $x \approx \pm 350\,c/\omega_{\rm pe}.$  For the high-$\beta_{\rm i}$ case, the large and small boxes show some discrepancy beyond $x \approx \pm 400\,c/\omega_{\rm pe},$ which is an effect of the large primary island extending from the domain boundary into the outflow region.
	

\section{Appendix B:  Outflow versus periodic boundary conditions}\label{sec:outvper}
\begin{figure}[h] 
		\centering
		\includegraphics[width=\textwidth]{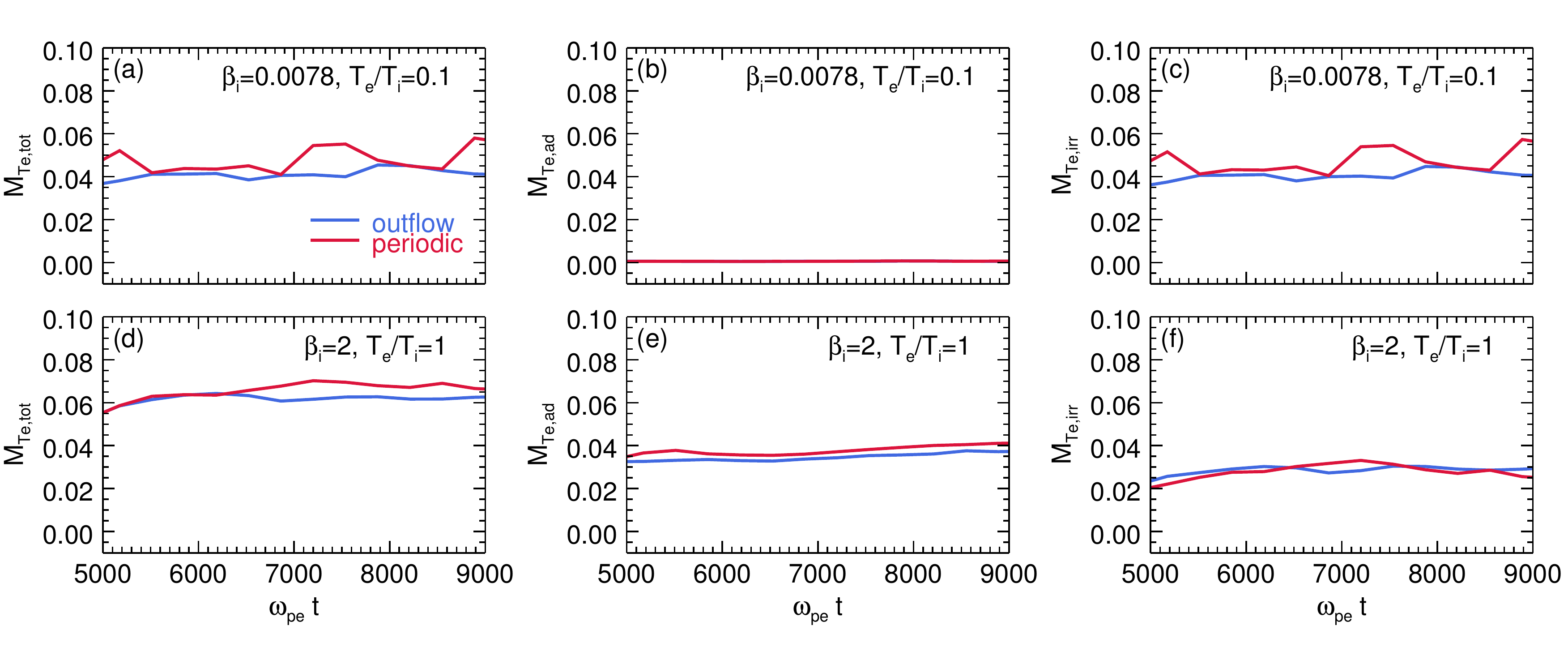} \\ 	
			\caption{Comparison between outflow (blue) and periodic (red) simulations with $\sigma_w=0.1$ and $\mime=25$. We show the time evolution of 
			(a): electron total heating fraction, $M_{T\rm e,tot}$ for $\beta_{\rm i}=0.0078,\, T_{\rm e}/T_{\rm i}=0.1;$
			(b): electron adiabatic heating fraction, $M_{T\rm e,ad}$ for $\beta_{\rm i}=0.0078, \,T_{\rm e}/T_{\rm i}=0.1;$
			(c): electron irreversible heating fraction, $M_{T\rm e,irr}$ for $\beta_{\rm i}=0.0078, \,T_{\rm e}/T_{\rm i}=0.1;$
			(d): electron total heating fraction, $M_{T\rm e,tot}$ for $\beta_{\rm i}=2, \, T_{\rm e}/T_{\rm i}=1$
			(e): electron adiabatic heating fraction, $M_{T\rm e,ad}$ for $\beta_{\rm i}=2, \, T_{\rm e}/T_{\rm i}=1$
			(f): electron irreversible heating fraction, $M_{T\rm e,irr}$ for $\beta_{\rm i}=2, \, T_{\rm e}/T_{\rm i}=1$
			The heating fractions are shown in the interval $t = 5 \times 10^{3}\,\omega_{\rm pe}^{-1}$ -- $9 \times 10^{3} \,\omega_{\rm pe}^{-1},$ which corresponds to $t \approx 0.36\,t_{\rm A}$ -- $0.64\,t_{\rm A}$ for the periodic simulations and $t \approx 0.6\,t_{\rm A}$ -- $1\,t_{\rm A}$ for the outflow ones.
			The curves have been shifted in time to account for slightly different onsets of reconnection in periodic vs. outflow cases, due to different initialization of the current sheet. \label{fig:outvper_mte}} 
	\end{figure}

We have compared the results of our main simulations, which are periodic in $x$, to a second set that employs outflow boundary conditions, similar to what is described in \citet{Sironi2016}.  In Fig.~\ref{fig:outvper_mte}, we show the time evolution of the electron heating fractions $M_{T\rm e,tot}$, $M_{T\rm e,ad}$, and $M_{T\rm e,irr}$ in a low-$\beta_{\rm i}$ simulation (Fig.~\ref{fig:outvper_mte}(a)--(c)) and a high-$\beta_{\rm i}$ case (Fig.~\ref{fig:outvper_mte}(d)--(f)), for both outflow (blue) and periodic (red) boundary conditions.  
For the periodic simulations the domain size is $L_{x} = 4318\,c/\omega_{\rm pe},$ whereas for the outflow runs $L_{x} \approx 2600\,c/\omega_{\rm pe}.$
Up to $\sim1$ Alfv\'{e}nic crossing time, which corresponds to $t \approx 1.4 \times 10^{4}\,\omega_{\rm pe}^{-1}$ for the periodic simulations and $t \approx 8.5 \times 10^{3}\, \omega_{\rm pe}^{-1}$ for the outflow runs, we find good agreement between the periodic and outflow simulations.
At later times, the pile-up of particles and magnetic flux in the primary magnetic island sitting at the boundary leads to the eventual suppression of reconnection in periodic simulations, whereas the outflow runs can be evolved for multiple Alfv\'{e}nic crossing times.  

\begin{figure}[h] 
		\centering
		\includegraphics[width=0.45\textwidth,trim={0cm 0cm 0cm 7cm},clip]{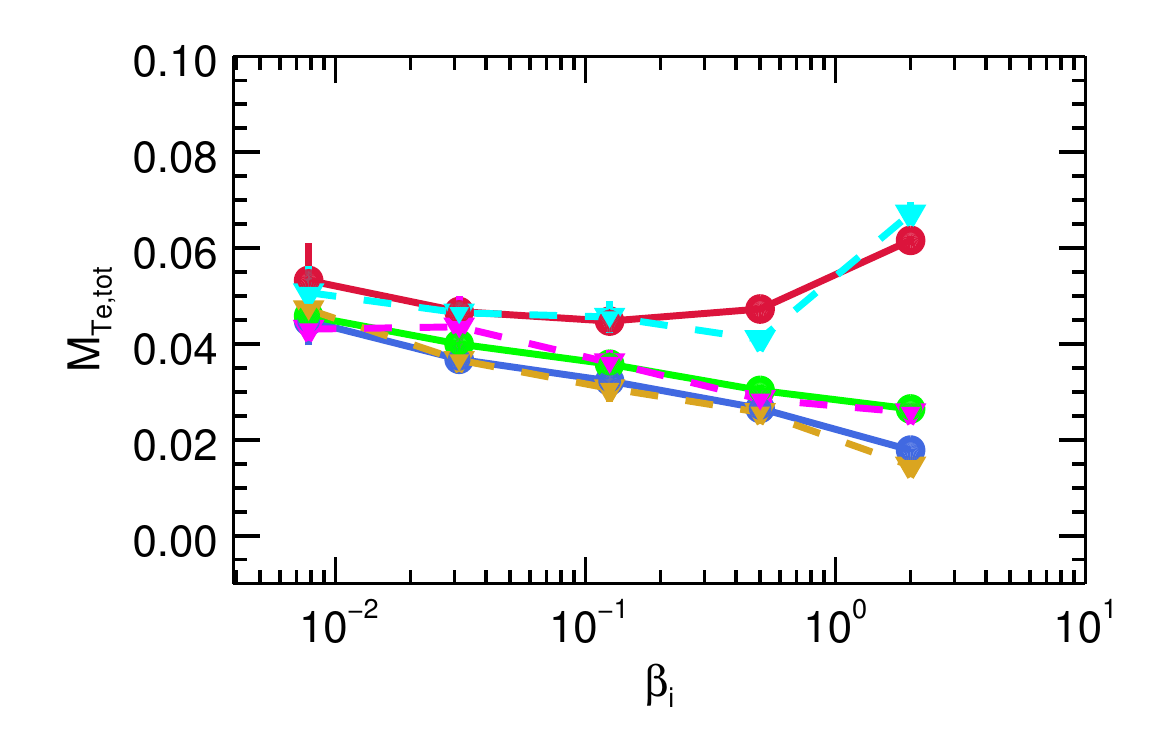} \\ 	
			\caption{Comparison of outflow and periodic simulations (with $\sigma_w=0.1$ and $\mime=25$), in terms of the dependence of $M_{T\rm e,tot}$ on $\beta_{\rm i}$ and $T_{\rm e}/T_{\rm i}$.  Circles connected by solid lines show the results of periodic simulations, and inverted triangles connected by dashed lines indicate outflow simulations.  For periodic runs, blue, green, and red correspond to runs with upstream temperature ratios $T_{\rm e}/T_{\rm i}=0.1,0.3,$ and $1$; for periodic, dark yellow, magenta, and cyan denote $T_{\rm e}/T_{\rm i}=0.1,0.3,$ and $1$.  \label{fig:outvper_beta}} 
	\end{figure}

In Fig.~\ref{fig:outvper_beta}, we compare the  dependence of the electron total heating fraction $M_{T\rm e,tot}$ on $\beta_{\rm i}$ and $T_{\rm e}/T_{\rm i}$ for periodic and outflow simulations with $m_{\rm i}/m_{\rm e}=25$ and $\sigma_{w}=0.1$.  The periodic simulations are indicated by blue, green, and red circles, corresponding to upstream temperature ratios of $T_{\rm e}/T_{\rm i}=0.1, 0.3,$ and $1$, respectively.  The results of outflow simulations are shown by dark yellow ($T_{\rm e}/T_{\rm i}=0.1$), magenta ($T_{\rm e}/T_{\rm i}=0.3$), and cyan ($T_{\rm e}/T_{\rm i}=1$) triangles.  The points corresponding to periodic runs are connected by solid lines, whereas the outflow cases are linked by dashed lines.  With regard to the $\beta_{\rm i}$- and $T_{\rm e}/T_{\rm i}$-dependence of the electron total heating fraction, $M_{T\rm e,tot}$, the outflow and periodic cases show good agreement.  The agreement for adiabatic and irreversible heating fractions is also good.
	
\section{Appendix C:  Convergence with respect to spatial resolution}\label{sec:compconvergence} 
To properly capture the electron physics, adequate spatial resolution of the electron skin depth $c/\omega_{\rm pe}$, or equivalently, temporal resolution of the inverse electron plasma frequency $\omega_{\rm pe}^{-1}$, is necessary.  In most of our simulations, we use $c/\omega_{\rm pe} = 4$ cells; since we fix $c = 0.45$ cells/timestep, the temporal resolution in our simulations is $\Delta t \approx 0.1\,\omega_{\rm pe}^{-1}$. In this appendix, we show that even at finer spatial (also, temporal) resolution, i.e. $c/\omega_{\rm pe} = 8$ cells $\rightarrow \Delta t \approx 0.05\,\omega_{\rm pe}^{-1}$, the heating fractions are essentially unchanged relative to those obtained in simulations with $c/\omega_{\rm pe}=4$ cells.

\begin{figure*}[h] 
		\centering
		\includegraphics[width=\textwidth]{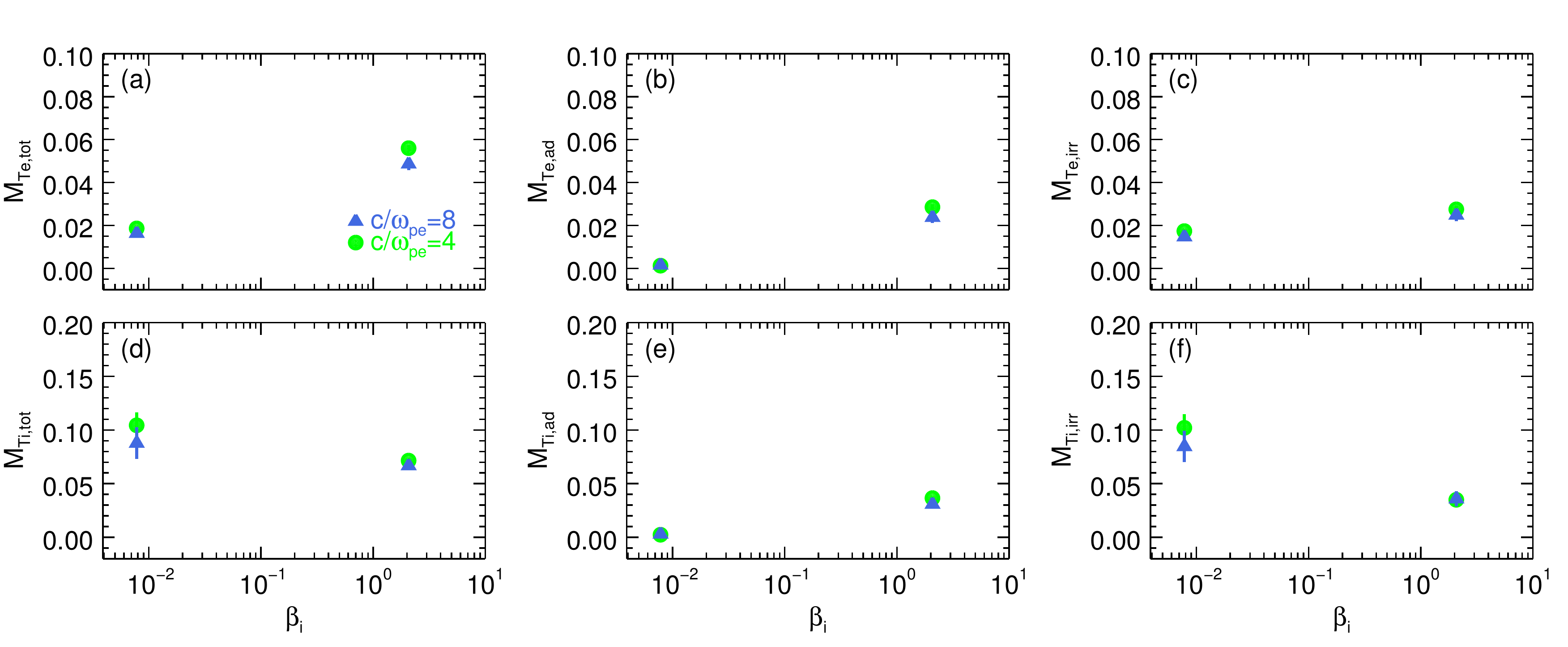} \\ 
			\caption{Comparison between simulations with $\comp=4$ cells $\rightarrow \Delta t \approx 0.1 \,\omega_{\rm pe}^{-1}$ (green circles) and $\comp=8$ cells $\rightarrow \Delta t \approx 0.05 \,\omega_{\rm pe}^{-1}$ (blue triangles) of the following heating fractions; (a): electron total, $M_{T\rm e,tot}$; (b): electron adiabatic, $M_{T\rm e,ad}$; (c): electron irreversible, $M_{T\rm e,irr}$; (d): proton total, $M_{T\rm i,tot}$; (e): proton adiabatic, $M_{T\rm i,ad}$; (f): proton irreversible, $M_{T\rm i,irr}$. We present a low-$\betai$ case with $\beta_{\rm i}=0.0078,\,T_{\rm e}/T_{\rm i}=1$, and a high-$\betai$ case with $\beta_{\rm i}=2,\, T_{\rm e}/T_{\rm i}=1;$ in both cases, we employ the realistic mass ratio $m_{\rm i}/m_{\rm e}=1836$ and $\sigma_{w}=0.1.$} \label{fig:compbeta} 
	\end{figure*}

In Fig. \ref{fig:compbeta}, we show the heating fractions for electrons (panels (a), (b), and (c)) and protons (panels (d), (e), and (f)).  For the cases $\betai=0.0078, \teti=1$ and $\betai=2, \teti=1$, we compare a simulation with $\comp=4$ cells (denoted by green circles) to one with $\comp=8$ cells (indicated by blue triangles).  In both sets of simulations, we employ $\mime=1836$ and magnetization $\sigma_{w}=0.1.$  To ensure that the simulations with $\comp=8$ cells contain the same number of electron skin depths as those with $\comp=4$ cells, it is necessary to double the size of the simulation domain in $x$ (in units of cells).  For the simulations with $\comp=4$ cells, we use $L_{x}\approx 8000$ cells, and for $\comp=8$ cells, we use $L_{x} \approx 1.6\times10^{4}$ cells; in both cases, the physical extent of the domain in $x$ is $L_{x} \approx 4318\,\comp$.  For both choices of the spatial resolution, the electron heating fractions (total, adiabatic, and irreversible) show good agreement.  The proton heating fractions show good agreement, too.

\section{Appendix D:  Control of numerical heating}\label{sec:ppc}
In simulations with high $\beta_{\rm i}$ and low temperature ratios, numerical effects can lead to an artificial increase in the upstream electron temperature, at the expense of protons.   The rate of numerical heating is proportional to the temperature difference between the two species, hence the high-$\beta_{\rm i}$ simulations with $T_{\rm e}/T_{\rm i}=0.1$ exhibit the strongest degree of numerical heating \citep{Melzani2013}.  As the temperature difference between electrons and protons in the upstream and downstream regions is not necessarily the same, the rate of numerical heating in the two regions may be different. If not adequately kept under control, this can affect our measured heating efficiencies.

\begin{figure}[h] 
		\centering
		\includegraphics[width=0.45\textwidth,trim={0 0.5cm 0 1cm},clip]{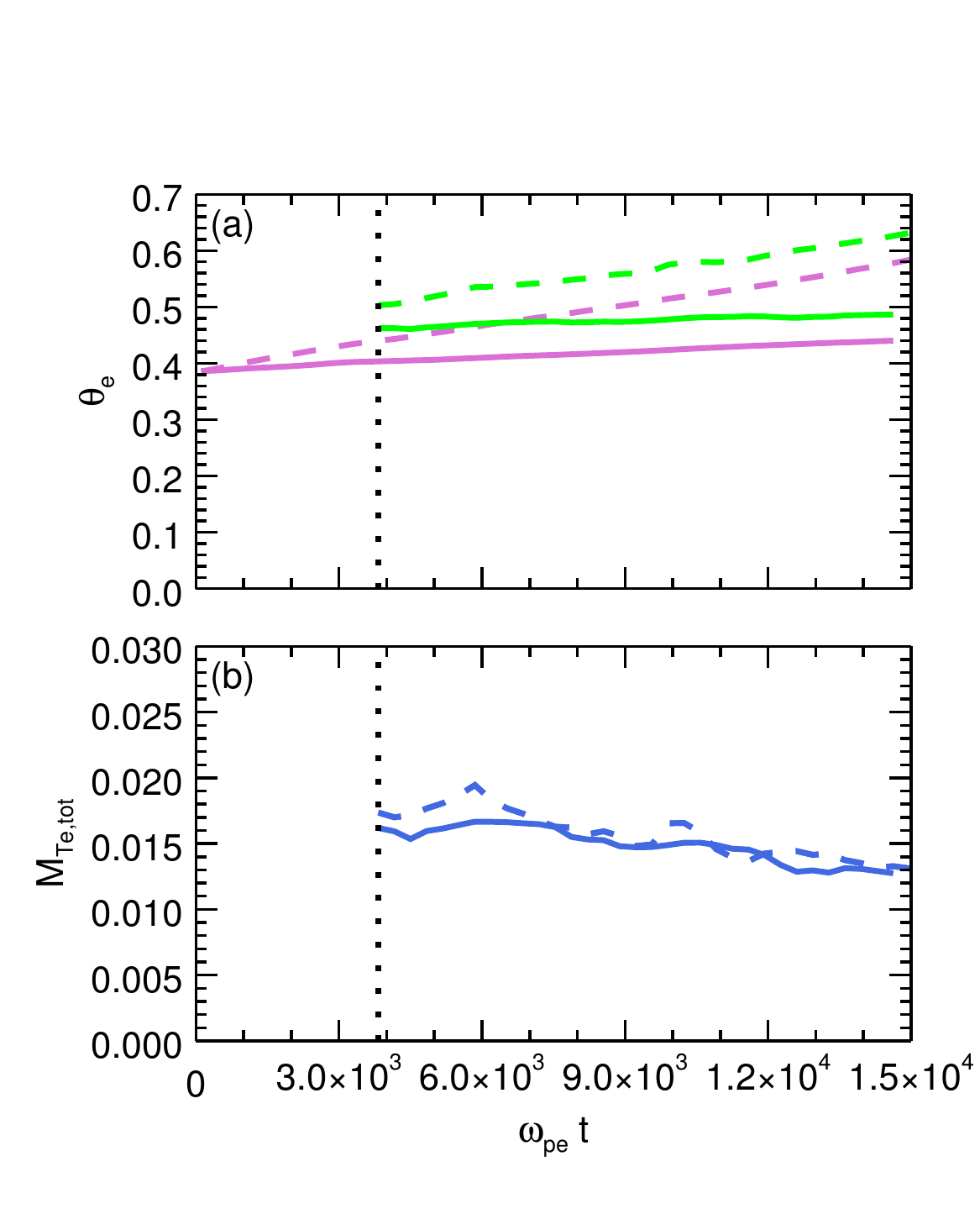} \\ 	
			\caption{Comparison of two simulations with $N_{\rm ppc}=16$ (dashed lines) and $N_{\rm ppc}=64$ (solid lines), having the same physical parameters: $\beta_{\rm i}=2, T_{\rm e}/T_{\rm i}=0.1, \sigma_{w}=0.1,$ and $m_{\rm i}/m_{\rm e}=25.$ We present the time evolution of (a): dimensionless electron temperature, $\theta_{e}$ in the upstream (magenta) and downstream (green); (b): total electron heating fraction, $M_{T\rm e, tot}$.  The upstream and downstream regions show an increase in electron temperature as time evolves,  caused by numerical heating.  The impact of numerical heating is significantly reduced by employing $N_{\rm ppc}=64.$  The measured value of $M_{T\rm e,tot}$ is, however, largely unaffected by numerical heating (panel (b)).    \label{fig:ppc}} 
	\end{figure}

In Fig.~\ref{fig:ppc}, we compare two simulations with $m_{\rm i}/m_{\rm e}=25, \sigma_{w}=0.1,\beta_{\rm i}=2,$ and $T_{\rm e}/T_{\rm i}=0.1$, which is the case where numerical heating is the most serious.  One has $N_{\rm ppc}=16$ (dashed lines), and the other $N_{\rm ppc}=64$ (solid lines).  In both cases, the size of the domain is $L_{x} =4318\, c/\omega_{\rm pe}$.  
In panel (a), we show the time evolution of the dimensionless electron temperature in the upstream (magenta) and downstream (green) for $N_{\rm ppc}=16$ (dashed) and $N_{\rm ppc}=64$ (solid). The vertical black dotted line indicates the time at which primary reconnection wavefronts recede past the region selected for our measurements (see Section \ref{sec:technique}).
The dimensionless electron temperature in both the upstream and downstream increases with time, however the amount of numerical heating is significantly less with $N_{\rm ppc}=64$ than with $N_{\rm ppc}=16.$  For example, the former shows a shift in downstream temperature (green) from $t \approx 4 \times 10^{3}\,\omega_{\rm pe}^{-1}$ to $ 1.5 \times 10^{4}\,\omega_{\rm pe}^{-1} $ of only $\Delta \theta_{\rm e} \approx 0.02,$ but for $N_{\rm ppc}=16$ the temperature shift is about six times larger.  The magenta lines show the analogous comparison for upstream temperatures.  For both choices of $N_{\rm ppc},$ the initial value of dimensionless electron temperature in the upstream is the same, but by $t = 1.5 \times 10^{4}\,\omega_{\rm pe}^{-1} ,$ they differ by  $\Delta \theta_{\rm e} \approx 0.15.$

In panel Fig.~\ref{fig:ppc}(b), we show the time evolution of the total electron heating fraction $M_{T\rm e,tot}$ for $N_{\rm ppc}=64$ (solid blue) and 16 (dashed blue).  Although numerical heating can significantly shift the measured values of dimensionless temperature in the downstream and upstream (panel (a)), we find that the  heating fractions are much less sensitive to the value of $N_{\rm ppc}$, with $N_{\rm ppc}=16$ already giving good results.  The heating fractions we measure are proportional to the difference between the downstream and upstream temperatures (or internal energy per particle), and it appears that the numerical heating in the downstream and upstream regions nearly cancels out in the difference.  Although we use $N_{\rm ppc}=64$ in simulations with $\beta_{\rm i} = 2,$ the agreement with the $N_{\rm ppc}=16$ case demonstrates that the impact of numerical heating is negligible for our measured heating fractions.  

We have tested the effect of numerical heating in a small box $(L_{x} \approx 1080\, c/\omega_{\rm pe})$ with up to $N_{\rm ppc}=256,$  however the difference (as regard to heating fractions) with respect to simulations with $N_{\rm ppc}=64,$ our standard choice for all $\beta_{\rm i}=2$ simulations, is again, negligible.

\section{Appendix E:  Anisotropy in the downstream}\label{sec:aniso}
We characterize the anisotropy in our simulations with ratios of the diagonal components of the stress-energy tensor,
\begin{align}
r_{x}&=T_{xx}/T_{\rm tot} \label{eq:rx} \\ 
r_{y}&=T_{yy}/T_{\rm tot} \label{eq:ry} \\
r_{z}&=T_{zz}/T_{\rm tot}, \label{eq:rz}
\end{align}
as seen in the fluid rest frame;  here, $T_{\rm tot} = (T_{xx} + T_{yy} + T_{zz})/3.$  As we show below, we typically measure anisotropies on the order of $5-10\%$ in the downstream, i.e., the reconnected plasma is nearly isotropic.

\begin{figure*}[h] 
		\centering
		\includegraphics[width=\textwidth,trim={0cm 3cm 0cm 0cm},clip]{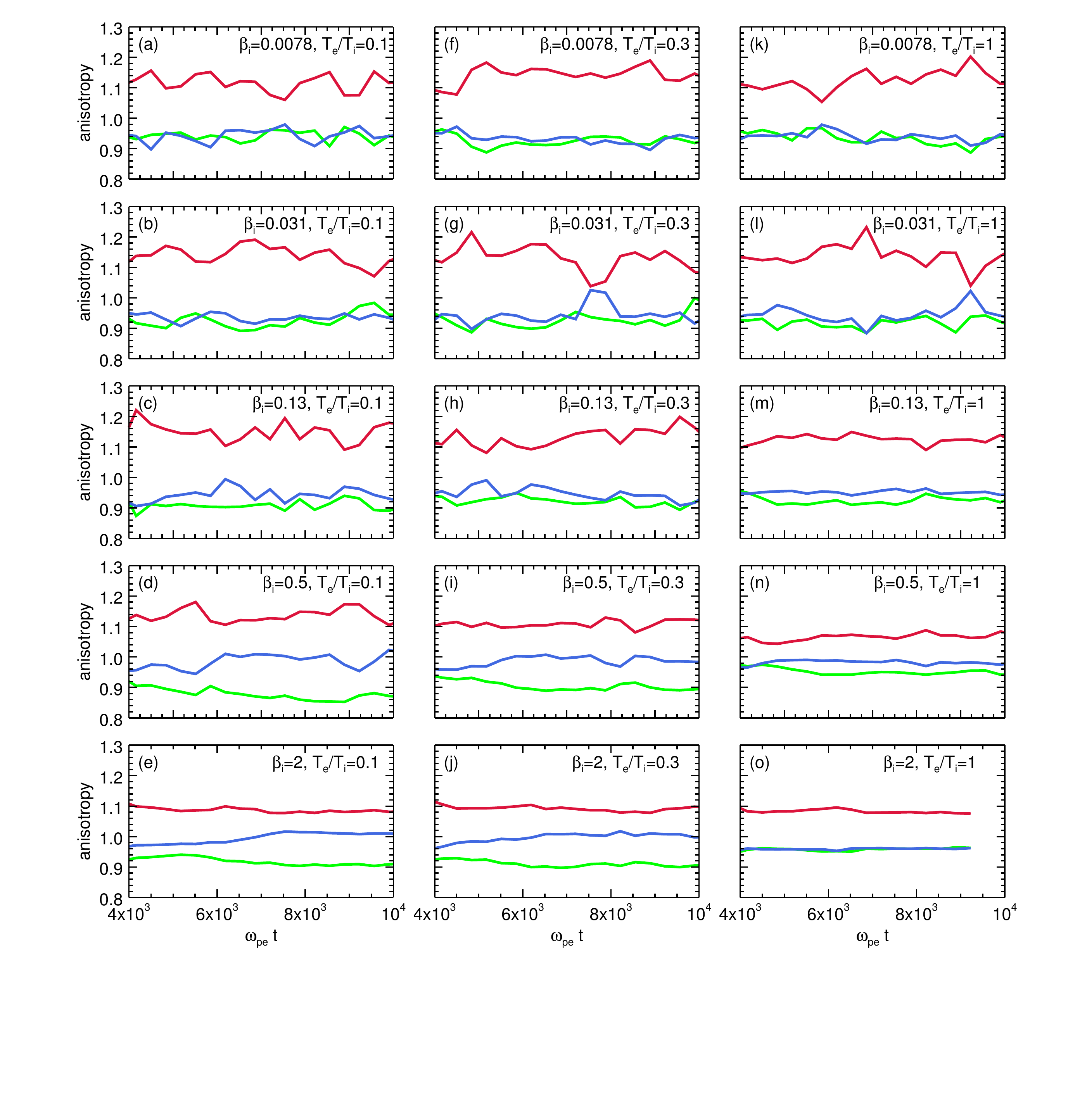} \\ 
			\caption{Time evolution of the anisotropy ratios in the reconnection downstream for a range of $\betai$ (increasing from top to bottom, as indicated in the legends) and $\teti$ (0.1 in the left column, 0.3 in the middle column, 1 in the right column). Here,  $\sigma_w=0.1$ and $\mime=25$.
			Red, green, and blue curves correspond to the ratios $r_{x}, r_{y},$ and $r_{z}$ (Eqs.~\ref{eq:rx}--\ref{eq:rz}).  Time evolution is shown starting at $\omega_{\rm pe}t=4 \times 10^{3},$ at which point the downstream region used for our heating measurements reaches a quasi-steady state.   \label{fig:aniso}}
	\end{figure*}

In Fig.~\ref{fig:aniso}, we show for $\sigma_w=0.1$ and $\mime=25$ the time evolution of the anisotropy ratios $r_{x}$ (red), $r_{y}$ (green), and $r_{z}$ (blue), for three temperature ratios $T_{\rm e}/T_{\rm i}=0.1,0.3,$ and $1$, and five values of $\beta_{\rm i}=0.0078, 0.031,0.13, 0.5,$ and $2$ ($\beta_{\rm i}$ and $T_{\rm e}/T_{\rm i}$ of the respective simulation are indicated at the top of each panel).  From top to bottom, $\beta_{\rm i}$ increases; from left to right, $T_{\rm e} / T_{\rm i}$ increases.  The temporal evolution starts from  $\omega_{\rm pe}t=4 \times 10^{3},$ when the downstream region reaches a quasi-steady state.  We find that the downstream pressures along the two directions transverse to the outflow ($\mathbf{\hat{y}}$ and $\mathbf{\hat{z}}$) are nearly identical, and slightly smaller than the pressure along the outflow direction ($\mathbf{\hat{x}}$, in our setup), which agrees with the findings of \citet{Shay2014}.  

\section{Appendix F:  Convergence of the layer width when varying the initial sheet thickness}\label{sec:recl_conv}%
\begin{figure*}[h] 
		\centering
		\includegraphics[width=0.45\textwidth]{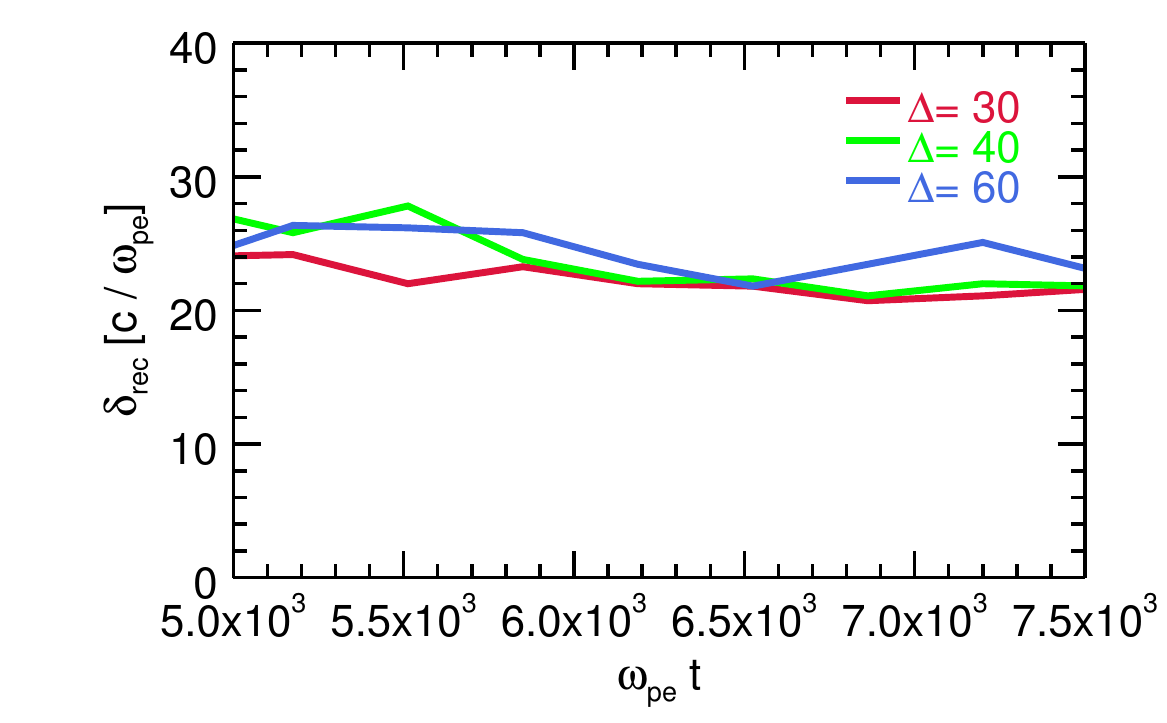} \\
			\caption{Time evolution of the reconnection layer width $\delta_{\rm rec}$ for a simulation with box size $L_{x}=2159\,c/\omega_{\rm pe}$, $\beta_{\rm i}=2,$ $T_{\rm e}/T_{\rm i}=1, \sigma_{w}=0.1,$ and $m_{\rm i}/m_{\rm e}=25.$  The value of $\delta_{\rm rec}$ is measured at $215\,c/\omega_{\rm pe}$ from the center. It does not depend on the choice of the initial current sheet thickness, $\Delta=30, 40, 60\,c/\omega_{\rm pe},$ shown by the red, green, and blue curves, respectively.  \label{fig:recl}} 
	\end{figure*}

In Fig.~\ref{fig:betainout}(e), we showed the $T_{\rm e}/T_{\rm i}$- and $\beta_{\rm i}$-dependence of the reconnection layer width $\delta_{\rm rec}$. 
As mentioned in Section \ref{sec:setup}, we set the initial current sheet thickness to be $\Delta = 40\,c/\omega_{\rm pe}$. A natural question is whether the measured value of $\delta_{\rm rec}$ is affected by the sheet thickness at initialization, or by the self-consistent reconnection physics alone.  To demonstrate that the measured values of $\delta_{\rm rec}$ do not depend on the initial current sheet thickness $\Delta,$ we show in Fig.~\ref{fig:recl} the time evolution of $\delta_{\rm rec}$ for $\Delta = 30$ (red), $40$ (green), and  $60\,c/\omega_{\rm pe}$ (blue).  Here, the box size is $L_{x}=2159\,c/\omega_{\rm pe}$, $\beta_{\rm i}=2,$ $T_{\rm e}/T_{\rm i}=1, \sigma_{w}=0.1,$ and $m_{\rm i}/m_{\rm e}=25.$  The reconnection width is measured at $215\,c/\omega_{\rm pe}$ downstream of the central X-point.  The $\Delta=40,\,60\,c/\omega_{\rm pe}$ curves have been shifted in time to account for the delayed onset of reconnection caused by the thicker initial current sheet.
The time evolution of $\delta_{\rm rec}$ in Fig. \ref{fig:recl} is shown starting at $t = 5000\,\omega_{\rm pe}^{-1} ,$ beyond which $\delta_{\rm rec}$ reaches a quasi-steady value. The three simulations converge to a similar value $\delta_{\rm rec} \approx 25\,c/\omega_{\rm pe}$, independent of the current sheet thickness at initialization.  

The values in this plot should not be directly compared to those in panel (e) of Fig.~\ref{fig:betainout}. Here, we extract $\delta_{\rm rec}$ at  a distance of $\approx215\,c/\omega_{\rm pe}$ downstream of the central X-point (in order to avoid the influence of the primary island sitting at the boundary), whereas in the larger box used in Fig.~\ref{fig:betainout}, $\delta_{\rm rec}$ was measured at $430\,c/\omega_{\rm pe}$ from the center.  Still, the results from the two experiments yield the same opening angle for the reconnection outflow.

	\section{Appendix G:  Heating efficiencies in terms of internal energy per particle}\label{sec:appmu}
\begin{figure*}[h] 
		\centering
		\includegraphics[width=\textwidth]{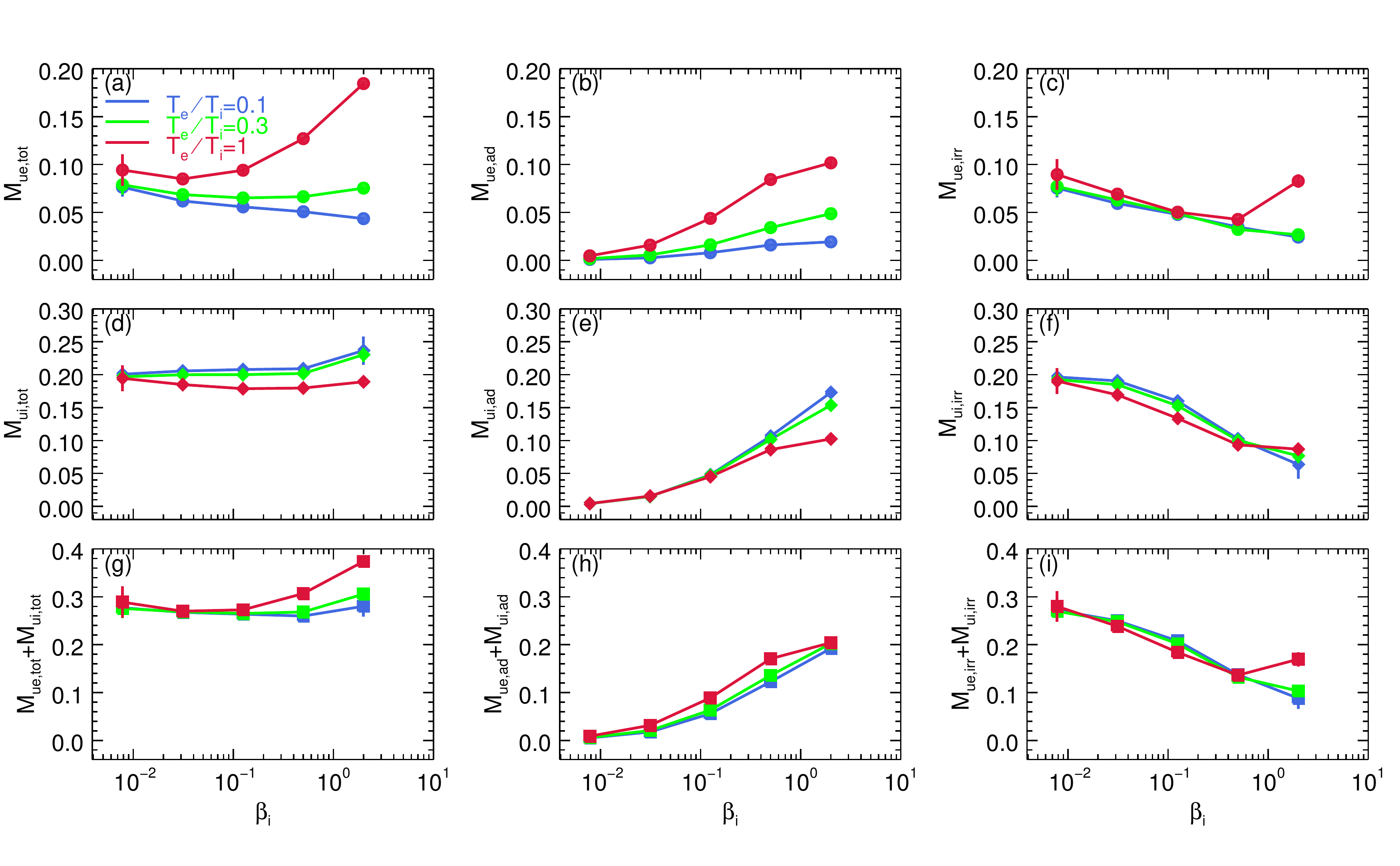} \\
			\caption{The layout here is analogous to Fig.~\ref{fig:mttt}, but for the internal energies, $u_{\rm e}, u_{\rm i}$ instead of temperatures, $T_{\rm e}, T_{\rm i}$. Plasma $\beta_{\rm i}$- and $T_{\rm e}/T_{\rm i}$-dependence of various heating efficiencies; (a): electron total, $M_{u\rm e, tot}$; (b): electron adiabatic, $M_{u\rm e,ad}$; (c): electron irreversible, $M_{u\rm e,irr}$; (d): proton total, $M_{u\rm i, tot}$; (e): proton adiabatic, $M_{u\rm i,ad}$; (f): proton irreversible, $M_{u\rm i,irr}$; (g): electron and proton total, $M_{u\rm e, tot}+M_{u\rm i, tot}$; (h): electron and proton adiabatic, $M_{u\rm e,ad}+M_{u\rm i,ad}$; (i): electron and proton irreversible, $M_{u\rm e,irr}+M_{u\rm i,irr}$.  The simulations shown here use a mass ratio $m_{\rm i}/m_{\rm e}=25$ and magnetization $\sigma_{w}=0.1$.  As in earlier plots, blue, green, and red points correspond to simulations with upstream $T_{\rm e}/T_{\rm i}$ ratios of $0.1, 0.3,$ and $1$.  \label{fig:mu}} 
	\end{figure*}

In the main body of the text, we phrased most of the heating fractions in terms of differences in temperature between downstream and upstream, but they can also be expressed in terms of differences in internal energy per particle.  In Fig.~\ref{fig:mu}, which is analogous to Fig.~\ref{fig:mttt}, we show the $T_{\rm e}/T_{\rm i}$- and $\beta_{\rm i}$-dependence of:
	electron heating fractions $M_{u\rm e, tot}, M_{u\rm e,ad}, M_{u\rm e,irr}$ (panels (a), (b), (c));  
	ion heating fractions $M_{u\rm i, tot}, M_{u\rm i,ad}, M_{u\rm i,irr}$ (panels (d), (e), (f));
	and total particle heating fractions $M_{u\rm e, tot} + M_{u\rm i, tot}, M_{u\rm e,ad} + M_{u\rm i,ad}, M_{u\rm e,irr} + M_{u\rm i,irr}$ (panels (g), (h), (i)).  
	As before, blue, green, and red lines denote temperature ratios $T_{\rm e}/T_{\rm i}=0.1,0.3,$ and $1$, and the simulations have $m_{\rm i}/m_{\rm e}=25$ and $\sigma_{w}=0.1$.
	Since the protons here are non-relativistic in both the upstream and downstream,  the points in panels (d) of Figs.~\ref{fig:mu} and \ref{fig:mttt} typically differ by a factor of $\hat{\gamma}_{\rm i}-1=2/3$ (excluding the $\beta_{\rm i}=2$ cases, for which the protons are mildly relativistic, with $\theta_{\rm i,up} \approx 0.4$), where $\hat{\gamma}_{\rm i}=5/3$ is the adiabatic index for a non-relativistic gas.
	The relationship between the two options for measuring the heating fractions of electrons, $M_{T\rm e,tot}$ and $M_{u\rm e,tot}$ in panels (a) of Figs.~\ref{fig:mu} and \ref{fig:mttt}, is not as simple because the electrons can be non-, trans- or ultra-relativistic.  For example, at $\beta_{\rm i}=2, T_{\rm e}/T_{\rm i}=1,$ the upstream and downstream dimensionless electron temperatures are $\theta_{\rm e,up} \approx \theta_{\rm e,down} \approx 10,$ and the adiabatic index is $\hat{\gamma}_{\rm e}\approx4/3$ in both the upstream and downstream.  The ratio of $M_{T\rm e,tot}$ to $M_{u\rm e,tot}$ is then $M_{T\rm e,tot}/M_{u\rm e,tot} \approx 1/3 = \hat{\gamma}_{\rm e} - 1$ for $\hat{\gamma}_{\rm e}=4/3.$  However, at low $\beta_{\rm i},$ electrons are less relativistic, and the ratio $M_{T\rm e,tot}/M_{u\rm e,tot}$ is typically larger because the adiabatic index is larger. Still, we remark that all of the conclusions presented in the paper hold when the heating efficiencies are measured using the internal energy per particle, rather than the temperature.
		
\clearpage
\bibliographystyle{apj}

\end{document}